\documentclass[epsfig,12pt]{article}
\usepackage{makeidx}
\usepackage{amsmath}
\usepackage{amsfonts}
\usepackage{amssymb}
\usepackage{graphicx}
\usepackage{accents}

\input epsf.sty
\newtheorem{theorem}{Theorem}
\newtheorem{acknowledgement}[theorem]{Acknowledgement}

\newtheorem{corollary}[theorem]{Corollary}

\newtheorem{proposition}[theorem]{Proposition}

\makeatletter \@addtoreset{equation}{section}

\input{tcilatex}
\begin{document}

\title{\rightline{\mbox{\small
{Lab/UFR-HEP0802/GNPHE/0802}}} \vspace{0.5cm} \textbf{BPS and non BPS} 
\textbf{7D Black Attractors }\\
\textbf{in M- Theory on K3}}
\author{ El Hassan\ Saidi{\small \thanks{%
h-saidi@fsr.ac.ma}} \\
%EndAName
{\small 1. \textit{Lab/UFR-Physique des Hautes Energies, Facult\'{e}
Sciences, Rabat,} Morocco,}\\
{\small 2. \textit{Groupement Nationale de Physique des Hautes Energies,
GNPHE, }}\\
{\small \textit{point focal, LPHE, Facult\'{e} Sciences, Rabat, Morocco}. }}
\maketitle

\begin{abstract}
We study the BPS and non BPS black attractors in 7D $\mathcal{N}=2$\
supergravity embedded in 11D M-theory compactified on K3.\ Combining Kahler
and complex moduli in terms of $SO\left( 3\right) $ representations, we
build the Dalbeault like (DL) basis for the second cohomology of K3 and set
up the fundamental relations of the special "hyperKahler" geometry of the
underlying moduli space of the 7D theory. We study the attractor eqs of the
7D black branes by using the method of the criticality of the effective
potential and also by using the extension of the so called 4D new attractor
approach to 7D $\mathcal{N}=2$ supergravity. A comment, regarding a 6D/7D
correspondence, along the line of Ceresole-Ferrara-Marrani used for 4D/5D 
\cite{FC}, is made.

\ \ \newline
\textbf{Key words}: \emph{11D M- theory on K3 and 7D N=2 supergravity,
Attractor eqs and 7D black branes, Special hyperkahler geometry, 6D/7D
correspondence.}
\end{abstract}

\tableofcontents

\section{Introduction}

\qquad The study of\ black attractors \textrm{\cite{BA1}-\cite{BA4}} in the
framework of\ compactifications of 10D superstrings and 11D M- theory has
been a subject of great interest. New classes of solutions to the attractor
equations (AEs) corresponding to BPS and non-BPS horizon geometries have
been obtained \textrm{\cite{BNB1}-\cite{BNB9}}; and many results regarding
extremal BPS and non BPS black holes in 4D extended supergravity theories
and higher dimensional space times have been derived both in the absence and
in the presence of fluxes \textrm{\cite{4D1}}-\textrm{\cite{4D65}; }see also 
\textrm{\cite{F2}} and refs therein. Several features of special Kahler
geometry (SKG) \textrm{\cite{SK1}-\cite{SK8}}, governing the physics of
extremal 4D black holes, have been uplifted to higher dimensions; in
particular to 5D and 6D with the underlying special real (SRG) and special
quaternionic$\footnote{%
In this paper, we will use the conventional notions: SRG, SKG, SHG, SQG.
They should be put in one to one correspondences with the number of real
scalars in the abelian vector multiplets of the non chiral $\mathcal{N}=2$
supersymmetric theory in 5D, 4D, 7D and 6D respectively.}$ (SQG) geometries
respectively \textrm{\cite{5D1}}-\textrm{\cite{5D4}}.

In this paper, we contribute to this matter; in particular to the issue
concerning the extremal 7D black attractors as well as to the special
hyperKahler geometry$\footnote{%
the \emph{"special hyperkahler geometry"} (SHG) should be understood in the
sense it has three Kahler 2-forms $\Omega _{a}=\left( \Omega _{1},\Omega
_{2},\Omega _{3}\right) $ with an $SO\left( 3\right) $ symmetry.}${\small \ (%
}SHG) underlying the physics of these extremal 7D black objects. More
precisely, we study the BPS and non BPS black attractors in 7D $\mathcal{N}%
=2 $ supergravity embedded in 11D M- theory compactified on K3 by using both
the criticality condition method as well as the so called "\emph{new
attractor}" approach introduced by Kallosh in the framework of 4D $\mathcal{N%
}=2$ supergravity and which we generalize here to the 7D theory. \newline
One of the key steps of this study is based on the use the $SO\left(
3\right) \times SO\left( 19\right) $ isotropy symmetry of the moduli space
of K3 
\begin{equation}
\boldsymbol{M}_{7D}^{\mathcal{N}=2}=\frac{SO\left( 3,19\right) }{SO\left(
3\right) \times SO\left( 19\right) }\times SO\left( 1,1\right) ,  \label{mo}
\end{equation}%
to build a real $22$ dimensional \emph{"Dalbeault like"} basis 
\begin{equation}
\left\{ \Omega _{a},\Omega _{I}\right\} _{I=1,...,19}^{a=1,2,3}  \label{dl}
\end{equation}%
for the second real cohomology group $H^{2}\left( K3,R\right) $. The real 2-
forms $\Omega _{a}$ and $\Omega _{I}$ transform respectively in the
representations $\left( \underline{3},\underline{1}\right) $ and $\left( 
\underline{1},\underline{19}\right) $\ of the $SO\left( 3\right) \times
SO\left( 19\right) $\ isotropy group of the moduli space $\boldsymbol{M}%
_{7D}^{N=2}$. The $\Omega _{a}$ and $\Omega _{I}$ may be compared with the
complex $\left( 1+h^{2,1}\right) $ Dalbeault basis 
\begin{equation*}
\begin{tabular}{llll}
$\Omega ^{\left( 3,0\right) }$ & , & $\Omega _{i}^{\left( 2,1\right) }$ & ,
\\ 
$\Omega ^{\left( 0,3\right) }$ & , & $\Omega _{i}^{\left( 1,2\right) }$ & ,%
\end{tabular}%
\qquad i=1,...,n=h^{2,1}\text{ },
\end{equation*}%
of $H^{3}\left( CY3,R\right) $ used in the compactification of type IIB
superstring on CY threefolds. \newline
With the $\left\{ \Omega _{a},\Omega _{I}\right\} $ basis at hand, we set up
the fundamental relations of the SHG of eq(\ref{mo}). We also study the
attractor equations for 7D black holes and black 3- branes. The solutions of
these eqs are obtained in the two above mentioned ways namely by directly
solving the critically conditions of the black brane potential and also by
extending the Kallosh \emph{new attractor }approach of 4D supergravity to
the 7D supersymmetric theory.

Recall that in the case of extremal black hole (BH) in 4D $\mathcal{N}=2$
supergravity realized in terms of 10D type IIB superstring on Calabi-Yau
threefolds, the BH effective scalar potential $\mathcal{V}_{BH}^{N=2}\left(
z,\overline{z},q,p\right) =\mathcal{V}_{BH}^{N=2}$ is given by the following
positive function,%
\begin{equation}
\mathcal{V}_{BH}^{N=2}=e^{\mathcal{K}}\left( \left\vert \mathcal{Z}%
\right\vert ^{2}+\sum_{i,j=1}^{n_{v}-1}\mathrm{g}^{i\overline{j}}\mathcal{Z}%
_{i}\overline{\mathcal{Z}}\overline{_{j}}\right) \geq 0.  \label{1}
\end{equation}%
where $n_{v}=\left( 1+h^{2,1}\right) $ is the number of 1-form gauge fields.
The function $\mathcal{K}=\mathcal{K}\left( z,\overline{z}\right) $ and $%
\mathrm{g}_{i\overline{j}}\sim \mathrm{\partial }_{i}\mathrm{\partial }_{%
\overline{j}}\mathcal{K}$ are respectively the Kahler potential and the
metric of the moduli space $\boldsymbol{M}_{4D}^{\mathcal{N}=2}$ of the 4D
supersymmetric theory. The function $\mathcal{Z}$ ($\overline{\mathcal{Z}}$)
is the holomorphic (antiholomorphic) central charge ($\mathcal{N}=2$
superpotential) and $\mathcal{Z}_{i}=D_{i}\mathcal{Z}$ is the matter central
charges given by the covariant derivative $\mathcal{Z}$ with respect to the
Kahler transformations. The (geometric) charge $\mathcal{Z}$ and the matter
ones $\mathcal{Z}_{i}$ are functions depending on the electric/magnetic
charges of the black hole and the moduli $z_{i}$ and $\overline{z}_{%
\overline{i}}$ parameterizing $\boldsymbol{M}_{4D}^{\mathcal{N}=2}$.

Using the basis $\left\{ \Omega _{a},\Omega _{I}\right\} $ and the fluxes of
the 4-form field strength $\mathcal{F}_{4}$ through the 4-cycles $S_{\infty
}^{2}\times \Psi ^{\Lambda }$ with $\Psi ^{\Lambda }\in H_{2}\left(
K3,R\right) $ and the 2-sphere $S_{\infty }^{2}$ in the 7D space time, we
show, amongst others, that the 7D black hole (black 3-brane) potential reads
as 
\begin{equation}
\mathcal{V}_{BH}^{7d,N=2}=\sum_{a,b=1}^{3}K^{ab}\left( e^{2\sigma }\left[
Z_{a}Z_{b}-\frac{1}{3}\sum_{I,J=1}^{n_{v}-3}g_{ab}^{IJ}Z_{I}Z_{J}\right]
\right) \geq 0,\qquad   \label{2}
\end{equation}%
where $n_{v}=b_{2}\left( K3\right) =22$ is the number of Maxwell gauge
fields, $\sigma $ is the dilaton parameterizing the $SO\left( 1,1\right) $
factor of $\boldsymbol{M}_{7D}^{\mathcal{N}=2}$ and $g_{IJ}^{ab}\left( \phi
\right) $ is the metric of the moduli space $\frac{SO\left( 3,19\right) }{%
SO\left( 3\right) \times SO\left( 19\right) }$ with fixed value of the
dilaton$\footnote{%
Due to the factorization of the moduli space of the 7D theory, the
dependence in the dilaton appears as a multiplicative global factor.}$ ($%
d\sigma =0$). The fundamental relations of the SHG of $\boldsymbol{M}_{7D}^{%
\mathcal{N}=2}$ are given by%
\begin{equation}
\begin{tabular}{llllll}
$\mathcal{K}_{ab}$ & $=$ & $\int_{K3}\Omega _{a}\wedge \Omega _{b}$ & , & $%
a,b=1,2,3$ &  \\ 
$\mathcal{K}_{IJ}$ & $=$ & $\int_{K3}\Omega _{I}\wedge \Omega _{J}$ & , &  & 
\\ 
$G_{aIbJ}$ & $=$ & $\int_{K3}\left( D_{aI}\Omega ^{c}\wedge D_{bJ}\Omega
^{d}\right) K_{cd}$ & , & $I,J=1,...,19$ & 
\end{tabular}
\label{kkg}
\end{equation}%
The field matrices $\mathcal{K}_{ab}\left( \sigma ,\phi \right) =e^{-2\sigma
}K_{ab}\left( \phi \right) $ and $\mathcal{K}_{IJ}\left( \sigma ,\phi
\right) =e^{-2\sigma }K_{ab}\left( \phi \right) $ are symmetric real
matrices and the moduli space metric $G_{aIbJ}\left( \sigma ,\phi \right)
=e^{-2\sigma }g_{aIbJ}\left( \phi \right) $ with the remarkable
factorization,%
\begin{equation}
g_{aIbJ}=K_{IJ}\times K_{ab},  \label{kkk}
\end{equation}%
and the flat limit $g_{aIbJ}\rightarrow \eta _{IJ}\times \delta
_{ab}=-\delta _{IJ}\times \delta _{ab}$. Putting this relation back into (%
\ref{2}), we can bring it to the remarkable form%
\begin{equation}
\mathcal{V}_{BH}^{7D,N=2}=\left( \sum_{a,b=1}^{3}\mathcal{K}%
^{ab}Z_{a}Z_{b}+\sum_{I,J=1}^{n_{v}-3}G^{IJ}Z_{I}Z_{J}\right) \geq 0,\qquad 
\end{equation}%
with $G^{IJ}=-\mathcal{K}^{IJ}$ and $\mathcal{K}^{ab}$ and $\mathcal{K}^{IJ}$
are as in eqs(\ref{kkg}).

The functions $Z_{a}=Z_{a}\left( \phi ,\sigma \right) $ and $%
Z_{I}=D_{I}^{a}Z_{a}$ are respectively the geometric and matter central
charges in 7D $\mathcal{N}=2$ supergravity; they play a quite similar role
to the $\mathcal{Z}$ and $\mathcal{Z}_{i}=D_{i}\mathcal{Z}$ of the 4D $%
\mathcal{N}=2$ supergravity theory. Notice that the expression of the
effective potential $\mathcal{V}_{BH}^{7D,N=2}$ for general 7D $\mathcal{N}%
=2 $ supergravity has been first considered by Cecotti, Ferrara and
Girardello in $\mathrm{\cite{SK11}}$. In our present study, the eq(\ref{2})
deals with 7D $\mathcal{N}=2$ supergravity \emph{embedded} in 11D M- theory
on K3 with $K^{ab}$ and $g_{ab}^{IJ}$ as in eqs(\ref{kkg}); and concerns the
geometric derivation of the 7D black hole (3-brane) attractor solutions
associated with eq(\ref{mo}).

We also determine the attractor eqs for the extremal $7D$ black hole
(3-brane) by extending the Kallosh attractor approach. In this set up, the
attractor eqs read in terms of the dressed charges $Z_{a}$ and $Z_{I}$, the $%
\left\{ \Omega _{a},\Omega _{I}\right\} $ basis and the matrix potentials $%
\mathcal{K}^{ab}$ and $\mathcal{K}^{IJ}$ (\ref{kkg}) as follows,%
\begin{equation}
\mathcal{H}_{2}=\mathcal{K}^{ab}Z_{a}\Omega _{b}+\mathcal{K}^{IJ}Z_{I}\Omega
_{J},  \label{F2}
\end{equation}%
where $\mathcal{H}_{2}$ is the real 2-form field strength given by $\mathcal{%
H}_{2}=\sum_{1}^{22}p^{\Lambda }\alpha _{\Lambda }$ with $p^{\Lambda }$
being integers and $\left\{ \alpha _{\Lambda }\right\} $ defining the Hodge
basis of $H^{2}\left( K3,\mathbb{R}\right) $. By integration of this
relation over the 2-cycles $\Psi ^{\Lambda }\in H_{2}\left( K3,R\right) $,
dual to $\left\{ \alpha _{\Lambda }\right\} $, we get the explicit
expression form of the attractor eqs.

The organization of this paper is as follows: In section 2, we give some
useful materials regarding extremal 7D black attractors and the
parametrization of the moduli space (\ref{mo}) \ In section 3, we study the
7D black hole and the 7D black 3-brane by first deriving the criticality
conditions of the effective potential and then solving the corresponding
attractor eqs. In section 4, we analyze some useful features of fields and
fluxes in 7D $\mathcal{N}=2$ supergravity embedded in 11D M- theory on K3;
in particular the issue regarding the gauge fields and matter
representations with respect to 7D $\mathcal{N}=2$ supersymmetry as well as
the $SO\left( 3\right) \times SO\left( 19\right) $ isotropy symmetry of the
moduli space (\ref{mo}). In section 5, we derive the basis $\left\{ \Omega
_{a},\Omega _{I}\right\} $ by using physical arguments and describe the
deformation tensor $\Omega _{aI}^{b}=D_{aI}\Omega ^{b}$ of the metric of K3.
In section 6, we derive the fundamental relations of the special
"hyperkahler" geometry of 11D M- theory on K3. In section 7, we develop the
new attractor approach for the case of 7D $\mathcal{N}=2$ supergravity
embedded in 11D M- theory on K3; and rederive the attractor eqs of the 7D
black hole and black 3- brane. In section 8, we give a conclusion and make a
discussion on 6D/7D correspondence along the field theoretical line of
Ceresole-Ferrara-Marrani used in \textrm{\cite{FC}} to deal with the 4D/5D
correspondence. In the appendix, we revisit the fundamental relations SKG of
4D $\mathcal{N}=2$ supergravity. This appendix completes the analysis of
sub-section 5.1 and allows to make formal analogies with the SHG relation
underlying 7D theory.

\section{Black attractors in 7D Supergravity}

\qquad We start by giving useful generalities on the various kinds of the
extremal 7D black attractors in $\mathcal{N}=2$ supergravity theory. Then we
describe\ the parametrization of the moduli space $\boldsymbol{M}_{7D}^{N=2} 
$. This step is important for the field theoretic derivation of the $%
H^{2}\left( K3,R\right) $ basis $\left\{ \Omega _{a},\Omega _{I}\right\} $
to be considered in section 5.

\subsection{Extremal 7D black attractors}

\qquad Generally speaking, there are different kinds of $\ $extended
supergravity theories in 7D space time \textrm{\cite{7DS1}-\cite{7DS3}}; the
most familiar ones \textrm{\cite{7DS1}} have $2\times 2^{3}=8+8$ conserved
supersymmetric charges captured by two real \emph{eight components} $%
SO\left( 1,6\right) $ spinors \textrm{Q}$_{\alpha }^{1}$ and \textrm{Q}$%
_{\alpha }^{2}$ that are rotated under the $USP\left( 2,R\right) $
automorphism\ group of the underlying 7D $\mathcal{N}=2$ superalgebra. A
particular class of these theories is given by the compactifications of 10D
superstrings and 11D M-theory. There, the matter fields have an
interpretation in terms of the coordinates of the moduli space of the
compactified theory. Below, we will focus our attention mainly on the 7D $%
\mathcal{N}=2$ supergravity \emph{embedded }11D M-theory on K3 with a moduli
space given by eq(\ref{mo}). Like in the case of black holes in 4D and 5D
dimensions, the 7D effective theory$\footnote{%
More precisely, the correspondence is as $4D\leftrightarrow 6D$ and $%
5D\leftrightarrow 7D$. The first ones have dyonic attractors, the second
ones haven't.}$ has also extremal BPS and non BPS black attractors that we
want to study here.

From the view of the field theory set up, we generally consider the 7D
extremal black attractors that are static, spherically and asymptotically
flat background solutions of 7D $\mathcal{N}=2$ supergravity. These
solutions breaks half ($\frac{1}{2}$BPS) or the total \emph{sixteen}
supersymmetric charges. \newline
In this case, we distinguish four basic kinds of\ extremal 7D \emph{black
p-brane} attractors related amongst others by the usual electric/magnetic
duality captured by the identity,%
\begin{equation}
p+p^{\prime }=3.
\end{equation}%
These \emph{black p-branes}, which may be BPS or non BPS states, are
classified as follows:

\ \newline
(\textbf{1}) a magnetic 7D black hole, ($0$-brane) with $22$ magnetic
charges $\left\{ p_{\Lambda }\right\} $,\newline
(\textbf{2}) an magnetic 7D black string, ($1$-brane), with a magnetic
charge g$_{0}$,\newline
(\textbf{3}) a electric 7D black membrane, ($2$-brane) with an electric
charge q$_{0}$,\newline
(\textbf{4}) an electric 7D black 3- brane, ($3$-brane) with $22$ electric
charges $\left\{ q_{\Lambda }\right\} $.

\ \newline
These asymptotically flat, static and spherical black $p$-branes have also
near horizons geometries given by the product of $AdS_{p+2}$ with the real
sphere $S^{5-p}$,%
\begin{equation}
AdS_{p+2}\times S^{5-p}\qquad \text{{\small with}\qquad }p=0,1,2,3,4.
\end{equation}%
Below we shall mainly deal with the magnetic 7D black hole and its dual
electric 7D black 3-brane. As we will see later on, these two solutions can
be elegantly embedded in M-theory compactification on K3. \newline
The magnetic F-string and its dual electric black membrane can be also
considered in the M-theory framework. They correspond respectively to M5
wrapping K3 (4-cycle) and M2 filling two space directions in the 7D space
time (0-cycle in K3).\newline
As noticed above, the extremal 7D black hole and 7D black 3-brane attractors
have either electric charges $\left\{ q_{\Lambda }\right\} $ or magnetic
charges $\left\{ p_{\Lambda }\right\} $. These charges stabilize the static
moduli at horizon of the attractor.%
\begin{equation}
\varphi ^{m}=\varphi ^{m}\left( r_{h},p_{\Lambda }\right) ,\qquad m=1,...,58%
\text{,}  \label{58}
\end{equation}%
where $r$ stands for the radial coordinate of the 7D space time and $r_{h}$
is the horizon radius: $r_{h}\equiv r_{horizon}$. The relation (\ref{58})
follows as the solution of the attractor eqs given by the minimization of
the effective attractor potential (\ref{2}) or also by using eq(\ref{F2}).

\subsection{Useful properties of $\boldsymbol{M}_{7D}^{N=2}$}

\qquad We first describe the self couplings of the scalars of the 7D $%
\mathcal{N}=2$ supergravity. Then, we make comments regarding the matrix
parametrization of the moduli space $\boldsymbol{M}_{7D}^{N=2}$. These
properties are useful to fix the ideas and they are also relevant for the
analysis to be developed in sections 5, 6 and 7.

\subsubsection{Metric of moduli space}

\qquad In eq(\ref{58}), the \emph{fifty eight} field variables $\varphi
^{m}\left( x\right) =\varphi ^{m}\left( x_{0},...,x_{6}\right) $ with $%
m=1,...,58$, are the real scalar fields of the 7D $\mathcal{N}=2$
supergravity embedded in 11D M-theory on K3. At the level of the
supergravity component fields Lagrangian density $\mathcal{L}_{7D}^{\mathcal{%
N}=2}$, these 7D scalar fields have typical self interactions involving the
space time field derivatives $\left( \partial _{\mu }\varphi ^{m}\right) $.
These interactions appear in $\mathcal{L}_{7D}^{\mathcal{N}=2}$ as follows,%
\begin{equation}
\mathcal{L}_{7D}^{\mathcal{N}=2}=-\frac{1}{2}\sqrt{-\mathrm{G}}\mathcal{R}-%
\frac{1}{2}\sum_{\mu ,\nu =0}^{6}\sqrt{-\mathrm{G}}\mathrm{G}^{\mu \nu
}\left( \sum_{n,m=1}^{58}G_{mn}\left[ \varphi \right] \partial _{\mu
}\varphi ^{m}\left( x\right) \partial _{\nu }\varphi ^{n}\left( x\right)
\right) +....
\end{equation}%
In this relation, the $7\times 7$ real matrix $\mathrm{G}_{\mu \nu }\left(
x\right) $ is the metric of the 7D space time with scalar curvature $%
\mathcal{R}$; and the $58\times 58$ real matrix $G_{mn}\left[ \varphi \right]
$ is the metric of the moduli space $\boldsymbol{M}_{7D}^{N=2}$ of the 11D
M-theory on K3.\newline
The field variables $\varphi ^{m}$ can be then imagined as real local
coordinates of the moduli space $\boldsymbol{M}_{7D}^{N=2}$ and the local
field coupling $G_{mn}$ as the symmetric metric of $\boldsymbol{M}_{7D}^{N=2}
$,%
\begin{equation}
dl^{2}=\sum_{m,n=1}^{58}G_{mn}d\varphi ^{m}d\varphi ^{n},\qquad   \label{ll}
\end{equation}%
with $d\varphi ^{m}=dx^{\mu }\left( \partial _{\mu }\varphi ^{m}\right) $
and $G_{mn}=G_{mn}\left( \varphi \right) $. Like in the case of the 4D $%
\mathcal{N}=2$ supergravity theory embedded in 10D type IIB superstring on
CY3s, it happens that the specific properties of the field metric, 
\begin{equation}
G_{mn}=G_{mn}\left[ \varphi \left( x\right) \right] ,  \label{met}
\end{equation}%
play also an important role in the study of BPS and non BPS 7D black
attractors. It is then interesting to give some useful properties regarding
this metric and the way it may be handled.\newline
First, notice that because of the factorization property of the moduli space 
$\boldsymbol{M}_{7D}^{N=2}$ 
\begin{equation}
\begin{tabular}{llll}
$\boldsymbol{M}_{7D}^{N=2}$ & $=$ & $G_{0}\times \left( G/H\right) $ & , \\ 
$G_{0}$ & $=$ & $SO\left( 1,1\right) $ & , \\ 
$G$ & $=$ & $SO\left( 3,19\right) $ & , \\ 
$H$ & $=$ & $H_{1}\times H_{2}$ & ,%
\end{tabular}%
\qquad   \label{gf}
\end{equation}%
and because of the isotropy symmetry of $\boldsymbol{M}_{7D}^{N=2}$ 
\begin{equation}
H_{1}\times H_{2}=SO\left( 3\right) \times SO\left( 19\right) ,
\end{equation}%
it is convenient to split the \emph{58} local coordinates $\varphi ^{m}$, in 
$SO\left( 3\right) \times SO\left( 19\right) $ representations, like%
\begin{equation}
\varphi ^{m}=\left( \sigma ,\phi ^{aI}\right) ,\qquad a=1,2,3;\text{\/\qquad 
}I=1,...,19\text{,}  \label{sf}
\end{equation}%
where $\left( aI\right) $ is a double index. In this splitting, the dilaton $%
\sigma $ is an isosinglet of $SO\left( 3\right) \times SO\left( 19\right) $;
it will be put aside. The $\phi ^{aI}$'s are in the $\left( \underline{3},%
\underline{19}\right) $ bi-fundamental, $\phi _{a}^{I}$ in $\left( 
\underline{3}^{t},\underline{19}\right) $ and so on; they will be discussed
below.\newline
Notice also that in the coordinate frame (\ref{sf}), the length element $%
dl^{2}$ (\ref{ll}) reads as follows 
\begin{equation}
dl^{2}=G_{\sigma \sigma }d\sigma d\sigma +2G_{\sigma \left( aI\right)
}d\sigma d\phi ^{aI}+G_{\left( aI\right) \left( bJ\right) }d\phi ^{aI}d\phi
^{bJ},  \label{dll}
\end{equation}%
and the local field metric tensor $G_{mn}$ decomposes like 
\begin{equation}
G_{mn}=\left( 
\begin{array}{cc}
G_{\sigma \sigma } & G_{\sigma \left( bJ\right) } \\ 
G_{\left( aI\right) \sigma } & G_{\left( aI\right) \left( bJ\right) }%
\end{array}%
\right) .
\end{equation}%
We will see later on that the $G_{\sigma \sigma }$, $G_{\left( aI\right)
\sigma }$ and $G_{\left( aI\right) \left( bJ\right) }$ component fields of
the metric read as%
\begin{equation}
\begin{tabular}{llll}
$G_{\sigma \sigma }$ & $=$ & $e^{-2\sigma }$ & , \\ 
$G_{\sigma \left( bJ\right) }$ & $=$ & $G_{\left( aI\right) \sigma }=0$ & ,
\\ 
$G_{\left( aI\right) \left( bJ\right) }$ & $=$ & $e^{-2\sigma }g_{\left(
aI\right) \left( bJ\right) }$ & ,%
\end{tabular}%
\end{equation}%
where the $57\times 57$ real matrix $g_{\left( aI\right) \left( bJ\right) }$
is a function of the field coordinates $\phi ^{aI}$, 
\begin{equation}
g_{\left( aI\right) \left( bJ\right) }=g_{\left( aI\right) \left( bJ\right)
}\left( \phi \right) .
\end{equation}%
To deal to the metric tensor of $\frac{SO\left( 3,19\right) }{SO\left(
3\right) \times SO\left( 19\right) }$, we will also use the following
relations 
\begin{equation}
\begin{tabular}{ll}
$g_{IJ}^{ab}=K^{ac}K^{ad}g_{\left( cI\right) \left( dJ\right) }$ & , \\ 
$g_{ab}^{IJ}=K^{IK}K^{JL}g_{\left( aK\right) \left( bL\right) }$ & ,%
\end{tabular}%
\end{equation}%
they will be rederived rigourously later on,. In these relations, the
symmetric matrices $K^{ab}$ and $K^{IK}$ appear then as field metric tensors
to rise and lower the corresponding indices. For simplicity, we will drop
out the brackets for the bi-fundamentals $\left( aI\right) $, $\left(
bJ\right) $; and write $g_{\left( aI\right) \left( bJ\right) }$ simply as $%
g_{aIbJ}$. \newline
One of the remarkable results to be derived in this paper is that the metric
tensor $g_{IJ}^{ab}$ of $\frac{SO\left( 3,19\right) }{SO\left( 3\right)
\times SO\left( 19\right) }$\ factorizes as 
\begin{equation}
g_{IJ}^{ab}\sim K^{ab}\times K_{IJ},  \label{faq}
\end{equation}%
where $K_{ab}$ and $K_{IJ}$ are as in eqs(\ref{kkg}).\newline
Notice moreover that performing a general coordinate transformation from a
curved coordinate frame $\left\{ \phi ^{m}\right\} $ to an inertial one $%
\left\{ \xi ^{\underline{m}}\right\} $; 
\begin{equation}
\phi ^{m}\qquad \rightarrow \qquad \xi ^{\underline{m}}\left( \phi \right)
,\qquad m=aI,  \label{gc}
\end{equation}%
and putting back into eq(\ref{dll}), we can usually rewrite the local field
metric (\ref{met}) as 
\begin{equation}
\begin{tabular}{llll}
$g_{mn}\left( \phi \right) $ & $=$ & $\dsum\limits_{k,l=1}^{57}\mathrm{\eta }%
_{\underline{k}\underline{l}}\left( \xi \right) \left( \frac{\partial \xi ^{%
\underline{k}}}{\partial \phi ^{m}}\right) \left( \frac{d\xi ^{\underline{l}}%
}{\partial \phi ^{n}}\right) $ & , \\ 
$\mathrm{\eta }_{\underline{k}\underline{l}}\left( \xi \right) $ & $=$ & $%
\dsum\limits_{m,n=1}^{57}g_{mn}\left( \phi \right) \left( \frac{\partial
\phi ^{m}}{\partial \xi ^{\underline{k}}}\right) \left( \frac{\partial \phi
^{n}}{d\xi ^{\underline{l}}}\right) $ & ,%
\end{tabular}%
\end{equation}%
or equivalently like%
\begin{equation}
\begin{tabular}{llllll}
$G_{mn}\left( \phi \right) $ & $=$ & $\dsum\limits_{\underline{i},\underline{%
j}=1}^{57}E_{m}^{\underline{i}}E_{n}^{\underline{j}}\eta _{\underline{i}%
\underline{j}}$ & $\qquad ,\qquad $ & $E^{\underline{i}}=\dsum%
\limits_{m=1}^{57}E_{m}^{\underline{i}}d\varphi ^{m}$ & , \\ 
$\eta _{\underline{i}\underline{j}}\left( \xi \right) $ & $=$ & $%
\dsum\limits_{m,n=1}^{57}E_{\underline{i}}^{m}E_{\underline{j}}^{n}G_{mn}$ & 
$\qquad ,\qquad $ & $E^{m}=\dsum\limits_{\underline{i}=1}^{57}E_{\underline{i%
}}^{m}d\xi ^{\underline{i}}$ & ,%
\end{tabular}%
\qquad   \label{fac}
\end{equation}%
where $E_{m}^{\underline{i}}=E_{m}^{\underline{i}}\left( \varphi ,\xi
\right) $ is the vielbein with the usual properties; in particular%
\begin{equation}
\dsum\limits_{\underline{i}=1}^{57}E_{m}^{\underline{i}}E_{\underline{i}%
}^{n}=\delta _{m}^{n},\qquad \dsum\limits_{m=1}^{57}E_{\underline{i}%
}^{m}E_{m}^{\underline{j}}=\delta _{\underline{i}}^{\underline{j}}.
\end{equation}%
Below, we shall think about the inertial coordinate frame $\left\{ \xi ^{%
\underline{m}}\right\} $ as the local coordinate of the tangent flat space $%
R^{3,19}$ and about $\mathrm{\eta }_{\underline{m}\underline{n}}$ as the
corresponding flat metric%
\begin{equation}
\mathrm{\eta }_{\underline{m}\underline{n}}=\left( 
\begin{array}{cc}
+\delta _{\underline{a}\underline{b}} & 0_{3\times 19} \\ 
0_{19\times 3} & -\delta _{\underline{I}\underline{J}}%
\end{array}%
\right) .
\end{equation}%
\newline
The factorization (\ref{fac}) can be also done for the metric $g_{aIbJ}$ and
its inverse $g^{cKdL}$. We have%
\begin{equation}
\begin{tabular}{llll}
$g_{aIbJ}$ & $=$ & $\mathrm{\eta }_{\underline{c}\underline{d}}$ $E_{aI}^{%
\underline{c}\underline{K}}$ $E_{bJ}^{\underline{d}\underline{L}}$ $\mathrm{%
\eta }_{\underline{K}\underline{L}}$ & , \\ 
$g_{aIbJ}$ $g^{cKdL}$ & $=$ & $\delta _{a}^{c}$ $\delta _{b}^{d}$ $\delta
_{I}^{K}$ $\delta _{J}^{L}$ & , \\ 
$g^{cKdL}$ & $=$ & $\mathrm{\eta }^{\underline{c}\underline{d}}$ $E_{%
\underline{c}\underline{d}}^{cK}$ $E_{\underline{K}\underline{L}}^{dL}$ $%
\mathrm{\eta }^{\underline{K}\underline{L}}$ & , \\ 
$E_{aI}^{\underline{c}\underline{K}}$ $E_{\underline{c}\underline{K}}^{bJ}$
& $=$ & $\delta _{a}^{b}$ $\delta _{I}^{J}$ & , \\ 
$E_{\underline{a}\underline{I}}^{cK}$ $E_{cK}^{\underline{b}\underline{J}}$
& $=$ & $\delta _{\underline{a}}^{\underline{b}}$ $\delta _{\underline{I}}^{%
\underline{J}}$ & ,%
\end{tabular}
\label{ge}
\end{equation}%
with%
\begin{equation}
E_{aI}^{\underline{c}\underline{K}}=E_{aI}^{\underline{c}\underline{K}%
}\left( \phi ,\xi \right) ,\qquad \phi \equiv \left( \phi ^{bJ}\right)
,\qquad \xi \equiv \left( \xi ^{\underline{b}\underline{J}}\right) ,
\end{equation}%
and $\left( aI\right) $ (resp. $\left( \underline{a}\underline{I}\right) $ )
referring to the curved (resp. inertial) coordinate indices and $E_{aI}^{%
\underline{c}\underline{K}}$ to the vielbein linking the two frames.\newline
Moreover, because of the $SO\left( 3\right) \times SO\left( 19\right) $
isotropy symmetry of $\boldsymbol{M}_{7D}^{\mathcal{N}=2}$, it also useful
to introduce the \emph{"small"} vielbeins $e_{a}^{\underline{c}}$, $e_{I}^{%
\underline{K}}$ and their inverses,%
\begin{equation}
\begin{tabular}{llllll}
$e_{a}^{\underline{c}}e_{\underline{c}}^{b}=\delta _{a}^{b}$ & $\quad ,\quad 
$ & $e_{\underline{a}}^{c}e_{c}^{\underline{b}}=\delta _{\underline{a}}^{%
\underline{b}}$ & $\quad ,\quad $ & $e_{a}^{\underline{c}}=e_{a}^{\underline{%
c}}\left( \phi ,\xi \right) $ & , \\ 
&  &  &  &  &  \\ 
$e_{I}^{\underline{K}}e_{\underline{K}}^{J}=\delta _{I}^{J}$ & $\quad ,\quad 
$ & $e_{\underline{I}}^{K}e_{K}^{\underline{J}}=\delta _{\underline{I}}^{%
\underline{J}}$ & $\quad ,\quad $ & $e_{I}^{\underline{K}}=e_{I}^{\underline{%
K}}\left( \phi ,\xi \right) $ & .%
\end{tabular}
\label{vie}
\end{equation}%
With these $e_{a}^{\underline{c}}$ and $e_{I}^{\underline{K}}$ vielbeins, we
can build new geometrical objects; in particular the following ones,%
\begin{eqnarray}
K_{ab} &=&e_{a}^{\underline{c}}e_{b}^{\underline{d}}\eta _{\underline{a}%
\underline{d}},\qquad \mathrm{\eta }_{\underline{a}\underline{b}}=e_{%
\underline{a}}^{c}e_{\underline{b}}^{d}K_{cd},  \notag \\
&&  \label{kk} \\
K_{IJ} &=&e_{I}^{\underline{K}}e_{J}^{\underline{L}}\eta _{\underline{K}%
\underline{L}},\qquad \mathrm{\eta }_{\underline{I}\underline{J}}=e_{%
\underline{I}}^{K}e_{\underline{J}}^{L}K_{KL},  \notag
\end{eqnarray}%
where $K_{ab}$ and $K_{IJ}$ are precisely the matrices used in eqs(\ref{faq}%
). All these relations will be rigourously rederived later on in the SHG set
up.

\subsubsection{Matrix formulation}

\qquad In the above analysis, we have used \emph{58 }$=$ \emph{1+57} curved
coordinates $\left\{ \sigma ,\phi _{bJ}\right\} $ to parameterize the moduli
space $SO\left( 1,1\right) \times \frac{SO\left( 3,19\right) }{SO\left(
3\right) \times SO\left( 19\right) }$. These \emph{58} field coordinate
variables are independent variables; but exhibit non linear interactions
captured by the metric tensor $G_{mn}$ of the moduli space. \newline
A different, but equivalent, way to deal with the parametrization of $%
\boldsymbol{M}_{7D}^{N=2}$ is to consider a constrained linear matrix
formulation. This formulation is useful in the analysis of the criticality
conditions of the 7D black attractor potential and in the study SHG of the
moduli space vacua of 7D $\mathcal{N}=2$ supergravity. Let us give some
details on this approach.\newline
The idea of the matrix formulation is based on siting in a local patch $%
\mathcal{U}$ of the curved moduli space $\boldsymbol{M}_{7D}^{N=2}$, do the
calculations we need; and then use general coordinate transformations (\ref%
{gc}) to cover $\boldsymbol{M}_{7D}^{N=2}$.\newline
To begin, consider a local patch $\mathcal{U}$ of the group manifold $%
SO\left( 1,1\right) \times SO\left( 3,19\right) $ together with a real
matrix $R=\ln M$ where, 
\begin{equation}
M\in SO\left( 1,1\right) \times SO\left( 3,19\right) .
\end{equation}%
The matrix $R$, or equivalently $M$, captures too much degrees of freedom as
needed by $\boldsymbol{M}_{7D}^{N=2}$ since, 
\begin{equation}
\dim \left[ SO\left( 1,1\right) \times SO\left( 3,19\right) \right] =1+\frac{%
22\times 21}{2},
\end{equation}%
that is $232$ is real degrees of freedom. The reduction of this number down
to $1+57$ is ensured by gauging out the degrees of freedom associated with
the isotropy sub-symmetry $SO\left( 3\right) \times SO\left( 19\right)
\subset SO\left( 3,19\right) $. This means that the matrix $M$ should obey
the identifications,%
\begin{equation}
M\equiv \mathcal{O}^{t}M\mathcal{O},  \label{193}
\end{equation}%
with $\mathcal{O}\in SO\left( 3\right) \times SO\left( 19\right) .$

(\textbf{a}) \emph{constraint eqs}\newline
Because of the property (\ref{gf}), the matrix $M$ factorizes as the tensor
product%
\begin{equation}
M=P\otimes L\qquad
\end{equation}%
with%
\begin{equation}
P\text{ }\in \text{ }SO\left( 1,1\right) \text{ \ }\subset \text{ \ }%
End\left( \mathbb{R}^{1,1}\right) \qquad
\end{equation}%
and%
\begin{equation}
L\text{ }\in \text{ }SO\left( 3,19\right) \text{ \ }\subset \text{ \ }%
End\left( \mathbb{R}^{3,19}\right) \qquad .
\end{equation}%
The $2\times 2$ real matrix $P$ and the $22\times 22$ real matrix $L$
satisfy the orthogonality group relations,%
\begin{equation}
P^{t}\mathrm{\eta }_{{\small 2\times 2}}P=\mathrm{\eta }_{{\small 2\times 2}%
},  \label{pn}
\end{equation}%
\begin{equation}
L^{t}\mathrm{\eta }_{{\small 22\times 22}}L=\mathrm{\eta }_{{\small 22\times
22}},  \label{ln}
\end{equation}%
where 
\begin{equation}
\mathrm{\eta }_{{\small 2\times 2}}={\small diag}\left( {\small +1,-1}%
\right) \qquad ,\qquad \mathrm{\eta }_{{\small 22\times 22}}={\small diag}%
\left( {\small +++,-\cdots -}\right)
\end{equation}%
are respectively the metric tensors of the flat $\mathbb{R}^{1,1}$ and $%
\mathbb{R}^{3,19}$ spaces.

(\textbf{b}) \emph{solving eq}(\ref{pn})\newline
The orthogonality constraint equation $P^{t}\mathrm{\eta }_{{\small 2\times 2%
}}P=\mathrm{\eta }_{{\small 2\times 2}}$ is solved like%
\begin{equation}
P\left( \sigma \right) =\left( 
\begin{array}{cc}
\cosh \sigma & \sinh \sigma \\ 
\sinh \sigma & \cosh \sigma%
\end{array}%
\right) =e^{\sigma J},\qquad \sigma \in \mathbb{R},
\end{equation}%
with $\sigma J=\ln P$ and 
\begin{equation}
J=\left( 
\begin{array}{cc}
0 & 1 \\ 
1 & 0%
\end{array}%
\right) ,
\end{equation}%
being the generator of $SO\left( 1,1\right) $.\newline
The condition $L^{t}\mathrm{\eta }L=\mathrm{\eta }$ and the $SO\left(
3\right) \times SO\left( 19\right) $ isotropy symmetry require however more
analysis. Below, we give details

(\textbf{c}) \emph{solving the condition }(\ref{ln})\emph{\ }\newline
First notice that the \emph{the condition }$L^{t}\mathrm{\eta }L=\mathrm{%
\eta }$ on the matrix $L_{\underline{\Lambda }}^{\underline{\Sigma }}$ can
be interpreted in terms of invariance of vector norms in $\mathbb{R}^{3,19}$%
. The matrix $L_{\underline{\Lambda }}^{\underline{\Sigma }}$ rotates real
vectors $\mathrm{v}^{\underline{\Upsilon }}$ of $\mathbb{R}^{3,19}$, 
\begin{equation}
L_{\underline{\Upsilon }}^{\underline{\Sigma }}:\mathrm{v}^{\underline{%
\Upsilon }}\in \mathbb{R}^{3,19}\qquad \rightarrow \qquad L_{\underline{%
\Upsilon }}^{\underline{\Sigma }}\mathrm{v}^{\underline{\Upsilon }}\in 
\mathbb{R}^{3,19}
\end{equation}%
Invariance of the norm $\left\Vert \mathrm{v}^{\underline{\Upsilon }%
}\right\Vert $ requires the condition (\ref{ln}); i.e $L\in SO\left(
3,19\right) $. \newline
Then, we use the $\left( 3,19\right) $ signature of the $\mathbb{R}^{3,19}$
space to decompose the real matrix $L$ as follows%
\begin{equation}
L=\left( 
\begin{array}{cc}
A & B \\ 
C & D%
\end{array}%
\right) ,\qquad L^{t}=\left( 
\begin{array}{cc}
A^{t} & C^{t} \\ 
B^{t} & D^{t}%
\end{array}%
\right) ,  \label{sp}
\end{equation}%
with $A$ ($A^{t}$) and $D$ ($D^{t}$) being respectively $3\times 3$ and $%
19\times 19$ \emph{invertible} square matrices ($\det A\det D\neq 0$); while 
$B$ $\left( C^{t}\right) $ and $C$ $\left( B^{t}\right) $\ are $3\times 19$
and $19\times 3$ rectangular matrices (bi-fundamentals). \newline
Next, we put (\ref{sp}) back into $L^{t}\mathrm{\eta }L=\mathrm{\eta }$ to
end with the following constraint eqs on the sub-matrices A, B, C and D:%
\begin{equation}
\begin{tabular}{llll}
$A^{t}B=C^{t}D$ & $\qquad ,\qquad $ & $B^{t}A=D^{t}C$ & , \\ 
$C^{t}C=A^{t}A-I_{3}$ & $\qquad $,$\qquad $ & $B^{t}B=D^{t}D-I_{19}$ & ,%
\end{tabular}
\label{abcd}
\end{equation}%
where $I_{d}$ stands for identity matrix in \emph{d- dimensions}. \newline
Observe that these constraint relations are invariant under transposition
since 
\begin{equation}
\left( L^{t}\mathrm{\eta }L\right) ^{t}=L^{t}\mathrm{\eta }L,\qquad \mathrm{%
\eta }^{t}=\mathrm{\eta .}
\end{equation}%
The constraint eq(\ref{ln}) and eqs(\ref{abcd}) capture then 
\begin{equation}
\frac{22\times 23}{2}=243,
\end{equation}%
conditions restricting the initial $484$ initial number of degrees of
freedom down to 
\begin{equation}
484-253=231=\dim SO\left( 3,19\right) .
\end{equation}%
In the language of $SO\left( 3,19\right) $ group representations, the matrix 
$L$ corresponds to the reducible representation \underline{$22$}$\times 
\underline{22}^{t}$ which decomposes as 
\begin{equation}
\underline{22}\times \underline{22}^{t}=\left[ \underline{22}\times 
\underline{22}^{t}\right] _{s}\oplus \left[ \underline{22}\times \underline{%
22}^{t}\right] _{a}.  \label{as}
\end{equation}%
The constraint relation $L^{t}\mathrm{\eta }L=\mathrm{\eta }$ corresponds to
setting the \emph{symmetric} part as in eqs(\ref{abcd}). The latter may be
solved in different manners. A particular way to do it is to choose the
matrices A and D as follows%
\begin{equation}
\begin{tabular}{lllll}
$A$ & $=\lambda I_{3}$ & $\qquad ,\qquad $ & $\lambda =\sqrt{\left( 1+\frac{%
\alpha ^{2}}{3}\right) }$ & , \\ 
$D$ & $=\varrho I_{19}$ & $\qquad ,\qquad $ & $\varrho =\sqrt{\left( 1+\frac{%
\alpha ^{2}}{19}\right) }$ & ,%
\end{tabular}%
\qquad  \label{lr}
\end{equation}%
where $\alpha $ is a non zero real number to be identified as the norm of $B$%
. Then solve the constraint eqs(\ref{abcd}) as follows:%
\begin{equation}
\begin{tabular}{lll}
$C^{t}=\frac{\lambda }{\varrho }B=\sqrt{\frac{19\left( 3+\alpha ^{2}\right) 
}{3\left( 19+\alpha ^{2}\right) }}B$ & $\qquad ,\qquad $ & $Tr\left(
B^{t}B\right) =\alpha ^{2}.$%
\end{tabular}
\label{bc}
\end{equation}%
From this solution, we see that the degrees of freedom of the sub-matrices
A, C and D are completely expressed in terms of those \emph{57} degree of
freedom captured by B.\newline

(\textbf{d}) \emph{Gauging out} $SO\left( 3\right) \times SO\left( 19\right) 
$ \emph{isotropy}\newline
To get the appropriate constraint relations that fix the $SO\left( 3\right)
\times SO\left( 19\right) $ isotropy symmetry of the moduli space,\emph{\ }%
it is interesting to use the $\left( 3,19\right) $ signature of $\mathbb{R}%
^{3,19}$ to decompose the $SO\left( 3,19\right) $ vectors 
\begin{equation}
\underline{22}=\left( \underline{3},\underline{1}\right) \oplus \left( 
\underline{1},\underline{19}\right) \qquad ,\qquad \underline{22}^{t}=\left( 
\underline{3}^{t},\underline{1}\right) \oplus \left( \underline{1},%
\underline{19}^{t}\right) .  \label{ss}
\end{equation}%
Then, compute the two terms of eq(\ref{as}). We have%
\begin{eqnarray}
\left[ \underline{22}\times \underline{22}^{t}\right] _{s} &=&\left( \left[ 
\underline{3}\times \underline{3}^{t}\right] _{s},\underline{1}\right)
\oplus \left( \underline{1},\left[ \underline{19}\times \underline{19}^{t}%
\right] _{s}\right)  \notag \\
&&\oplus \left[ \left( \underline{3}\times ,\underline{19}^{t}\right) \oplus
\left( \underline{3}^{t},\underline{19}\right) \right] ,
\end{eqnarray}%
and%
\begin{eqnarray}
\left[ \underline{22}\times \underline{22}^{t}\right] _{a} &=&\left( \left[ 
\underline{3}\times \underline{3}^{t}\right] _{a},\underline{1}\right)
\oplus \left( \underline{1},\left[ \underline{19}\times \underline{19}^{t}%
\right] _{a}\right)  \notag \\
&&\oplus \left[ \left( \underline{3}\times ,\underline{19}^{t}\right)
\ominus \left( 3^{t},\underline{19}\right) \right] .  \label{st}
\end{eqnarray}%
In this set up, the constraint eqs(\ref{abcd}) and (\ref{193}) split as
follows%
\begin{equation}
\begin{tabular}{llll}
$\left[ \underline{3}\times \underline{3}^{t}\right] _{s}$ & $\qquad
\rightarrow \qquad $ & $\text{identiy }\lambda I_{3}$ & , \\ 
$\left[ 19\times \underline{19}^{t}\right] _{s}$ & $\qquad \rightarrow
\qquad \text{\ }$ & $\text{identiy }\varrho I_{19}$ & ,%
\end{tabular}
\label{smm}
\end{equation}%
and 
\begin{equation}
\left( \underline{3},\underline{19}^{t}\right) \equiv \left[ \left( 
\underline{3}^{t},\underline{19}\right) \right] ^{t}.
\end{equation}%
Notice in passing that the $SO\left( 3\right) \times SO\left( 19\right) $
isotropy symmetry can be usually used to set%
\begin{equation}
\begin{tabular}{llll}
$\left[ \underline{3}\times \underline{3}^{t}\right] _{a}$ & $\qquad
\rightarrow \qquad $ & $\text{0}$ & , \\ 
$\left[ \underline{19}\times \underline{19}^{t}\right] _{a}$ & $\qquad
\rightarrow \qquad \text{\ }$ & $\text{0}$ & .%
\end{tabular}%
\   \label{an}
\end{equation}%
Eqs(\ref{an}) reduce the previous $231=\dim SO\left( 3,19\right) $ number of
degrees of freedom down to 
\begin{equation}
231-\dim SO\left( 3\right) -\dim SO\left( 19\right) ,
\end{equation}%
that is $231-3-171=57$.\newline
To conclude this section, notice that a typical matrix $M$ of the coset $%
SO\left( 1,1\right) \times \frac{SO\left( 3,19\right) }{SO\left( 3\right)
\times SO\left( 19\right) }$ can be usually put in the form\footnote{%
Notice that the factorization $M\left( \sigma ,\xi \right) =e^{-\sigma
}L\left( \xi \right) $ takes regular values for $\sigma $ finite and is
singular for $\sigma \rightarrow \infty $. This difficulty will be avoided
by restricting the analysis to $\sigma $ finite.} 
\begin{equation}
M_{\underline{\Lambda }\underline{\Sigma }}\left( \sigma ,\xi \right)
=e^{-\sigma }L_{\underline{\Lambda }\underline{\Sigma }}\left( \xi \right) .
\end{equation}%
where $\sigma $ stands for the dilaton. The matrix $L_{\underline{\Lambda }%
\underline{\Sigma }}\left( \xi \right) $ obeys the orthogonality constraint
eq(\ref{ln}) and gauge symmetries under $SO\left( 3\right) \times SO\left(
19\right) $ transformations.\newline
Two ways to deal with these constraints:\newline
(\textbf{i}) solve the constraint eqs as we have done here above to find at
the end that the propagating degrees of freedom captured by $L_{\underline{%
\Lambda }}^{\underline{\Sigma }}$ are given by 
\begin{equation}
L_{\underline{\Lambda }}^{\underline{\Sigma }}=\left( 
\begin{array}{cc}
\lambda I_{3} & \varrho \mathrm{B} \\ 
\lambda \mathrm{B}^{t} & \varrho I_{19}%
\end{array}%
\right) ,  \label{ql}
\end{equation}%
with $\lambda $ and $\varrho $ as in eqs(\ref{lr}-\ref{bc}). This way of
doing is interesting from the view that it allows to fix the ideas; it will
be also used later on to motivate the basis $\left\{ \Omega _{a},\Omega
_{I}\right\} $ (\ref{dl}) for the second real cohomology of K3. As we will
see in section 5, the field moduli captured by eq(\ref{ql}) can be
interpreted as the periods, 
\begin{equation}
\begin{tabular}{llllllll}
$\lambda $ $\mathrm{\eta }_{\underline{a}}^{\underline{b}}$ & $\ \ \ \sim $ $%
\ \ \ $ & $\int_{B^{\underline{b}}}\Omega _{\underline{a}}$ & $\qquad
,\qquad $ & $\lambda $ $\mathrm{\xi }_{\underline{a}}^{\underline{I}}$ & $\
\ \ \sim $ $\ \ \ $ & $\int_{B^{\underline{I}}}\Omega _{\underline{a}}$ & ,
\\ 
$\varrho $ $\mathrm{\eta }_{\underline{I}}^{\underline{J}}$ & $\ \ \ \sim $ $%
\ \ \ $ & $\int_{B^{\underline{J}}}\Omega _{\underline{I}}$ & $\qquad
,\qquad $ & $\varrho $ $\mathrm{\xi }_{\underline{I}}^{\underline{a}}$ & $\
\ \ \sim $ $\ \ \ $ & $\int_{B^{\underline{a}}}\Omega _{\underline{I}}$ & ,%
\end{tabular}
\label{lrr}
\end{equation}%
where the 2-cycle basis $\left\{ B^{\underline{b}},B^{\underline{J}}\right\} 
$ is the dual of $\left\{ \Omega _{a},\Omega _{I}\right\} $. The symbols $%
\mathrm{\eta }_{\underline{a}}^{\underline{b}}$ and $\mathrm{\eta }_{%
\underline{I}}^{\underline{J}}\ $designate respectively the $3\times 3$ and $%
19\times 19$ identity matrices; i.e$\ \mathrm{\eta }_{\underline{a}}^{%
\underline{b}}=\mathrm{\delta }_{\underline{a}}^{\underline{b}}$, $\mathrm{%
\eta }_{\underline{I}}^{\underline{J}}=\mathrm{\delta }_{\underline{I}}^{%
\underline{J}}$.\newline
(\textbf{ii}) use a manifestly matrix formulation based on the matrix $L_{%
\underline{\Lambda }}^{\underline{\Sigma }}=\left( L_{\underline{a}}^{%
\underline{\Sigma }},L_{\underline{I}}^{\underline{\Sigma }}\right) $
constrained as%
\begin{equation}
\begin{tabular}{llllll}
$\eta _{\underline{\Lambda }\underline{\Sigma }}L_{\underline{c}}^{%
\underline{\Lambda }}L_{\underline{d}}^{\underline{\Sigma }}$ & $=\eta _{%
\underline{c}\underline{d}}$ & \qquad ,\qquad & $L_{\underline{c}}^{%
\underline{\Lambda }}$ & $\equiv U_{\underline{c}}^{\underline{d}}L_{%
\underline{d}}^{\underline{\Lambda }}$ & , \\ 
$\eta _{\underline{\Lambda }\underline{\Sigma }}L_{\underline{K}}^{%
\underline{\Lambda }}L_{\underline{L}}^{\underline{\Sigma }}$ & $=\eta _{%
\underline{K}\underline{L}}$ & \qquad ,\qquad & $L_{\underline{I}}^{%
\underline{\Lambda }}$ & $\equiv U_{\underline{I}}^{\underline{J}}L_{%
\underline{J}}^{\underline{\Lambda }}$ & ,%
\end{tabular}
\label{qq}
\end{equation}%
but without solving the constraints explicitly. These constraint eqs will be
fulfilled by requiring full gauge invariance at the level of physical
observables. This way of doing is powerful; we will use it in what follows
to study the extremal 7D black attractors.

\section{Black hole and black 3-brane}

\qquad In this section, we first study explicitly the BPS and non BPS black
holes in $\mathcal{N}=2$ 7D supergravity theory. Then, we give the key
relations for their dual 7D BPS and non BPS black 3-branes.

\subsection{Extremal 7D black holes}

\qquad In the 11D M-theory set up, 7D black holes are realized by wrapping a
M2 brane on the 2- cycles of K3. Since $\dim H^{2}\left( K3,\mathbb{R}%
\right) =b_{2}\left( K3\right) =22$, the 7D $\mathcal{N}=2$ supergravity has 
$U^{22}\left( 1\right) $ abelian gauge symmetry and the black hole has \emph{%
22} magnetic charges $p^{\Lambda }=\left( p^{1},...,p^{22}\right) $; but no
electric charges $q_{\Lambda }$. \newline
The magnetic charges $\left\{ p^{\Lambda }\right\} $ are given by the
integral of the real 4-form flux density $\mathcal{F}_{4}$ through the 4-
cycles basis $S_{\infty }^{2}\times \Psi ^{\Lambda }$, 
\begin{equation}
p^{\Lambda }=\int_{S_{\infty }^{2}}\left( \int_{\Psi ^{\Lambda }}\mathcal{F}%
_{4}\right) ,\qquad \Lambda =1,...,22.  \label{flu}
\end{equation}%
In this relation, the real 4- form $\mathcal{F}_{4}$ is the gauge invariant
field strength associated the RR gauge field 3-form $\mathcal{C}_{3}$ of the
M2 brane; i.e 
\begin{equation}
\mathcal{F}_{4}=\emph{d}\mathcal{C}_{3}.
\end{equation}%
The 2- cycle basis $\left\{ \Psi ^{\Lambda }\right\} $ is a basis of 2-
cycles of K3, dual to the Hodge 2-forms $\mathbf{\alpha }_{\Lambda }$, and
the compact real surface $S_{\infty }^{2}$ is a large radius 2- sphere
contained in the 7D space time. For simplicity, we shall use the
normalization 
\begin{equation}
\int_{S_{\infty }^{2}}d^{2}s=1,
\end{equation}%
where the factor $\frac{1}{4\pi }$ has been absorbed in the measure $d^{2}s$%
. The field moduli $\varphi _{\text{{\small h}}}^{m}$, at the horizon $%
r=r_{h}$ of the the \emph{static} and \emph{spherical} 7D black hole
attractor, are determined by the charges $p^{\Lambda }$ of the black hole%
\begin{equation}
\varphi _{\text{{\small h}}}^{m}\equiv \varphi ^{m}\left( r_{\text{{\small h}%
}},p^{1},...,p^{22}\right) .
\end{equation}%
The explicit relation between $\varphi _{\text{{\small h}}}^{m}$ and the
charges $\left\{ p^{\Lambda }\right\} $ can be determined by solving the
criticality condition of the effective scalar potential eq(\ref{2}); it will
be given later on.

\subsubsection{Black hole potential}

\qquad Here, we give the explicit expression of the black hole potential in
two coordinate frames of the moduli space. First in the inertial coordinate
frame $\left\{ \xi ^{\underline{m}}\right\} $ where most of the calculations
will be done. Then, we give the results in the curved frame $\left\{ \varphi
^{m}\right\} $ by using general coordinates transformations on the moduli
space.

\paragraph{\textbf{(1)} \emph{Inertial frame }\newline
}

\qquad In the inertial coordinates frame$\footnote{%
Because of the factorization of the moduli space $\boldsymbol{M}_{7D}^{N=2}$
as $\frac{SO\left( 3,19\right) }{SO\left( 3\right) \times SO\left( 19\right) 
}$ times $SO\left( 1,1\right) $, we will mainly deal with the first factor
and thinking about $\xi ^{0}$ as just the dilaton $\sigma $. The constraint
eq coming from the factor $SO\left( 1,1\right) $ does bring anything new; it
will be solved as in eq(\ref{xxi}) and implemented directly.}$ $\left\{ \xi
^{\underline{m}}=\left( \xi ^{0},\xi ^{\underline{a}\underline{I}}\right)
\right\} $ of $\boldsymbol{M}_{7D}^{\mathcal{N}=2}$, the black hole
effective potential is given by the simple relation,%
\begin{equation}
\mathcal{V}_{BH}^{7d,N=2}=\sum_{a=1}^{3}\mathcal{Z}_{\underline{a}}\mathcal{Z%
}^{\underline{a}}+\sum_{I=1}^{19}\mathcal{Z}_{\underline{I}}\mathcal{Z}^{%
\underline{I}}.  \label{vef}
\end{equation}%
As required by supersymmetry, this function is a positive scalar potential
induced by the central charges $\mathcal{Z}_{\underline{a}}$ and $\mathcal{Z}%
_{\underline{I}}$ of the 7D $\mathcal{N}=2$ supergravity theory. The central
charges $\mathcal{Z}_{\underline{a}}$ and $\mathcal{Z}_{\underline{I}}$ are
real functions on moduli space, 
\begin{equation}
\mathcal{Z}_{\underline{a}}=\mathcal{Z}_{\underline{a}}\left( p_{\Lambda
},\xi ^{\underline{m}}\right) ,\qquad \mathcal{Z}_{\underline{I}}=\mathcal{Z}%
_{\underline{I}}\left( p_{\Lambda },\xi ^{\underline{m}}\right) ,
\end{equation}%
describing respectively the "geometric" and "matter" dressed charges. Their
explicit expression are given by the following dressed magnetic charges 
\begin{equation}
\begin{tabular}{lll}
$\mathcal{Z}^{\underline{a}}$ & $=\sum_{\Lambda =1}^{22}p^{\underline{%
\Lambda }}\mathcal{L}_{\underline{\Lambda }}^{\underline{a}}$ & , \\ 
$\mathcal{Z}^{\underline{I}}$ & $=\sum_{\Lambda =1}^{22}p^{\underline{%
\Lambda }}\mathcal{L}_{\underline{\Lambda }}^{\underline{I}}$ & .%
\end{tabular}
\label{zl}
\end{equation}%
The underlined indices $\underline{\Lambda },$ $\underline{a}$ and $%
\underline{I}$ refer to the inertial (flat) coordinates frame $\left\{ \xi
\right\} $; they are lowered and raised by the respective flat metric
tensors $\eta _{\underline{\Upsilon }\underline{\digamma }}$, $\eta _{%
\underline{a}\underline{b}}$ and $\eta _{\underline{I}\underline{J}}$ of the
flat spaces $\mathbb{R}^{3,19}$, $\mathbb{R}^{3}$ and $\mathbb{R}^{0,19}$ ,%
\begin{equation}
\eta _{\underline{\Upsilon }\underline{\digamma }}=\eta _{\underline{a}%
\underline{b}}\text{ }\oplus \text{ }\eta _{\underline{I}\underline{J}%
},\qquad \eta _{\underline{a}\underline{b}}=+\delta _{\underline{a}%
\underline{b}},\qquad \eta _{\underline{I}\underline{J}}=-\delta _{%
\underline{I}\underline{J}}.
\end{equation}%
In (\ref{zl}), the $\mathcal{L}_{\underline{\Lambda }}^{\underline{a}}$ and $%
\mathcal{L}_{\underline{\Lambda }}^{\underline{I}}$ are local field living
on $\boldsymbol{M}_{7D}^{\mathcal{N}=2}$;%
\begin{equation}
\mathcal{L}_{\underline{\Lambda }}^{\underline{a}}=\mathcal{L}_{\underline{%
\Lambda }}^{\underline{a}}\left( \sigma ,\xi _{\underline{b}\underline{J}%
}\right) ,\ \qquad \mathcal{L}_{\underline{\Lambda }}^{\underline{I}}=%
\mathcal{L}_{\underline{\Lambda }}^{\underline{I}}\left( \sigma ,\xi _{%
\underline{b}\underline{J}}\right) ,\ 
\end{equation}%
with the factorization, (see \emph{footnote 6}),%
\begin{equation}
\begin{tabular}{llll}
$\mathcal{L}_{\underline{\Lambda }}^{\underline{a}}=e^{-\sigma }L_{%
\underline{\Lambda }}^{\underline{a}}$ & $\qquad ,\qquad $ & $L_{\underline{%
\Lambda }}^{\underline{a}}=L_{\underline{\Lambda }}^{\underline{a}}\left(
\xi _{\underline{b}\underline{J}}\right) $ & , \\ 
$\mathcal{L}_{\underline{\Lambda }}^{\underline{I}}=e^{-\sigma }L_{%
\underline{\Lambda }}^{\underline{I}}$ & $\qquad ,$ & $L_{\underline{\Lambda 
}}^{\underline{I}}=L_{\underline{\Lambda }}^{\underline{I}}\left( \xi _{^{%
\underline{b}\underline{J}}}\right) $ & ,%
\end{tabular}
\label{LL}
\end{equation}%
where the dependence in the dilaton is completely factorized as $e^{-\sigma
} $. The fields $L_{\underline{\Lambda }}^{\underline{a}}$ and $L_{%
\underline{\Lambda }}^{\underline{I}}$ live mainly on the group manifold 
\begin{equation}
\frac{SO\left( 3,19\right) }{SO\left( 3\right) \times SO\left( 19\right) },
\end{equation}%
and capture \emph{57} propagating degrees of freedom. These matrices should
be thought of as the matrices $L$\emph{\ } of eq(\ref{sp}) constrained as,%
\begin{equation}
\begin{tabular}{llll}
$L_{\underline{a}}^{\underline{\Upsilon }}\eta _{\underline{\Upsilon }%
\underline{\digamma }}L_{\underline{b}}^{\underline{\digamma }}$ & $=$ & $%
\eta _{\underline{a}\underline{b}}$ & , \\ 
$L_{\underline{a}}^{\underline{\Upsilon }}\eta _{\underline{\Upsilon }%
\underline{\digamma }}L_{\underline{J}}^{\underline{\digamma }}$ & $=$ & $0$
& , \\ 
$L_{\underline{I}}^{\underline{\Upsilon }}\eta _{\underline{\Upsilon }%
\underline{\digamma }}L_{\underline{J}}^{\underline{\digamma }}$ & $=$ & $%
\eta _{\underline{I}\underline{J}}$ & .%
\end{tabular}
\label{lz}
\end{equation}%
A representation of the tensors $L_{\underline{a}}^{\underline{\Upsilon }}$
and $L_{\underline{I}}^{\underline{\Upsilon }}$ in terms of the coordinates $%
\xi _{\underline{a}}^{\underline{I}}$, solving the above orthogonality
constraint eqs, is given by (\ref{ql}).\newline

(\textbf{a}) \emph{Special properties of} $\mathcal{V}_{BH}^{7D,N=2}$\newline
The black hole potential $\mathcal{V}_{BH}^{7d,N=2}$ and its constituents
exhibit a set of remarkable features. We list below the useful ones:\newline
(\textbf{i}) \emph{isotropy symmetry}:\qquad \newline
The dressed central charges $\mathcal{Z}_{\underline{a}}$ and $\mathcal{Z}_{%
\underline{I}}$ behave as real vectors under the $SO\left( 3\right) \times
SO\left( 19\right) $ gauge isotropy symmetry of the moduli space (\ref{md}):%
\begin{equation}
\mathcal{Z}_{\underline{a}}\text{ \ \ }\sim \text{ \ \ }\left( \underline{3},%
\underline{1}\right) ,\qquad \mathcal{Z}_{\underline{I}}\text{ \ \ }\sim 
\text{ \ \ }\left( \underline{1},\underline{19}\right) .
\end{equation}%
They are defined up to $SO\left( 3\right) \times SO\left( 19\right) $ gauge
transformations,%
\begin{equation}
\begin{tabular}{llll}
$\mathcal{Z}_{\underline{a}}$ & $\equiv $ & $U_{\underline{a}}^{\underline{b}%
}\mathcal{Z}_{\underline{b}}\qquad $ & $,$ \\ 
$\mathcal{Z}_{\underline{I}}$ & $\equiv $ & $V_{\underline{I}}^{\underline{J}%
}\mathcal{Z}_{\underline{J}}$ & ,%
\end{tabular}
\label{inc}
\end{equation}%
where $U$ and $V$ are local orthogonal matrices; $U_{\underline{a}}^{%
\underline{b}}=U_{\underline{a}}^{\underline{b}}\left( \xi \right) $ and $V_{%
\underline{I}}^{\underline{J}}=V_{\underline{I}}^{\underline{J}}\left( \xi
\right) $ with $U_{\underline{a}}^{\underline{c}}U_{\underline{c}}^{%
\underline{b}}=\delta _{\underline{a}}^{\underline{b}}$, and $V_{\underline{I%
}}^{\underline{K}}V_{\underline{K}}^{\underline{J}}=\delta _{\underline{I}}^{%
\underline{J}}$; they can be thought of as%
\begin{equation}
\begin{tabular}{lllll}
$U\left( \xi \right) $ & $=$ & $\exp \left( \dsum\limits_{a=1}^{3}T^{%
\underline{a}}\mathsf{\theta }_{\underline{a}}\left( \xi \right) \right) $ & 
$\in SO\left( 3\right) $ & , \\ 
$V\left( \xi \right) $ & $=$ & $\exp \left[ \dsum\limits_{I=1}^{3}L^{%
\underline{I}}\mathsf{\vartheta }_{\underline{I}}\left( \xi \right) \right] $
& $\in SO\left( 19\right) $ & ,%
\end{tabular}%
\end{equation}%
where $\mathsf{\theta }_{\underline{a}}\left( \xi \right) $ and $\mathsf{%
\vartheta }_{\underline{I}}\left( \xi \right) $ are the gauge group
parameters and $T^{\underline{a}}$ and $L^{\underline{I}}$ the generators of 
$SO\left( 3\right) $ and $SO\left( 19\right) $ respectively. In the case $T^{%
\underline{a}}$, we have the following coordinate realization,%
\begin{equation}
T_{\underline{a}}\sim \varepsilon _{\underline{a}\underline{b}\underline{c}%
}\eta ^{\underline{b}\underline{d}}\xi ^{\underline{c}\underline{I}}\frac{%
\partial }{\partial \xi ^{\underline{d}\underline{I}}},\qquad \varepsilon _{%
\underline{a}\underline{b}\underline{c}}=
\end{equation}%
where $\varepsilon _{\underline{a}\underline{b}\underline{c}}$ is the usual
3d completely antisymmetric tensor. A quite similar relation can be written
down for the $L_{\underline{I}}$ generators. \newline
(\textbf{ii}) \emph{dressed matter charges}\newline
The geometric dressed charges $\mathcal{Z}_{\underline{a}}$ and the matter
ones $\mathcal{Z}_{\underline{I}}$ are not completely independent. They are
related to each others in a quite similar manner as in 4D $\mathcal{N}=2$
supergravity theory embedded in 10D type IIB superstring on CY3. In the 7D
theory, the dressed charges $\mathcal{Z}_{\underline{a}}$ and $\mathcal{Z}_{%
\underline{I}}$ are related as follows%
\begin{equation}
\begin{tabular}{lll}
$\mathcal{Z}_{\underline{I}}$ & $=\eta ^{\underline{a}\underline{b}}D_{%
\underline{a}\underline{I}}\mathcal{Z}_{\underline{b}}$ & , \\ 
$D_{\underline{a}\underline{I}}$ & $=\partial _{\underline{a}\underline{I}%
}-A_{\underline{a}\underline{I}}$ & , \\ 
$\partial _{\underline{a}\underline{I}}$ & $=\frac{\partial }{\partial \xi ^{%
\underline{a}\underline{I}}}$ & ,%
\end{tabular}
\label{cde}
\end{equation}%
where the gauge connection 
\begin{equation}
A_{\underline{a}\underline{I}}=A_{\underline{a}\underline{I}}\left( \xi
\right)
\end{equation}%
is needed to compensate terms like $\eta ^{\underline{a}\underline{b}}U_{%
\underline{a}}^{\underline{c}}\left( \partial _{\underline{c}\underline{I}%
}U_{\underline{b}}^{\underline{d}}\right) $ and $\eta ^{\underline{I}%
\underline{J}}V_{\underline{I}}^{\underline{K}}\left( \partial _{\underline{a%
}\underline{K}}V_{\underline{J}}^{\underline{K}}\right) $ arising from the
gauge transformations (\ref{inc}).\newline
Notice moreover that, using eq(\ref{cde}), we can rewrite the black hole
potential as follows%
\begin{equation}
\mathcal{V}_{BH}^{7D,N=2}=\sum_{a,b=1}^{3}\eta ^{\underline{a}\underline{b}%
}\left( \mathcal{Z}_{\underline{a}}\mathcal{Z}_{\underline{b}%
}-\sum_{c,d=1}^{3}\eta ^{\underline{c}\underline{d}}\left[
\sum_{I,J=1}^{19}\eta ^{\underline{I}\underline{J}}\left( D_{\underline{a}%
\underline{I}}\mathcal{Z}_{\underline{b}}\right) \left( D_{\underline{c}%
\underline{J}}\mathcal{Z}_{\underline{d}}\right) \right] \right) .
\label{317}
\end{equation}%
Clearly this expression is invariant under the gauge change (\ref{inc})
since $D_{\underline{a}\underline{I}}\mathcal{Z}_{\underline{b}}$ transform
in covariant manner. Using the following relation, which will be derived in
section 5, 
\begin{equation}
D_{\underline{a}\underline{I}}\mathcal{Z}_{\underline{b}}=\frac{1}{3}\eta _{%
\underline{a}\underline{b}}\mathcal{Z}_{\underline{I}},  \label{318}
\end{equation}%
and putting back into eq(\ref{317}) as well as using the identity 
\begin{equation}
\mathcal{Z}_{\underline{I}}=D_{\underline{c}\underline{I}}\mathcal{Z}^{%
\underline{c}}.  \label{319}
\end{equation}%
we rediscover (\ref{vef}).\newline
(\textbf{iii}) \emph{gauge invariant }$\mathcal{I}_{+}$\emph{: the Weinhold
potential\qquad }\newline
The existence of two kinds of dressed charges geometric and matter combined
with the $SO\left( 3\right) \times SO\left( 19\right) $ isotropy symmetry
induce an interesting property. We distinguish two kinds of gauge invariants,%
\begin{equation}
\mathcal{I}_{1}=\eta ^{\underline{a}\underline{b}}\mathcal{Z}_{\underline{a}}%
\mathcal{Z}_{\underline{b}}\qquad ,\qquad \mathcal{I}_{2}=\eta ^{\underline{I%
}\underline{J}}\mathcal{Z}_{\underline{I}}\mathcal{Z}_{\underline{J}%
}=-\delta ^{\underline{I}\underline{J}}\mathcal{Z}_{\underline{I}}\mathcal{Z}%
_{\underline{J}}\text{ \ },
\end{equation}%
or equivalently%
\begin{equation}
\mathcal{I}_{\pm }=\mathcal{I}_{1}\mp \mathcal{I}_{2}.
\end{equation}%
The \emph{Weinhold} potential $\mathcal{V}_{BH}^{7D,N=2}$ is one of these
invariants namely $\mathcal{I}_{+}$. This is a positive number as required
by supersymmetry. It is invariant under the $SO\left( 3\right) \times
SO\left( 19\right) $ gauge symmetry (\ref{inc}). \newline
The other gauge invariant $\mathcal{I}_{-}$, which reads as follows,%
\begin{equation}
\sum_{a,b=1}^{3}\eta ^{\underline{a}\underline{b}}\left( \mathcal{Z}_{%
\underline{a}}\mathcal{Z}_{\underline{b}}+\sum_{c,d=1}^{3}\eta ^{\underline{c%
}\underline{d}}\left[ \sum_{I,J=1}^{19}\eta ^{\underline{I}\underline{J}%
}\left( D_{\underline{a}\underline{I}}\mathcal{Z}_{\underline{b}}\right)
\left( D_{\underline{c}\underline{J}}\mathcal{Z}_{\underline{d}}\right) %
\right] \right) =p^{2},  \label{320}
\end{equation}%
has an indefinite sign and will be interpreted later on in terms of a gauge
invariant constraint eq needed by the matrix formulation.\newline
(\textbf{iv}) \emph{behaviors of} $\mathcal{V}_{BH}^{7D,N=2}$\newline
Using (\ref{xxi}), the black hole effective potential (\ref{vef}) can be put
in the remarkable factorization%
\begin{equation}
\mathcal{V}_{BH}^{7D,N=2}\left( \sigma ,\xi \right) =e^{-2\sigma }\mathcal{V}%
\left( \xi \right) ,  \label{po}
\end{equation}%
with $\mathcal{V}\left( \xi \right) $, having no dependence in $\sigma $,
given by%
\begin{equation}
\mathcal{V}\left( \xi \right) =\left( \sum_{a,b=1}^{3}\delta ^{\underline{a}%
\underline{b}}Z_{\underline{a}}Z_{\underline{b}}+\sum_{I,J=1}^{19}\delta ^{%
\underline{I}\underline{J}}Z_{\underline{I}}Z_{\underline{J}}\right) .
\label{poo}
\end{equation}%
Notice that the potential $\mathcal{V}_{BH}^{7D,N=2}\left( \sigma ,\xi
\right) $ has a very special dependence on the dilaton $\sigma $. According
to the values of this field, we distinguish the three following particular
cases:\newline
$\left( \mathbf{\alpha }\right) $ \emph{case} $\sigma \rightarrow 0$:\qquad 
\newline
For finite values of $\sigma $ (\emph{see also footnote} \emph{5}), say
around $\sigma _{0}=0$, the behavior of the black hole potential is
dominated by the factor $\mathcal{V}\left( \xi _{\underline{b}\underline{J}%
}\right) $; i.e 
\begin{equation}
\mathcal{V}_{BH}^{7d,N=2}\left( \sigma ,\xi \right) \sim \mathcal{V}\left(
\xi \right) .  \label{dom}
\end{equation}%
$\left( \mathbf{\beta }\right) $ \emph{case }$\sigma \rightarrow -\infty $%
:\qquad \newline
In this case the behavior of the black hole potential is dominated by the
factor $e^{+2\left\vert \sigma \right\vert }$ and $\mathcal{V}_{BH}^{7d,N=2}$
could be approximated as follows%
\begin{equation}
\mathcal{V}_{BH}^{7d,N=2}\left( \sigma ,\xi \right) =\mathcal{V}%
_{0}e^{+2\left\vert \sigma \right\vert },
\end{equation}%
where $\mathcal{V}_{0}$ is some fixed value extremizing eq(\ref{poo}). In
the 11D M-theory compactification set up, this case corresponds to a K3
manifold with large volume; 
\begin{equation}
Vol\left( K3\right) \rightarrow \infty \text{ },
\end{equation}%
but small metric deformations.\newline
$\left( \mathbf{\gamma }\right) $ \emph{case} $\sigma \rightarrow +\infty $%
:\qquad \newline
Here the behavior of the black hole potential is dominated by the factor $%
e^{-2\left\vert \sigma \right\vert }$ and $\mathcal{V}_{BH}^{7d,N=2}$ might
be approximated as follows%
\begin{equation}
\mathcal{V}_{BH}^{7d,N=2}\left( \sigma ,\xi \right) =e^{-2\left\vert \sigma
\right\vert }\mathcal{V}_{0}.
\end{equation}%
This case corresponds to compactifying M-theory on a K3 manifold with small
volume%
\begin{equation}
Vol\left( K3\right) \rightarrow 0.
\end{equation}

(\textbf{b}) \emph{gauge invariant }$\mathcal{I}_{-}$\emph{: the constraint
eqs}\newline
The constraint eqs(\ref{lz}) combine altogether as follows 
\begin{equation}
\eta _{\underline{\Upsilon }\underline{\digamma }}L_{\underline{\Lambda }}^{%
\underline{\Upsilon }}L_{\underline{\Sigma }}^{\underline{\digamma }}=\delta
_{\underline{a}\underline{b}}L_{\underline{\Lambda }}^{\underline{a}}L_{%
\underline{\Sigma }}^{\underline{b}}-\delta _{\underline{I}\underline{J}}L_{%
\underline{\Lambda }}^{\underline{I}}L_{\underline{\Sigma }}^{\underline{J}%
}=\eta _{\underline{\Lambda }\underline{\Sigma }},  \label{lls}
\end{equation}%
and show that $L_{\underline{\Lambda }}^{\underline{\Upsilon }}$ is not an
arbitrary $22\times 22$ matrix; but an orthogonal matrix of $SO\left(
3,19\right) $. Eqs(\ref{lls}) fix the undesired degrees of freedom.\newline
It turns out that these constraint relations are gauge invariant under the $%
SO\left( 3\right) \times SO\left( 19\right) $ isotropy symmetry. They also
play an important role in the study of the criticality condition of 7D black
hole and in the underlying "hyperKahler" special geometry.\newline
Let us show how these constraints can be brought to the form $\mathcal{I}%
_{-} $ and how they are used in the solving of the criticality condition.%
\newline
Multiplying both sides of (\ref{lls}) by the bare magnetic charges $p^{%
\underline{\Lambda }}$ and $p^{\underline{\Sigma }}$; then using eqs(\ref{zl}%
), which we rewrite as follow%
\begin{equation}
\begin{tabular}{llll}
$Z^{\underline{a}}=p^{\underline{\Lambda }}L_{\underline{\Lambda }}^{%
\underline{a}}\left( \xi \right) $ & $\qquad ,\qquad $ & $\mathcal{Z}^{%
\underline{a}}=e^{-\sigma }Z^{\underline{a}}\left( \xi \right) $ & , \\ 
$Z^{\underline{I}}=p^{\underline{\Lambda }}L_{\underline{\Lambda }}^{%
\underline{I}}\left( \xi \right) $ & $\qquad ,\qquad $ & $\mathcal{Z}^{%
\underline{I}}=e^{-\sigma }Z^{\underline{I}}\left( \xi \right) $ & ,%
\end{tabular}
\label{xxi}
\end{equation}%
we obtain the following remarkable relation between the dressed charges%
\begin{equation}
\sum_{a,b=1}^{3}\delta ^{\underline{a}\underline{b}}Z_{\underline{a}}Z_{%
\underline{b}}-\sum_{I,J=1}^{19}\delta ^{\underline{I}\underline{J}}Z_{%
\underline{I}}Z_{\underline{J}}=p^{2},\qquad  \label{inf}
\end{equation}%
with%
\begin{equation}
p^{2}=\eta _{\underline{\Lambda }\underline{\Sigma }}p^{\underline{\Lambda }%
}p^{\underline{\Sigma }}=\delta ^{\underline{a}\underline{b}}p_{\underline{a}%
}p_{\underline{b}}-\delta ^{\underline{I}\underline{J}}p_{\underline{I}}p_{%
\underline{J}}.  \label{imf}
\end{equation}%
Eq(\ref{inf}), which reads also as 
\begin{equation}
p^{2}=e^{2\sigma }\mathcal{I}_{-},  \label{nf}
\end{equation}%
has no definite sign since it can be positive, zero or negative. It is
manifestly gauge invariant. \newline
There is two basic ways to deal with this constraint relation. The first way
is to solve it as 
\begin{equation}
\sum_{I,J=1}^{19}\delta ^{\underline{I}\underline{J}}Z_{\underline{I}}Z_{%
\underline{J}}=-p^{2}+\sum_{a,b=1}^{3}\delta ^{\underline{a}\underline{b}}Z_{%
\underline{a}}Z_{\underline{b}}.  \label{alp}
\end{equation}%
Then substitute back into eq(\ref{poo}) to end with the black hole potential
factor%
\begin{equation}
\mathcal{V}\left( \xi \right) =\left( -p^{2}+2\sum_{a,b=1}^{3}\delta ^{%
\underline{a}\underline{b}}Z_{\underline{a}}Z_{\underline{b}}\right) .
\end{equation}%
Since from (\ref{alp}), we should have 
\begin{equation}
-p^{2}+\sum_{a,b=1}^{3}\delta ^{\underline{a}\underline{b}}Z_{\underline{a}%
}Z_{\underline{b}}\geq 0
\end{equation}%
then we have 
\begin{equation}
\mathcal{V}\left( \xi \right) \geq \sum_{a,b=1}^{3}\delta ^{\underline{a}%
\underline{b}}Z_{\underline{a}}Z_{\underline{b}}\geq 0.
\end{equation}%
Moreover seen that $\delta \mathcal{V}\left( \xi \right) =2\sum \left(
\delta ^{\underline{a}\underline{b}}Z_{\underline{a}}\delta Z_{\underline{b}%
}\right) $, the critical points of the black hole potential factor $\delta ^{%
\underline{a}\underline{b}}Z_{\underline{a}}Z_{\underline{b}}$ is completely
controlled by the zeros of $\delta ^{\underline{a}\underline{b}}\left[ Z_{%
\underline{a}}\delta Z_{\underline{b}}\right] $. \newline
The second way to approach eq(\ref{inf}) is to keep it is; and use the
Lagrange multiplier method to deal with it. The Lagrange multiplier method
method as well as comments on the entropies for dual pairs of black
attractors in 6D and 7D will be exposed in \textrm{\cite{BHS}.}\newline
Expressing the variation of eq(\ref{inf}) as, 
\begin{equation}
\sum_{a=1}^{3}Z^{\underline{a}}T_{\underline{a}}=\sum_{I=1}^{19}Z^{%
\underline{I}}T_{\underline{I}},  \label{zdz}
\end{equation}%
where the metric $\delta ^{\underline{a}\underline{b}}$ and $\delta ^{%
\underline{I}\underline{J}}$ have been used and where we have set $T_{%
\underline{a}}=\delta Z_{\underline{a}}$ and $T_{\underline{I}}=\delta Z_{%
\underline{I}}$, then we have the following results:

\begin{theorem}
\ \ \ \newline
Denoting by $T_{\underline{a}}=\delta Z_{\underline{a}}$ and $T_{\underline{I%
}}=\delta Z_{\underline{I}}$ as in eq(\ref{zdz}), then:\newline
the $SO\left( 3\right) $ scalar $Z^{\underline{a}}T_{\underline{a}}=0$ if $%
Z^{\underline{I}}T_{\underline{I}}=0;$ that is the $Z^{\underline{I}}$ and $%
T_{\underline{I}}$ are normal real vectors in R$^{19}$. This happens in
particular for:\newline
(\textbf{i}) $Z_{\underline{I}}=0$ $\ \forall $ $I\in \mathcal{I}=\left\{
1,...,19\right\} $ whatever the $T_{\underline{I}}$'s are, \newline
(\textbf{ii}) $T_{\underline{I}}=0$ $\ \forall $ $I\in \mathcal{I}=\left\{
1,...,19\right\} $ whatever the $Z_{\underline{I}}$'s are,\newline
(\textbf{iii}) $Z_{\underline{I}}=0$\ \ for $I\in \mathcal{J}\subset 
\mathcal{I}$ and $T_{\underline{I}}=0$ \ for $I\in \mathcal{I}/\mathcal{J}$.%
\newline
\ \ \newline
Inversely, the $SO\left( 19\right) $ scalar $Z^{\underline{I}}T_{\underline{I%
}}=0$ if $Z^{\underline{a}}T_{\underline{a}}=0$, that is the $Z^{\underline{a%
}}$ and $T_{\underline{a}}$ are normal vectors in R$^{3}$. In particular:%
\newline
(\textbf{iv}) $Z_{\underline{a}}=0$ $\ \forall $ $a\in \mathcal{I}=\left\{
1,2,3\right\} $ whatever the $T_{\underline{a}}$'s are, \newline
(\textbf{v}) $T_{\underline{a}}=0$ $\ \forall $ $a\in \mathcal{I}=\left\{
1,2,3\right\} $ whatever the $Z_{\underline{a}}$'s are,\newline
(\textbf{vi}) $Z_{\underline{a}}=0$ \ for $I\in \mathcal{J}\subset \mathcal{I%
}$ and $T_{\underline{a}}=0$\ \ for $I\in \mathcal{I}/\mathcal{J}.$
\end{theorem}

\ \ \newline
Notice that the variation of $Z^{\underline{a}}Z_{\underline{a}}$ can be
gauge covariantly expanded as 
\begin{equation}
\sum_{a}\left( Z^{\underline{a}}\delta Z_{\underline{a}}\right)
=\sum_{a,b,I}\left( Z^{\underline{a}}D_{\underline{b}\underline{I}}Z_{%
\underline{a}}\right) \nabla \xi ^{\underline{b}\underline{I}}.
\end{equation}%
By using the identities (\ref{318}-\ref{319}), we can bring this variation
to the form%
\begin{equation}
\sum_{b=1}^{3}\left( Z_{\underline{b}}\delta Z^{\underline{b}}\right)
=\sum_{I=1}^{19}Z_{\underline{I}}\nabla Z^{\underline{I}},\qquad \nabla Z^{%
\underline{I}}=\sum_{b=1}^{3}Z_{\underline{b}}\nabla \xi ^{\underline{b}%
\underline{I}},
\end{equation}%
or equivalently%
\begin{equation}
\sum_{b=1}^{3}\left( Z_{\underline{b}}T^{\underline{b}}\right)
=\sum_{b=1}^{3}Z_{\underline{b}}\left( \sum_{I=1}^{19}Z_{\underline{I}%
}\nabla \xi ^{\underline{b}\underline{I}}\right) .
\end{equation}%
It follows from the two last relations the result:

\begin{corollary}
\ \newline
(\textbf{i}) If $Z_{\underline{a}}\neq 0$ \ $\forall $ $a\in \left\{
1,2,3\right\} $\ and $Z_{\underline{I}}=0$ \ $\forall $ $I\in \mathcal{I}%
=\left\{ 1,...,19\right\} $, then $T^{\underline{a}}=0$ $\forall $ $a$%
\newline
(\textbf{ii}) the potential factor $\left( \sum_{a,b=1}^{3}\delta ^{%
\underline{a}\underline{b}}Z_{\underline{a}}Z_{\underline{b}%
}+\sum_{I,J=1}^{19}\delta ^{\underline{I}\underline{J}}Z_{\underline{I}}Z_{%
\underline{J}}\right) $ has extremals for:\newline
($\mathbf{\alpha }$) $Z_{\underline{a}}=0$ $\ \forall $ $a\in \left\{
1,2,3\right\} $; $\forall $ $Z_{\underline{I}}$\newline
($\mathbf{\beta }$) $Z_{\underline{I}}=0$ $\ \forall $ $I\in \left\{
1,...,19\right\} $ ; $\forall $ $Z_{\underline{a}}$
\end{corollary}

\paragraph{(\textbf{2}) \emph{curved coordinates frame}\newline
}

In the curved coordinates frame $\left\{ \varphi ^{m}\right\} =\left\{
\varphi ^{0}=\sigma ;\phi ^{aI}\right\} $, the curved space relations
analogue of the above inertial frame ones are obtained, by using eqs(\ref%
{vie}), as follows:%
\begin{equation}
\begin{tabular}{llllllll}
$Z_{\underline{a}}$ & $=$ & $e_{\underline{a}}^{c}Y_{c}$ & $\qquad ,\qquad $
& $Y_{c}$ & $=$ & $e_{c}^{\underline{a}}Z_{\underline{a}}$ & , \\ 
$Z_{\underline{I}}$ & $=$ & $e_{\underline{I}}^{K}Y_{K}$ & $\qquad ,\qquad $
& $Y_{K}$ & $=$ & $e_{K}^{\underline{I}}Z_{\underline{I}}$ & ,%
\end{tabular}
\label{ZE}
\end{equation}%
where%
\begin{equation}
\begin{tabular}{llllllll}
$e_{\underline{a}}^{c}$ & $=$ & $e_{\underline{a}}^{c}\left( \xi ,\phi
\right) $ & $\qquad ,\qquad $ & $e_{\underline{I}}^{K}$ & $=$ & $e_{%
\underline{I}}^{K}\left( \xi ,\phi \right) $ & ,%
\end{tabular}%
\end{equation}%
are the vielbeins introduced previously (\ref{vie}). They allow to move from
the inertial frame to a generic curved one. Substituting the change (\ref{ZE}%
) back into $\delta ^{\underline{a}\underline{b}}Z_{\underline{a}}Z_{%
\underline{b}}$ and $\delta ^{\underline{I}\underline{J}}Z_{\underline{I}}Z_{%
\underline{J}}$, we get%
\begin{equation}
\begin{tabular}{llll}
$\delta ^{\underline{a}\underline{b}}Z_{\underline{a}}Z_{\underline{b}}$ & $%
= $ & $\delta ^{\underline{a}\underline{b}}e_{\underline{a}}^{c}e_{%
\underline{b}}^{d}Y_{c}Y_{d}$ & , \\ 
& $=$ & $K^{cd}Y_{c}Y_{d}$ & ,%
\end{tabular}
\label{daz}
\end{equation}%
and%
\begin{equation}
\begin{tabular}{llll}
$\delta ^{\underline{I}\underline{J}}Z_{\underline{I}}Z_{\underline{J}}$ & $%
= $ & $\delta ^{\underline{a}\underline{b}}e_{\underline{I}}^{K}e_{%
\underline{J}}^{L}Y_{K}Y_{L}$ & , \\ 
& $=$ & $K^{KL}Y_{K}Y_{L}$ & .%
\end{tabular}
\label{diz}
\end{equation}%
Notice that%
\begin{equation}
Y_{c}=Y_{c}\left( \phi \right) ,\qquad Y_{K}=Y_{K}\left( \phi \right) .
\end{equation}%
Similar relations can be written down by using the inverse vielbeins $e_{K}^{%
\underline{I}}$. Moreover, we have the following properties:\newline
(\textbf{i}) the effective potential (\ref{poo}) reads, in the curved
coordinates frame, as 
\begin{equation}
\mathcal{V}_{BH}^{7D,N=2}\left( \sigma ,\phi \right) =e^{-2\sigma }\mathcal{V%
}\left( \phi \right) ,  \label{sfr}
\end{equation}%
where now $\mathcal{V}\left( \phi \right) $ is given by%
\begin{equation}
\mathcal{V}\left( \phi \right) =\left(
\sum_{c,d=1}^{3}K^{cd}Y_{c}Y_{d}-\sum_{K,L=1}^{19}K^{KL}Y_{K}Y_{L}\right) ,
\label{va}
\end{equation}%
and where%
\begin{equation}
\begin{tabular}{lllll}
$K^{cd}=\eta ^{\underline{a}\underline{b}}e_{\underline{a}}^{c}e_{\underline{%
b}}^{d}$ & $\qquad ,\qquad $ & $K^{KL}=\eta ^{\underline{I}\underline{J}}e_{%
\underline{I}}^{K}e_{\underline{J}}^{L}$ & $.$ & 
\end{tabular}%
\end{equation}%
(\textbf{ii}) putting eqs(\ref{daz}-\ref{diz}) back into eq(\ref{inf}), we
get the gauge invariant constraint relation%
\begin{equation}
\sum_{c,d=1}^{3}K^{cd}Y_{c}Y_{d}+\sum_{K,L=1}^{19}K^{KL}Y_{K}Y_{L}=p^{2}.
\end{equation}%
The variation of this constraint eq gives%
\begin{equation}
K^{ab}Y_{a}\left( \mathcal{D}Y_{b}\right) =-K^{IJ}Y_{I}\left( \mathcal{D}%
Y_{J}\right) ,
\end{equation}%
with%
\begin{equation}
\begin{tabular}{llll}
$DY_{b}$ & $=$ & $\left[ \left( \delta Y_{b}\right) +\frac{1}{2}K_{bc}\left(
\delta K^{cd}\right) Y_{d}\right] $ & , \\ 
$DY_{J}$ & $=$ & $\left[ \left( \delta Y_{J}\right) +\frac{1}{2}K_{JK}\left(
\delta K^{KL}\right) Y_{L}\right] $ & .%
\end{tabular}%
\end{equation}%
(\textbf{iii}) by implementing the dilaton $\sigma $, the relations (\ref%
{daz}-\ref{diz}) can be also put in the form%
\begin{equation}
\begin{tabular}{llll}
$\delta ^{\underline{a}\underline{b}}\mathcal{Z}_{\underline{a}}\mathcal{Z}_{%
\underline{b}}$ & $=$ & $=K^{cd}\mathcal{Y}_{c}\mathcal{Y}_{d}=e^{+2\sigma
}K^{cd}Y_{c}Y_{d}$ & , \\ 
$\eta ^{\underline{I}\underline{J}}\mathcal{Z}_{\underline{I}}\mathcal{Z}_{%
\underline{J}}$ & $=$ & $K^{KL}\mathcal{Y}_{K}\mathcal{Y}_{L}=e^{+2\sigma
}K^{KL}Y_{K}Y_{L}$ & ,%
\end{tabular}%
\end{equation}%
where we have set%
\begin{equation}
\begin{tabular}{llll}
$\mathcal{Y}_{c}\left( \sigma ,\phi \right) $ & $=$ & $e^{-\sigma
}Y_{c}\left( \phi \right) $ & , \\ 
$\mathcal{Y}_{K}\left( \sigma ,\phi \right) $ & $=$ & $e^{-\sigma
}Y_{K}\left( \phi \right) $ & .%
\end{tabular}%
\end{equation}

\subsubsection{Criticality conditions}

\qquad In the inertial coordinate frame $\left\{ \xi \right\} $, the
critically condition of the black hole potential takes a simple form; it
reads as follows: 
\begin{equation}
\delta \mathcal{V}_{BH}^{7D,N=2}=2\left( \sum_{a,b=1}^{3}\delta ^{\underline{%
a}\underline{b}}\mathcal{Z}_{\underline{a}}\delta \mathcal{Z}_{\underline{b}%
}\right) +2\left( \sum_{I,J=1}^{19}\delta ^{\underline{I}\underline{J}}%
\mathcal{Z}_{\underline{I}}\mathcal{Z}_{\underline{J}}\right) =0.  \label{sd}
\end{equation}%
This variation can rewritten formally like%
\begin{equation}
2\left( \sum_{a,b=1}^{3}\delta ^{\underline{a}\underline{b}}\mathcal{Z}_{%
\underline{a}}\mathcal{T}_{\underline{b}}\right) +2\left(
\sum_{I,J=1}^{19}\delta ^{\underline{I}\underline{J}}\mathcal{Z}_{\underline{%
I}}\mathcal{T}_{\underline{J}}\right) =0.  \label{cc}
\end{equation}%
where, in general,%
\begin{equation}
\begin{tabular}{lll}
$\mathcal{T}^{\underline{a}}=\delta \mathcal{Z}^{\underline{a}}$ & $=\left( 
\frac{\partial \mathcal{Z}^{\underline{a}}}{\partial \xi ^{0}}\right) \delta
\xi ^{0}+\left( \frac{\partial \mathcal{Z}^{\underline{a}}}{\partial \xi ^{%
\underline{b}\underline{I}}}\right) \delta \xi ^{\underline{b}\underline{I}}$
& , \\ 
& $=\left( \frac{\partial \mathcal{Z}^{\underline{a}}}{\partial \xi ^{0}}%
\right) \delta \xi ^{0}+\left( D_{\underline{b}\underline{I}}\mathcal{Z}^{%
\underline{a}}\right) \nabla \xi ^{\underline{b}\underline{I}}$ & , \\ 
&  &  \\ 
$\mathcal{T}^{\underline{I}}=\delta \mathcal{Z}^{\underline{I}}$ & $=\left( 
\frac{\partial \mathcal{Z}^{\underline{I}}}{\partial \xi ^{0}}\right) \delta
\xi ^{0}+\left( \frac{\partial \mathcal{Z}^{\underline{I}}}{\partial \xi ^{%
\underline{b}\underline{J}}}\right) \delta \xi ^{\underline{b}\underline{J}}$
& , \\ 
& $\left( \frac{\partial \mathcal{Z}^{\underline{I}}}{\partial \xi ^{0}}%
\right) \delta \xi ^{0}+\left( D_{\underline{b}\underline{I}}\mathcal{Z}^{%
\underline{I}}\right) \nabla \xi ^{\underline{b}\underline{I}}$ & .%
\end{tabular}%
\end{equation}%
In the case of 7D $\mathcal{N}=2$ supergravity embedded in 11D M-theory on
K3, $\mathcal{Z}^{\underline{a}}$ and $\mathcal{Z}^{\underline{I}}$ are
respectively given by $e^{-\sigma }Z^{\underline{a}}\left( \xi \right) $ and 
$e^{-\sigma }Z^{\underline{I}}\left( \xi \right) $ eqs(\ref{xxi}). So we have%
\begin{equation}
\begin{tabular}{llll}
$\left( \frac{\partial \mathcal{Z}^{\underline{a}}}{\partial \sigma }\right)
=-e^{-\sigma }Z^{\underline{a}}$ & $,\qquad $ & $\left( \frac{\partial 
\mathcal{Z}^{\underline{a}}}{\partial \xi ^{\underline{b}\underline{J}}}%
\right) =e^{-\sigma }\left( \frac{\partial Z^{\underline{a}}}{\partial \xi ^{%
\underline{b}\underline{J}}}\right) $ & $,$ \\ 
&  &  &  \\ 
$\left( \frac{\partial \mathcal{Z}^{\underline{I}}}{\partial \sigma }\right)
=-e^{-\sigma }Z^{\underline{I}}$ & , & $\left( \frac{\partial \mathcal{Z}^{%
\underline{I}}}{\partial \xi ^{\underline{b}\underline{J}}}\right)
=e^{-\sigma }\left( \frac{\partial Z^{\underline{I}}}{\partial \xi ^{%
\underline{b}\underline{J}}}\right) $ & .%
\end{tabular}%
\end{equation}

\emph{Classification of solutions of eq}(\ref{cc})\newline
The above theorem and corollary show that the black hole solutions
associated with the critical points of eq(\ref{cc}) are of three kinds: a%
\emph{\ 1/2-BPS }and\emph{\ two non BPS black holes; }to which we refer to
as type 1 and type 2$\emph{.}$\newline
The non degenerate solutions of eq(\ref{cc}) with black hole effective
potential at horizon like 
\begin{equation}
\left( \mathcal{V}_{BH}^{7d,N=2}\right) _{horizon}>0,
\end{equation}%
and the \emph{Arnowitt-Deser-Misner} (ADM) mass $\mathcal{M}_{ADM}^{2}$
bounded like,%
\begin{equation}
\left( \sum_{a,b=1}^{3}\delta ^{\underline{a}\underline{b}}\mathcal{Z}_{%
\underline{a}}\mathcal{Z}_{\underline{b}}\right) \leq \mathcal{M}%
_{ADM}^{2}=\left( \sum_{a,b=1}^{3}\delta ^{\underline{a}\underline{b}}%
\mathcal{Z}_{\underline{a}}\mathcal{Z}_{\underline{b}}+\sum_{I,J=1}^{19}%
\delta ^{\underline{I}\underline{J}}\mathcal{Z}_{\underline{I}}\mathcal{Z}_{%
\underline{J}}\right) ,
\end{equation}%
are given by:\newline
(\textbf{1}) \emph{1/2- BPS state.}\newline
This black hole state has \emph{eight supersymmetries} and corresponds to,%
\begin{equation}
\left( \mathcal{Z}_{1},\mathcal{Z}_{2},\mathcal{Z}_{3}\right) \neq \left(
0,0,0\right) ,  \label{cr}
\end{equation}%
but%
\begin{equation}
\sum_{a,b=1}^{3}\delta ^{\underline{a}\underline{b}}\mathcal{Z}_{\underline{a%
}}\mathcal{T}_{\underline{b}}=0,
\end{equation}%
and 
\begin{equation}
\left( \mathcal{Z}_{I}\right) =\left( \mathcal{Z}_{1},...,\mathcal{Z}%
_{19}\right) =\left( 0,...,0\right) .  \label{cs}
\end{equation}%
In this case the ADM mass $\mathcal{M}_{ADM}^{2}$ saturates the bound 
\begin{equation}
\mathcal{M}_{ADM}\geq \sqrt{\left( \sum_{a,b=1}^{3}\delta ^{\underline{a}%
\underline{b}}\mathcal{Z}_{\underline{a}}\mathcal{Z}_{\underline{b}}\right) }%
.
\end{equation}%
At the event horizon, the critical ADM mass $\left( \mathcal{M}_{ADM}\right)
_{\text{h}}$ is obtained by extremizing the effective potential $\mathcal{V}%
_{BH}^{7D,N=2}$ with respect to the scalar moduli $\xi ^{\underline{m}}$. 
\newline
Using eq(\ref{cr}), we then have 
\begin{equation}
0<\left( \mathcal{M}_{ADM}\right) _{\text{BPS}}=\left( \mathcal{M}%
_{ADM}\right) _{\text{h}},
\end{equation}%
where we set 
\begin{equation}
\left( \mathcal{M}_{ADM}\right) _{\text{BPS}}=\sqrt{\left(
\sum_{a,b=1}^{3}\delta ^{\underline{a}\underline{b}}\mathcal{Z}_{\underline{a%
}}\mathcal{Z}_{\underline{b}}\right) _{\text{horizon}}}.
\end{equation}%
The lower bound of $\left( \mathcal{M}_{ADM}\right) _{\text{h}}$ is positive
definite. By using eq(\ref{inf}) and eq(\ref{zl}), we also have for the case 
$\eta _{\underline{\Lambda }\underline{\Sigma }}p^{\underline{\Lambda }}p^{%
\underline{\Sigma }}\neq 0$ and $p^{\underline{d}}\delta _{\underline{d}%
\underline{h}}p^{\underline{h}}$,%
\begin{equation}
\begin{tabular}{lll}
$\left( Z^{\underline{a}}\right) _{_{\text{horizon}}}$ & $=p^{\underline{a}}%
\sqrt{\frac{\left\vert \left( p^{\underline{b}}\delta _{\underline{b}%
\underline{c}}p^{\underline{c}}-p^{\underline{J}}\delta _{\underline{J}%
\underline{K}}p^{K}\right) \right\vert }{p^{\underline{d}}\delta _{%
\underline{d}\underline{h}}p^{\underline{h}}}}$ & , \\ 
&  &  \\ 
$\left( L_{\underline{\Lambda }}^{\underline{a}}\right) _{_{\text{horizon}}}$
& $=p_{\underline{\Lambda }}\left( \left\vert \left( p^{\underline{b}}\delta
_{\underline{b}\underline{c}}p^{\underline{c}}-p^{\underline{J}}\delta _{%
\underline{J}\underline{K}}p^{K}\right) \right\vert \right) ^{-1}Z^{%
\underline{a}}$ \ . & .%
\end{tabular}%
\end{equation}%
(\textbf{2}) \emph{non BPS state}: \emph{type 1 }\newline
This is a \emph{non supersymmetric} state corresponding to,%
\begin{equation}
\left( \mathcal{Z}_{I}\right) =\left( \mathcal{Z}_{1},...,\mathcal{Z}%
_{19}\right) \neq \left( 0,...,0\right) ,  \label{zzi}
\end{equation}%
and%
\begin{equation}
\sum \delta ^{\underline{I}\underline{J}}\mathcal{Z}_{\underline{I}}\mathcal{%
T}_{\underline{J}}=0,
\end{equation}%
and moreover 
\begin{equation}
\left( \mathcal{Z}_{\underline{a}}\right) =\left( \mathcal{Z}_{1},\mathcal{Z}%
_{2},\mathcal{Z}_{3}\right) =\left( 0,0,0\right) .
\end{equation}%
In this case the critical ADM mass $\left( \mathcal{M}_{ADM}\right) _{\text{h%
}}$ is given by:%
\begin{equation}
0\leq \left( \mathcal{M}_{ADM}\right) _{\text{h}}=\sqrt{\left(
\sum_{I,J=1}^{19}\delta ^{\underline{I}\underline{J}}\mathcal{Z}_{\underline{%
I}}\mathcal{Z}_{\underline{J}}\right) _{\text{horizon}}}.
\end{equation}%
Notice that since $\mathcal{Z}_{\underline{a}}$ and $\mathcal{Z}_{\underline{%
I}}$ are defined up to $SO\left( 3\right) \times SO\left( 19\right) $ gauge
symmetry eqs(\ref{fac}), we can usually perform a rotation to bring eq(\ref%
{zzi}) to the form 
\begin{equation}
\left( \mathcal{Z}_{\underline{I}}\right) =\left( \mathcal{Z}%
_{1},0,...,0\right) ,
\end{equation}%
with $\left( \mathcal{Z}_{\underline{I}}\right) _{I=1}\neq 0$ and all others 
$\mathcal{Z}_{\underline{I}}$ with $I\neq 1$ equal to zero. Similar
conclusion can made for $\mathcal{Z}_{\underline{a}}$ or both $\mathcal{Z}_{%
\underline{a}}$ and $\mathcal{Z}_{\underline{I}}$.\newline
(\textbf{3}) \emph{non BPS state}:\emph{\ type 2}\newline
This state is non supersymmetric and corresponds to%
\begin{equation}
\left( \mathcal{Z}_{\underline{a}}\right) \neq \left( 0,0,0\right) ,\qquad 
\text{i.e \qquad }\mathcal{Z}_{\underline{a}}\neq 0\text{ for some }a\in 
\mathcal{J}\subset \left\{ 1,2,3\right\} ,
\end{equation}%
and%
\begin{equation}
\sum_{a,b\in \mathcal{J}}\delta ^{\underline{a}\underline{b}}\mathcal{Z}%
\underline{_{a}}\mathcal{T}_{\underline{b}}=0
\end{equation}%
together with%
\begin{equation}
\left( \mathcal{Z}_{\underline{I}}\right) \neq \left( 0,...,0\right) ,\qquad 
\text{i.e \qquad }\mathcal{Z}_{\underline{I}}\neq 0\text{ for some }I\in 
\mathcal{J}^{\prime }\subset \left\{ 1,...,19\right\}
\end{equation}%
as well as%
\begin{equation}
\sum_{I,J\in \mathcal{J}^{\prime }}\delta ^{\underline{I}\underline{J}}%
\mathcal{Z}_{\underline{I}}\mathcal{T}_{\underline{J}}=0.
\end{equation}%
This configuration leads to%
\begin{equation}
0<\left( \mathcal{M}_{ADM}\right) _{\text{h}}=\sqrt{\left( \sum_{a,b\in 
\mathcal{J}}\delta ^{\underline{a}\underline{b}}\mathcal{Z}_{\underline{a}}%
\mathcal{Z}_{\underline{b}}\right) _{\text{horizon}}+\left( \sum_{I,J\in 
\mathcal{J}^{\prime }}\delta ^{\underline{I}\underline{J}}\mathcal{Z}_{%
\underline{I}}\mathcal{Z}_{\underline{J}}\right) _{\text{horizon}}}.
\label{cz}
\end{equation}%
For more details on this classification, see also the analysis of subsection
7.1.\newline
In the end of this discussion, notice that a similar and equivalent study
can be done for the criticality condition by using the curved coordinates
frame $\left\{ \varphi ^{m}\right\} $. The two methods are equivalent and
are related by the identities $Z_{\underline{a}}\left( \xi \right) =e_{%
\underline{a}}^{c}\left( \xi ,\phi \right) Y_{c}\left( \phi \right) $ and $%
Z_{\underline{I}}\left( \xi \right) =e_{\underline{I}}^{K}\left( \xi ,\phi
\right) Y_{K}\left( \phi \right) .$

\subsection{7D black 3- brane}

\qquad The 7D black \emph{3}- brane is realized by wrapping the M5 brane on
the 2- cycles of K3. The three remaining space directions fill part of the
seven space time dimensions. \newline
The \emph{3}- brane is \emph{electrically} charged under the $U^{22}\left(
1\right) $ gauge group symmetry of the $\mathcal{N}=2$ 7D supergravity
theory. The solutions for 7D black \emph{3}- brane are given by the dual of
the previous black hole ones.\newline
The electric charges 
\begin{equation}
q^{\Lambda }=\left( q^{1},...,q^{22}\right) ,
\end{equation}%
are given by the integral of the real 7-form flux density $\mathcal{F}_{7}$
through the basis of the 7- cycles $S_{\infty }^{5}\times \Psi ^{\Lambda }$, 
\begin{equation}
q^{\Lambda }=\int_{S_{\infty }^{5}}\left( \int_{\Psi ^{\Lambda }}\mathcal{F}%
_{7}\right) ,\qquad \Lambda =1,...,22,
\end{equation}%
where the real 5-sphere $S_{\infty }^{5}$ is normalized as, 
\begin{equation}
\int_{S_{\infty }^{5}}d^{5}s=1.
\end{equation}%
In the above relation, the real space time 7- form $\mathcal{F}_{7}$ is the
Hodge dual of the field strength $\mathcal{F}_{4}=\emph{d}\mathcal{C}_{3}$
considered previously.\newline
The black 3-brane potential 
\begin{equation}
\mathcal{V}_{\text{3-brane}}^{7D,N=2}
\end{equation}%
is obtained by dualizing the Weinhold potential of the 7D black hole (\ref%
{vef}). This scalar potential can be defined either by using the inertial
coordinates frame $\left\{ \xi \right\} $ or, in general, the curved one.

\subsubsection{Effective potential}

\qquad In the inertial coordinate frame $\left\{ \xi \right\} $, the black
3-brane potential $\mathcal{V}_{\text{3-brane}}^{7D,N=2}$ reads as follows, 
\begin{equation}
\mathcal{V}_{\text{3-brane}}^{7D,N=2}=\sum_{a,b=1}^{3}\delta ^{\underline{a}%
\underline{b}}\widetilde{\mathcal{Z}}_{\underline{a}}\widetilde{\mathcal{Z}}%
_{\underline{b}}+\sum_{I,J=1}^{19}\delta ^{\underline{I}\underline{J}}%
\widetilde{\mathcal{Z}}_{\underline{I}}\widetilde{\mathcal{Z}}_{\underline{J}%
},  \label{3b}
\end{equation}%
where $\widetilde{\mathcal{Z}}_{\underline{a}}$ and $\widetilde{\mathcal{Z}}%
_{\underline{I}}$ are the dressed electric charges dual to the dressed
magnetic $\mathcal{Z}_{\underline{a}}$ and $\mathcal{Z}_{\underline{I}}$.
They are given by,%
\begin{equation}
\begin{tabular}{llll}
$\widetilde{\mathcal{Z}}^{\underline{a}}=\sum_{\Lambda =1}^{22}q^{\Lambda }%
\widetilde{\mathcal{L}}_{\Lambda }^{\underline{a}}$ & $\qquad ,\qquad $ & $%
\widetilde{\mathcal{L}}_{\Lambda }^{\underline{a}}=\widetilde{\mathcal{L}}%
_{\Lambda }^{\underline{a}}\left( \xi \right) $ & , \\ 
$\widetilde{\mathcal{Z}}^{\underline{I}}=\sum_{\Lambda =1}^{22}q^{\Lambda }%
\widetilde{\mathcal{L}}_{\Lambda }^{\underline{I}}$ & $\qquad ,\qquad $ & $%
\widetilde{L}_{\Lambda }^{\underline{I}}=\widetilde{\mathcal{L}}_{\Lambda }^{%
\underline{I}}\left( \xi \right) $ & ,%
\end{tabular}%
\end{equation}%
where the $\widetilde{\mathcal{L}}_{\Lambda }^{\underline{a}}$ and $%
\widetilde{\mathcal{L}}_{\Lambda }^{\underline{I}}$ are related to the $L_{%
\underline{b}}^{\Lambda }$ and $L_{\underline{J}}^{\Lambda }$ of eqs(\ref{zl}%
) as follows 
\begin{equation}
\widetilde{\mathcal{L}}_{\Lambda }^{\underline{a}}\mathcal{L}_{\underline{b}%
}^{\Lambda }=\delta _{\underline{b}}^{\underline{a}},\qquad \qquad 
\widetilde{\mathcal{L}}_{\Lambda }^{\underline{I}}\mathcal{L}_{\underline{J}%
}^{\Lambda }=\delta _{\underline{J}}^{\underline{I}}.
\end{equation}%
The matrices $\widetilde{\mathcal{L}}_{\Lambda }^{\underline{a}}$ satisfy
constraint relations similar to those satisfied by given by $\mathcal{L}_{%
\underline{J}}^{\Lambda }$. In particular, the analogue of (\ref{zl}) reads
as%
\begin{equation}
\eta _{\underline{\Upsilon }\underline{\digamma }}\widetilde{\mathcal{L}}_{%
\underline{a}}^{\underline{\Upsilon }}\widetilde{\mathcal{L}}_{\underline{b}%
}^{\underline{\digamma }}=\eta _{\underline{a}\underline{b}},\qquad \eta _{%
\underline{\Upsilon }\underline{\digamma }}\widetilde{\mathcal{L}}_{%
\underline{I}}^{\underline{\Upsilon }}\widetilde{\mathcal{L}}_{\underline{J}%
}^{\underline{\digamma }}=\eta _{\underline{I}\underline{J}}\text{.}
\end{equation}%
We also have the factorization of the dilaton,%
\begin{equation}
\widetilde{\mathcal{L}}_{\Lambda }^{\underline{a}}=e^{+\sigma }\widetilde{L}%
_{\Lambda }^{\underline{a}},\qquad \widetilde{\mathcal{L}}_{\Lambda }^{%
\underline{I}}=e^{+\sigma }\widetilde{L}_{\Lambda }^{\underline{I}},
\end{equation}%
as well as%
\begin{equation}
\begin{tabular}{llll}
$\widetilde{\mathcal{Z}}^{\underline{a}}=e^{+\sigma }\widetilde{Z}^{%
\underline{a}}\ $ & $\qquad ,\qquad $ & $\widetilde{Z}^{\underline{a}%
}=\sum_{\Lambda =1}^{22}q^{\Lambda }\widetilde{L}_{\Lambda }^{\underline{a}%
}\ $ & , \\ 
$\widetilde{\mathcal{Z}}^{\underline{I}}=e^{+\sigma }\widetilde{Z}^{%
\underline{I}}$ & $\qquad ,\qquad $ & $\widetilde{Z}^{\underline{I}%
}=\sum_{\Lambda =1}^{22}q^{\Lambda }\widetilde{L}_{\Lambda }^{\underline{I}}$
& .%
\end{tabular}%
\end{equation}%
Putting these expressions back into eq(\ref{3b}), we obtain the factorization%
\begin{equation}
\mathcal{V}_{\text{3-brane}}^{7d,N=2}\left( \sigma ,\xi \right) =e^{+2\sigma
}\mathcal{V}_{\text{3}}\left( \xi \right) ,
\end{equation}%
with%
\begin{equation}
\mathcal{V}_{3}=\sum_{a,b=1}^{3}\delta ^{\underline{a}\underline{b}}%
\widetilde{Z}_{\underline{a}}\widetilde{Z}_{\underline{b}}+\sum_{I,J=1}^{19}%
\delta ^{\underline{I}\underline{J}}\widetilde{Z}_{\underline{I}}\widetilde{Z%
}_{\underline{J}}.
\end{equation}%
Moreover, using the usual electric/magnetic duality relation between the
electric and magnetic charges namely 
\begin{equation}
p_{\Lambda }q^{\Sigma }\sim \delta _{\Lambda }^{\Sigma },  \label{emd}
\end{equation}%
it is not difficult to check that we have the following relations, 
\begin{equation}
\widetilde{Z}^{\underline{a}}Z_{\underline{b}}\sim \delta _{\underline{b}}^{%
\underline{a}},\qquad \widetilde{Z}^{\underline{I}}Z_{\underline{J}}\sim
\delta _{\underline{J}}^{\underline{I}},
\end{equation}%
defining the duality between the dressed electric and magnetic charges.

\subsubsection{Criticality conditions}

\qquad The solutions of the criticality condition of eq(\ref{3b}) are quite
similar to those obtained for the 7D black hole. In fact they are precisely
the duals; and they may be obtained directly by making every where the
substitution 
\begin{equation}
e^{-\sigma }\rightarrow e^{+\sigma }\qquad ,\qquad Z^{\underline{a}%
}\rightarrow \widetilde{Z}_{\underline{a}}\qquad ,\qquad Z^{\underline{I}%
}\rightarrow \widetilde{Z}_{\underline{I}}.
\end{equation}%
The classification of the BPS and non BPS 3-branes is given by the dual of
eqs(\ref{cr}-\ref{cz}). Then, we have:\newline
(\textbf{1}) $\frac{1}{2}$\emph{BPS black 3-brane:}$\widetilde{\mathcal{Z}}_{%
\underline{a}}\neq \left( 0,0,0\right) ,$ $\widetilde{\mathcal{Z}}_{%
\underline{I}}=0,$ $\ \forall I.$\newline
This is a supersymmetric state preserving eight supersymmetric charges and
has a critical ADM mass as 
\begin{equation}
0<\left( \widetilde{\mathcal{M}}_{ADM}\right) _{\text{BPS}}=\left( 
\widetilde{\mathcal{M}}_{ADM}\right) _{\text{h}},
\end{equation}%
with 
\begin{equation}
\left( \widetilde{\mathcal{M}}_{ADM}\right) _{\text{BPS}}=\sqrt{\left(
\sum_{a,b=1}^{3}\delta ^{\underline{a}\underline{b}}\widetilde{\mathcal{Z}}_{%
\underline{a}}\widetilde{\mathcal{Z}}_{\underline{b}}\right) _{\text{3-brane
horizon}}}.
\end{equation}%
(\textbf{2}) \emph{non BPS 3-brane: type 1, }$\widetilde{\mathcal{Z}}_{%
\underline{a}}=\left( 0,0,0\right) ,$ $\widetilde{\mathcal{Z}}_{\underline{I}%
}\neq \left( 0,...,0\right) .$\newline
This is a non supersymmetric state with critical ADM mass $\left( \mathcal{M}%
_{ADM}\right) _{\text{h}}$ given by:%
\begin{equation}
0\leq \left( \widetilde{\mathcal{M}}_{ADM}\right) _{\text{h}}=\sqrt{\left(
\sum_{I,J=1}^{19}\delta ^{\underline{I}\underline{J}}\widetilde{\mathcal{Z}}%
_{\underline{I}}\widetilde{\mathcal{Z}}_{\underline{J}}\right) _{\text{%
3-brane horizon}}}.
\end{equation}%
(\textbf{3}) \emph{non BPS 3-brane: type 2, }$\widetilde{\mathcal{Z}}_{%
\underline{a}}\neq \left( 0,0,0\right) ,$ $\widetilde{\mathcal{Z}}_{%
\underline{I}}\neq \left( 0,...,0\right) $. \newline
Its critical ADM mass is given by%
\begin{equation}
\left( \widetilde{\mathcal{M}}_{ADM}\right) _{\text{h}}=\sqrt{\left(
\sum_{a,b\in \mathcal{J}^{\prime }}^{3}\delta ^{\underline{a}\underline{b}}%
\widetilde{\mathcal{Z}}_{\underline{a}}\widetilde{\mathcal{Z}}_{\underline{b}%
}\right) _{\text{3-brane horizon}}+\left( \sum_{I,J\in \mathcal{J}^{\prime
\prime }}^{19}\delta ^{\underline{I}\underline{J}}\widetilde{\mathcal{Z}}_{%
\underline{I}}\widetilde{\mathcal{Z}}_{\underline{J}}\right) _{\text{3-brane
horizon}}},
\end{equation}%
where \emph{some} (not all) of the geometric dressed charges as well as the
matter ones are equal to zero.

\section{Fields and fluxes in 7D supergravity}

\qquad In this section, we study the field content of the $7D$ $\mathcal{N}%
=2 $ supergravity. This analysis is not new; but it is useful for two
things: First to fix the ideas; in particular the issue regarding how the 7D
field spectrum is generated from 11D M-theory on K3. Second, it allows to
physically motivate the derivation of the Dalbeault like basis $\left\{
\Omega _{a},\Omega _{I}\right\} $ (\ref{dl}) of $H^{2}\left( K3,R\right) $
that we will develop in the next section.

We consider the $11D$- M-theory compactified on K3 determining an effective $%
7D$ $\mathcal{N}=2$ supergravity at Planck scale. Under compactification on
K3, the eleven dimensional 3-form gauge field $\mathcal{C}_{{\small MNP}}^{%
{\small 11D}}\left( \mathrm{x}^{Q}\right) =\mathcal{C}_{{\small MNP}}^{%
{\small 11D}},$ 
\begin{equation}
\mathcal{C}_{{\small MNP}}^{{\small 11D}}\equiv \mathcal{C}_{{\small MNP}}^{%
{\small 11D}}\left( x^{\mu },y^{i}\right) ,\qquad y\in K3,
\end{equation}%
with%
\begin{equation}
\begin{tabular}{llll}
$\mathrm{x}^{Q}$ & $=$ & $\left( x^{0},...,x^{10}\right) $ & , \\ 
$x^{\mu }$ & $=$ & $\left( x^{0},...,x^{6}\right) $ & , \\ 
$y^{i}$ & $=$ & $\left( x^{7},...,x^{10}\right) $ & ,%
\end{tabular}%
\end{equation}%
decomposes into:\newline
\textbf{(i)} a 7D space time real 3-form gauge field $\mathcal{C}_{\mu \nu
\rho }\left( x\right) $ (the membrane gauge field in 7D space time). It is
dual to a rank 2- tensor $B_{\mu \nu }$ field.\newline
\textbf{(ii)} twenty two (\emph{22}) 1- form gauge fields $\mathcal{A}%
^{\Lambda }$ ( 7D space time gauge particles). \newline
As these gauge particles play a central role in this study, let us give more
details.

\subsection{11D gauge 3-form on K3}

\qquad The $7D$ $\mathcal{N}=2$ supergravity theory we are considering here
is very special. It is the supersymmetric field theoretic limit of\ the 11D
M-theory on K3. \newline
This 7D theory has an abelian $U^{22}\left( 1\right) $ gauge symmetry
captured by \emph{22} Maxwell type gauge fields 
\begin{equation}
\mathcal{A}^{\Lambda }=dx^{\mu }\mathcal{A}_{\mu }^{\Lambda }\left( x\right)
,
\end{equation}%
with gauge transformation 
\begin{equation}
\mathcal{A}^{\Lambda }\rightarrow \mathcal{A}^{\Lambda }+d\left( \varepsilon
^{\Lambda }\right) .
\end{equation}%
The corresponding \emph{22} gauge invariant field strengths are%
\begin{equation}
\mathcal{F}_{2}^{\Lambda }=d\mathcal{A}^{\Lambda },\qquad \mathcal{G}%
_{5}^{\Lambda }=\text{ }^{\ast }\left( \mathcal{F}_{2}^{\Lambda }\right)
\qquad \Lambda =1,\cdots ,22.
\end{equation}%
where 
\begin{equation}
\mathcal{F}_{2}^{\Lambda }=dx^{\nu }dx^{\mu }\mathcal{F}^{\Lambda }{}_{[\mu
\nu ]}.
\end{equation}%
The gauge invariant 5- form $\mathcal{G}_{5}^{\Lambda }$ is the Hodge-dual
of $\mathcal{F}_{2}^{\Lambda }$ in seven dimension space time. For
simplicity, we shall drop out the sub-indices 2 and 5,%
\begin{equation}
\mathcal{F}_{2}^{\Lambda }\rightarrow \mathcal{F}^{\Lambda }\qquad ,\qquad 
\mathcal{G}_{5}^{\Lambda }\rightarrow \mathcal{G}^{\Lambda }.
\end{equation}%
The gauge fields $\mathcal{A}_{\mu }^{\Lambda }$ follow from the
compactification of the gauge 3- form 
\begin{equation}
\mathcal{C}_{3}^{{\small 11D}}=dx^{P}dx^{N}dx^{M}\mathcal{C}_{\left[ {\small %
MNP}\right] }^{{\small 11D}}.
\end{equation}%
Denoting by $\mathcal{H}_{4}^{{\small 11D}}$ the gauge invariant 4- form
field strength of $\mathcal{C}_{3}^{{\small 11D}}$ and by $\widetilde{%
\mathcal{H}}_{7}^{{\small 11D}}$ the 11D Hodge dual of $\mathcal{H}_{4}^{%
{\small 11D}}$, then the 7D gauge fields $\mathcal{A}^{\Lambda }$, $\mathcal{%
F}^{\Lambda }$ and $\mathcal{G}^{\Lambda }$\ can be defined as: 
\begin{equation}
\mathcal{A}^{\Lambda }=\int_{\Psi ^{\Lambda }}\mathcal{C}_{3},\qquad 
\mathcal{F}^{\Lambda }=\int_{\Psi ^{\Lambda }}\mathcal{H}_{4}^{{\small 11D}%
},\qquad \mathcal{G}^{\Lambda }=\int_{\Psi ^{\Lambda }}\widetilde{\mathcal{H}%
}_{7}^{{\small 11D}},  \label{lam}
\end{equation}%
where $\Psi ^{\Lambda }\in H_{2}\left( K3,R\right) $ is a real basis of
2-cycles.\newline
The integration of the field strength $\mathcal{F}^{\Lambda }$ (resp. $%
\mathcal{G}^{\Lambda }$) throughout the sphere $S_{\infty }^{2}$ (resp. $%
S_{\infty }^{5}$) give the magnetic (resp. electric) charges $p^{\Lambda }$
(resp. $q^{\Lambda }$), 
\begin{equation}
p^{\Lambda }=\int_{S_{\infty }^{2}}\mathcal{F}^{\Lambda }\qquad ,\qquad
q^{\Lambda }=\int_{S_{\infty }^{5}}\mathcal{G}^{\Lambda }.
\end{equation}%
Up on using eq(\ref{lam}), these magnetic and electric charges can be also
put in the following way by using 11D gauge fields and the second homology
basis $\left\{ \Psi ^{\Lambda }\right\} $ of K3,%
\begin{equation}
\begin{tabular}{llll}
$p^{\Lambda }$ & $=$ & $\int_{S_{\infty }^{2}}\left( \int_{\Psi ^{\Lambda }}%
\mathcal{H}_{4}^{{\small 11D}}\right) $ & , \\ 
$q^{\Lambda }$ & $=$ & $\int_{S_{\infty }^{5}}\left( \int_{\Psi ^{\Lambda }}%
\widetilde{\mathcal{H}}_{7}^{{\small 11D}}\right) $ & .%
\end{tabular}%
\end{equation}%
The magnetic charges $p^{\Lambda }$ and the electric ones $q_{\Lambda }$
obey the usual Dirac quantization (\ref{emd}).

\subsection{Two 7D $N=2$\ supersymmetric representations}

\qquad From the view of the 7D $\mathcal{N}=2$ supergravity, the \emph{22}
gauge fields $\mathcal{A}_{\mu }^{\Lambda }$ do not carry the same
supersymmetric quantum numbers. It happens that the $\mathcal{A}_{\mu
}^{\Lambda }$ and the corresponding field strengths $\mathcal{F}_{\mu \nu
}^{\Lambda }$ and $^{\ast }\left( \mathcal{F}_{\mu \nu }^{\Lambda }\right) $
split into \emph{triplets} and \emph{19-uplets} as shown below,%
\begin{equation}
\left( \mathcal{A}_{\mu }^{\Lambda }\right) =\left( \mathcal{A}_{\mu
}^{a}\right) \oplus \left( \mathcal{A}_{\mu }^{I}\right) ,  \label{dec}
\end{equation}%
and 
\begin{equation}
\left( \mathcal{F}_{\mu \nu }^{\Lambda }\right) =\left( \mathcal{F}_{\mu \nu
}^{a}\right) \oplus \left( \mathcal{F}_{\mu \nu }^{I}\right) ,
\end{equation}%
as well as%
\begin{equation}
\text{ }^{\ast }\left( \mathcal{F}_{\mu \nu }^{\Lambda }\right) =\text{ }%
^{\ast }\left( \mathcal{F}_{\mu \nu }^{a}\right) \oplus \text{ }^{\ast
}\left( \mathcal{F}_{\mu \nu }^{I}\right) .  \label{dac}
\end{equation}%
The component fields $\mathcal{A}_{\mu }^{a}$, $\mathcal{F}_{\mu \nu }^{a}$
and $^{\ast }\left( \mathcal{F}_{\mu \nu }^{a}\right) $ transform as real
vectors under $SO\left( 3\right) $; but like real scalars under $SO\left(
19\right) $. \newline
Similarly, the component fields $\mathcal{A}_{\mu }^{I}$, $\mathcal{F}_{\mu
\nu }^{I}$ and $^{\ast }\left( \mathcal{F}_{\mu \nu }^{I}\right) $ transform
as real scalars under $SO\left( 3\right) $; but like real vectors under $%
SO\left( 19\right) $.\newline
This property translates the fact that the \emph{22} abelian gauge fields
belong to two different 7D $N=2$ supersymmetric representations, namely the $%
7D$ $\mathcal{N}=2$ supergravity multiplet, denoted as, 
\begin{equation}
\mathcal{G}_{7D,\mathcal{N}=2},
\end{equation}%
and the $7D$ $\mathcal{N}=2$ gauge multiplets 
\begin{equation}
\left( \mathcal{V}_{7D,\mathcal{N}=2}\right) ^{I},\qquad I=1,...,19.
\end{equation}%
Below we comment briefly these two representations:

\subsubsection{Supergravity multiplet $\mathcal{G}_{7D,\mathcal{N}=2}$}

\qquad The component fields content of the $7D$ $\mathcal{N}=2$ supergravity
multiplet $\mathcal{G}_{7D,\mathcal{N}=2}$ reads as follows:%
\begin{equation}
\begin{tabular}{lllll}
$\text{Bosons}:$ & $G_{\mu \nu }\left( x\right) ,$ & $\mathcal{C}_{\mu \nu
\rho }\left( x\right) ,$ & $\mathcal{A}_{\mu }^{a}\left( x\right) ,$ & $%
\sigma \left( x\right) $ \\ 
&  &  &  &  \\ 
$\text{Fermions}:$ & $\mathrm{\psi }_{\mathrm{\alpha }\mu }^{1}\left(
x\right) ,$ & $\mathrm{\psi }_{\mathrm{\alpha }\mu }^{2}\left( x\right) ,$ & 
$\mathrm{\chi }_{\mathrm{\alpha }}^{1}\left( x\right) ,$ & $\mathrm{\chi }_{%
\mathrm{\alpha }}^{2}\left( x\right) $%
\end{tabular}
\label{go}
\end{equation}%
The first line refers to the 7D bosonic fields; they describe respectively
the 7D graviton $G_{\mu \nu }$, the 7D antisymmetric 3-form gauge field $%
\mathcal{C}_{\mu \nu \rho }$, the space time 1- form gauge fields triplet $%
\mathcal{A}_{\mu }^{a}$ and the 7D dilaton $\mathcal{\sigma }$. \newline
The second line refers to the 7D fermionic field partners namely:\newline
(\textbf{i}) two 7D gravitinos $\left( \mathrm{\psi }_{\mathrm{\alpha }\mu
}^{1},\mathrm{\psi }_{\mathrm{\alpha }\mu }^{2}\right) $ \newline
(\textbf{ii}) two 7D\ gravi-photinos $\left( \mathrm{\chi }_{\mathrm{\alpha }%
}^{1},\mathrm{\chi }_{\mathrm{\alpha }}^{2}\right) $:\newline
Both of these fermionic fields form isodoublets of the $USP_{R}\left( 2,%
\mathbb{R}\right) $\ automorphism symmetry$\footnote{%
The automorphism group $USP\left( 2,\mathbb{R}\right) $ of the 7D $\mathcal{N%
}=2$ superalgebra is related to the $SO\left( 3\right) $ isotropy symmetry
factor of the moduli space. The homomorphism is given by the usual relation $%
x^{\left( \alpha \beta \right) }=\sum x^{a}\left( \mathcal{\sigma }%
_{a}\right) ^{\alpha \beta }$ mapping the adjoint of $USP\left( 2,\mathbb{R}%
\right) $\ to the 3- vector of $SO\left( 3\right) $.}$ of the 7D $\mathcal{N}%
=2$ superalgebra.

\subsubsection{Abelian gauge supermultiplets}

\qquad The component fields content of the \emph{nineteen} $7D$ $\mathcal{N}%
=2$ abelian gauge supermultiplets $\mathcal{V}_{7D,\mathcal{N}=2}^{I}$ is
given by%
\begin{equation}
\begin{tabular}{lllll}
$\text{Bosons}:$ & $\mathcal{A}_{\mu }^{I}$ & $,$ & $\text{\ }\mathcal{\phi }%
^{aI}$ & , \\ 
&  &  &  &  \\ 
$\text{Fermions}:$ & $\mathrm{\lambda }_{\mathrm{\alpha }}^{1I}$ & $,$ & $%
\mathrm{\lambda }_{\mathrm{\alpha }}^{2I}$ & .%
\end{tabular}
\label{ga}
\end{equation}%
Each multiplet $\mathcal{V}_{7D,\mathcal{N}=2}$ consists of :\newline
(\textbf{i}) a $7D$ gauge field $\mathcal{A}_{\mu }$, which is a singlet
under the $USP\left( 2,\mathbb{R}\right) ,$\newline
(\textbf{ii}) two $7D$ fermions $\left( \mathrm{\lambda }_{\mathrm{\alpha }%
}^{1},\mathrm{\lambda }_{\mathrm{\alpha }}^{2}\right) $ forming an
isodoublet under the $USP\left( 2,\mathbb{R}\right) $ automorphism symmetry
of the $7D$ $\mathcal{N}=2$ superalgebra\newline
(\textbf{iii}) three 7D scalar fields\footnote{%
For simplicity, we shall refer to the gauge multiplet $\mathcal{V}_{7D}^{%
\mathcal{N}=2}$ as matter and to the gravity $\mathcal{G}_{7D}^{\mathcal{N}%
=2}$ as geometry.} $\mathcal{\phi }^{a}=\left( \phi ^{1},\phi ^{2},\phi
^{3}\right) $ forming an $USP\left( 2,\mathbb{R}\right) $ isotriplet. 
\newline
The gauge fields (\ref{ga}) capture different quantum numbers of the $%
SO\left( 3\right) \times SO\left( 19\right) $ isotropy symmetry of the
moduli space $\boldsymbol{M}_{7D}^{N=2}$ 
\begin{equation}
\boldsymbol{M}_{7D}^{N=2}=\mathcal{G}\times SO\left( 1,1\right) ,\qquad 
\mathcal{G}=\frac{SO\left( 3,19\right) }{SO\left( 3\right) \times SO\left(
19\right) },  \label{md}
\end{equation}%
where $SO\left( 3\right) $ should be thought of as the R- symmetry group $%
USP\left( 2,\mathbb{R}\right) $. For the matter multiplet ($\mathcal{V}_{7D,%
\mathcal{N}=2}$, see \emph{footnote 8}), we have%
\begin{equation}
\text{{\small Bosons}}:\qquad 
\begin{tabular}{lll}
$\mathcal{A}_{\mu }^{I}$ & $\quad \sim \quad $ & $\left( 1,19\right) $ \\ 
$\phi ^{aI}$ & $\quad \sim \quad $ & $\left( 3,19\right) $%
\end{tabular}
\label{fi}
\end{equation}%
and 
\begin{equation}
\text{{\small Fermions}}:\qquad 
\begin{tabular}{lll}
$\left( \mathrm{\lambda }_{\mathrm{\alpha }}^{1I},\mathrm{\lambda }_{\mathrm{%
\alpha }}^{2I}\right) $ & $\quad \sim \quad $ & $\left( 2,19\right) $%
\end{tabular}%
,  \label{lad}
\end{equation}%
where $\left( s,19\right) $, with $s=1,2,3$, refer to $SO\left( 3\right)
\times SO\left( 19\right) $ representations.\newline
A quite similar classification can made for the fields of the supergravity
multiplet $\mathcal{G}_{7D,\mathcal{N}=2}$. The quantum numbers of the
supergravity fields under the $SO\left( 3\right) \times SO\left( 19\right) $
isotropy symmetry is as follows:%
\begin{equation}
\text{Bosons}:\qquad 
\begin{tabular}{llll}
$G_{\mu \nu }$ & $\quad \sim \quad $ & $\left( 1,1\right) $ & , \\ 
$\mathcal{C}_{\mu \nu \rho }$ & $\quad \sim \quad $ & $\left( 1,1\right) $ & 
, \\ 
$\mathcal{A}_{\mu }^{a}$ & $\quad \sim \quad $ & $\left( 3,1\right) $ & , \\ 
$\sigma $ & $\quad \sim \quad $ & $\left( 1,1\right) $ & ,%
\end{tabular}
\label{ad}
\end{equation}%
and%
\begin{equation}
\text{Fermions}:\qquad 
\begin{tabular}{llll}
$\left( \mathrm{\psi }_{\mathrm{\alpha }\mu }^{1},\text{ }\mathrm{\psi }_{%
\mathrm{\alpha }\mu }^{2}\right) $ & $\quad \sim \quad $ & $\left(
2,1\right) $ & , \\ 
$\left( \mathrm{\chi }_{\mathrm{\alpha }}^{1},\text{ }\mathrm{\chi }_{%
\mathrm{\alpha }}^{2}\right) $ & $\quad \sim \quad $ & $\left( 2,1\right) $
& .%
\end{tabular}
\label{fa}
\end{equation}%
Notice that all the fields of $\mathcal{G}_{7D,\mathcal{N}=2}$ are scalar
under $SO\left( 19\right) $; but can be either isosinglets, isodoublets or
isotriplets under $SO\left( 3\right) \sim $ $USP\left( 2,\mathbb{R}\right) $.

In what follows, and in order to alleviate the notations, we shall drop out
the 7D spinor index $\mathrm{\alpha }$ (Roman character). We will use the
index $\alpha $ (\emph{in Math character}) to refer to the isospin 1/2
representation of the $USP\left( 2,\mathbb{R}\right) $ symmetry group. The 
\emph{two} gravitinos, the\emph{\ two} gravi-photinos and the \emph{nineteen}
gaugino doublets will be collectively written as follows%
\begin{equation}
\begin{tabular}{llll}
$\mathrm{\psi }_{\mathrm{\alpha }\mu }^{\beta }=\left( \mathrm{\psi }_{%
\mathrm{\alpha }\mu }^{1},\text{ }\mathrm{\psi }_{\mathrm{\alpha }\mu
}^{2}\right) $ & $\qquad \rightarrow \qquad $ & $\mathrm{\psi }_{\mu
}^{\beta }=\left( \mathrm{\psi }_{\mu }^{1},\text{ }\mathrm{\psi }_{\mu
}^{2}\right) $ & , \\ 
$\mathrm{\chi }_{\mathrm{\alpha }}^{\beta }=\left( \mathrm{\chi }_{\mathrm{%
\alpha }}^{1},\text{ }\mathrm{\chi }_{\mathrm{\alpha }}^{2}\right) $ & $%
\qquad \rightarrow \qquad $ & $\mathrm{\chi }^{\beta }=\left( \mathrm{\chi }%
^{1},\text{ }\mathrm{\chi }^{2}\right) $ & , \\ 
$\mathrm{\lambda }_{\mathrm{\alpha }}^{\beta I}=\left( \mathrm{\lambda }_{%
\mathrm{\alpha }}^{1I},\text{ }\mathrm{\lambda }_{\mathrm{\alpha }%
}^{2}\right) $ & $\qquad \rightarrow \qquad $ & $\mathrm{\lambda }^{\beta
I}=\left( \mathrm{\lambda }^{1I},\text{ }\mathrm{\lambda }^{2I}\right) $ & ,%
\end{tabular}
\label{fd}
\end{equation}%
where the space time spinor index $\mathrm{\alpha }$ has been dropped out.
We also have the relation between $USP\left( 2,\mathbb{R}\right) $ and $%
SO\left( 3,\mathbb{R}\right) ,$%
\begin{equation}
\phi ^{aI}=\sum_{\alpha ,\beta =1}^{2}\mathcal{\sigma }_{\alpha \beta
}^{a}\phi ^{\left( \alpha \beta \right) I}\quad ,\quad a=1,2,3,
\end{equation}%
where $\phi ^{\left( \alpha \beta \right) }$ stands for the symmetric part
of $\phi ^{\alpha \beta }$.

\section{Deriving the $\left\{ \Omega _{a},\Omega _{I}\right\} $ basis of $%
H^{2}\left( K3,\mathbb{R}\right) $}

\qquad In this section, we use physical arguments to construct one of the
basis tools to deal with the \emph{special hyperKahler geometry} (SHG) of
the 11D M- theory on K3. This construction concerns the derivation of a "%
\emph{Dalbeault like"} basis $\left\{ \Omega _{a},\Omega _{I}\right\} $ of
the second real cohomology of K3. This is a real 22 dimensional 2-form basis
of $H^{2}\left( K3,\mathbb{R}\right) $ 
\begin{equation}
\left\{ \Omega _{a},\Omega _{I}\right\} ,\qquad a=1,2,3,\qquad I=1,...,19,
\label{sa}
\end{equation}%
with the particularity of combining both the\ K3 Kahler 2-form%
\begin{equation}
\Omega ^{0}=\Omega ^{\left( 1,1\right) },
\end{equation}%
and the associated complex holomorphic and antiholomorphic 2-forms%
\begin{equation}
\Omega ^{+}=\Omega ^{\left( 2,0\right) },\qquad \Omega ^{-}=\Omega ^{\left(
0,2\right) },
\end{equation}%
in an $SO\left( 3\right) $ isotriplet 
\begin{equation}
\Omega ^{a}=\left( \Omega ^{+},\Omega ^{0},\Omega ^{-}\right) .
\end{equation}%
This operation corresponds naively to combining the Kahler $t_{I}\equiv
z_{I}^{0}$ and complex deformation $z_{I}^{\pm }$ moduli of the metric of K3
into \emph{19} isotriplets $\xi _{I}^{a}$ with $a=0,\pm $.\newline
The 19-uplet 2-forms $\Omega _{I}$, which turn out to be equal to the
covariant derivative of $\Omega ^{a}$; i.e 
\begin{equation}
\Omega _{I}=\mathcal{D}_{aI}\Omega ^{a},
\end{equation}
can be imagined as the real 2-form generating $SO\left( 3\right) $ spherical
deformations of the metric of K3.

To that purpose, we start by recalling some useful results on the special
Kahler geometry (SKG) of 10D type IIB superstring on CY3s; in particular the
role played by the Dalbeault basis of $H^{3}\left( CY3,R\right) $. Then, we
derive eq(\ref{sa}) by using constraint relations from 7D $\mathcal{N}=2$
supergravity theory. More analysis on the the special hyperKahler geometry
(SHG) set up using the basis (\ref{sa}) will be developed in the next
sections.

\subsection{General on SKG of CY3}

\qquad Following \textrm{\ref{f}}, the third real cohomology $H^{3}\left(
X_{3},\mathbb{R}\right) $ of the Calabi-Yau $X_{3}$ threefold can be
Hodge-decomposed along the third Dalbeault basis as follows,%
\begin{equation}
H^{3}\left( X_{3},\mathbb{R}\right) =H^{3,0}\left( X_{3}\right) \oplus
_{s}H^{2,1}\left( X_{3}\right) \oplus _{s}H^{1,2}\left( X_{3}\right) \oplus
_{s}H^{0,3}\left( X_{3}\right) ,
\end{equation}%
where the subscript \emph{s} stands for the semi-direct cohomological sum
due to non vanishing intersections. \newline
The above Hodge decomposition corresponds to make a change of basis from the
usual real symplectic basis$\footnote{%
In this subsection $\mathbf{\alpha }_{\Lambda }$ and $\mathbf{\beta }%
^{\Lambda }$ are 3-forms of $H^{3}\left( CY3\right) $; they should not be
confused with the Hodge basis of $H^{2}\left( K3\right) $ denoted by the
same letters.}$ of $H^{3}\left( X_{3},\mathbb{R}\right) $ namely,%
\begin{equation}
\begin{tabular}{lllll}
$\mathbf{\alpha }_{\Lambda }$ & , & $\mathbf{\beta }^{\Lambda }$ & , & $%
\Lambda =0,....,h^{2,1}$%
\end{tabular}
\label{cb}
\end{equation}%
to the Dalbeault basis 
\begin{equation}
\begin{tabular}{lllllll}
$\Omega _{3}$ & $,$ & $D_{i}\Omega _{3}$ & $,$ & $\overline{D}_{\overline{i}}%
\overline{\Omega }_{3}$ & $,$ & $\overline{\Omega }_{3}$%
\end{tabular}%
,  \label{db}
\end{equation}%
where $i=1,...,h^{1,2}\left( CY3\right) $. \newline
In the above relation, $\Omega _{3}\in H^{3,0}\left( CY3\right) $ and $%
\overline{\Omega }_{3}\in H^{0,3}\left( CY3\right) $ stand respectively for
the usual holomorphic and antiholomorphic 3-forms on the Calabi-Yau
threefold $X_{3}$. They are expressed in terms of $\mathbf{\alpha }_{\Lambda
}$ and $\mathbf{\beta }^{\Lambda }$ like%
\begin{equation}
\begin{tabular}{lll}
$\Omega _{3}\left( z\right) $ & $=X^{\Lambda }\left( z\right) \mathbf{\alpha 
}_{\Lambda }-F_{\Lambda }\left( z\right) \mathbf{\beta }^{\Lambda }$ & , \\ 
$\overline{\Omega }_{3}\left( \overline{z}\right) $ & $=\overline{X}%
^{\Lambda }\left( \overline{z}\right) \mathbf{\alpha }_{\Lambda }-\overline{F%
}_{\Lambda }\left( \overline{z}\right) \mathbf{\beta }^{\Lambda }$ & .%
\end{tabular}%
\end{equation}%
Here, the moduli space coordinate variables%
\begin{equation}
z=\left( z^{i}\right) ,\qquad \overline{z}=\left( \overline{z}^{\overline{i}%
}\right) ,\qquad z^{0}=1,
\end{equation}%
are the complex structure moduli describing the complex deformations of the
metric of $X_{3}$ and 
\begin{equation}
\left\{ X^{\Lambda }\left( z\right) \text{ },\text{ }F_{\Lambda }\left(
z\right) \right\} ,
\end{equation}%
with the property%
\begin{equation}
\left( X^{\Lambda },F_{\Lambda }\right) \rightarrow e^{f\left( z\right)
}\left( X^{\Lambda },F_{\Lambda }\right) ,\qquad \frac{\partial X^{\Lambda }%
}{\partial \overline{z}}=0,\qquad \frac{\partial F_{\Lambda }}{\partial 
\overline{z}}=0,
\end{equation}%
is a basis of symplectic holomorphic fundamental periods of $\Omega _{3}$\
around the 3-cycles $\left\{ A^{\Lambda },B_{\Lambda }\right\} $, 
\begin{equation}
X^{\Lambda }=\int_{A^{\Lambda }}\Omega _{3},\qquad F_{\Lambda
}=\int_{B_{\Lambda }}\Omega _{3}.
\end{equation}%
Recall that the set of real 3-forms $\left\{ \mathbf{\alpha }_{\Lambda },%
\mathbf{\beta }^{\Lambda }\right\} $ satisfy the symplectic structure%
\begin{equation}
\begin{tabular}{llll}
$\left\langle \mathbf{\alpha }_{\Lambda },\mathbf{\beta }^{\Sigma
}\right\rangle $ & $=$ & $\delta _{\Lambda }^{\Sigma }$ & , \\ 
$\left\langle \mathbf{\alpha }_{\Lambda },\mathbf{\alpha }_{\Sigma
}\right\rangle $ & $=$ & $0$ & , \\ 
$\left\langle \mathbf{\beta }^{\Lambda },\mathbf{\beta }^{\Sigma
}\right\rangle $ & $=$ & $0$ & ,%
\end{tabular}%
\end{equation}%
where the inner product of two 3- forms $F$ and $G$ is defined as%
\begin{equation}
\left\langle F,G\right\rangle =\int_{CY3}F\wedge G=-\left\langle
G,F\right\rangle .
\end{equation}%
By Poincar\'{e} duality of the 3-forms $\left\{ \mathbf{\alpha }_{\Lambda }%
\mathbf{,\beta }^{\Lambda }\right\} $ on the Calabi-Yau threefold, we also
have the set of real 3-cycles 
\begin{equation}
\left\{ A^{\Lambda },B_{\Lambda }\right\} \text{ \ ,}
\end{equation}%
dual to (\ref{cb}) and generating the third real homology $H_{3}\left( CY3,%
\mathbb{R}\right) $. The basis $\left\{ \mathbf{\alpha }_{\Lambda }\mathbf{%
,\beta }^{\Lambda }\right\} $ and its dual $\left\{ A^{\Lambda },B_{\Lambda
}\right\} $ satisfy%
\begin{equation}
\begin{tabular}{llll}
$\int_{A^{\Lambda }}\mathbf{\alpha }_{\Sigma }=\delta _{\Sigma }^{\Lambda }$
& $\qquad ,\qquad $ & $\int_{A^{\Lambda }}\mathbf{\beta }^{\Sigma }=0$ & ,
\\ 
$\int_{B_{\Lambda }}\mathbf{\alpha }_{\Sigma }=0$ & $\qquad ,$ & $%
\int_{B_{\Lambda }}\mathbf{\beta }^{\Sigma }=-\delta _{\Sigma }^{\Lambda }$
& .%
\end{tabular}
\label{alb}
\end{equation}%
We also have the following fundamental relations of special Kahler geometry%
\begin{equation}
\begin{tabular}{lll}
$\left\langle \Omega _{3},\overline{\Omega }_{3}\right\rangle $ & $=$ & $%
-ie^{-K}$ \\ 
$\left\langle D_{i}\Omega _{3},\overline{D}_{i}\overline{\Omega }%
_{3}\right\rangle $ & $=$ & $ig_{i\overline{j}}e^{-K}$%
\end{tabular}%
\end{equation}%
together with (\emph{see also the appendix})%
\begin{equation}
\begin{tabular}{lllll}
$\left\langle \Omega _{3},\Omega _{3}\right\rangle $ & $=$ & $\left\langle 
\overline{\Omega }_{3},\overline{\Omega }_{3}\right\rangle $ & $=0$ & , \\ 
$\left\langle \Omega _{3},\overline{D}_{i}\overline{\Omega }%
_{3}\right\rangle $ & $=$ & $\left\langle D_{i}\Omega _{3},\overline{\Omega }%
_{3}\right\rangle $ & $=0$ & .%
\end{tabular}%
\end{equation}%
Recall also that the Dalbeault basis (\ref{db}) of the cohomology of CY3 has
been shown to be particularly convenient to deal with the two following
things:\newline
\textbf{(1)} the SKG of the 10D type IIB superstring on CY3; in particular
in the study of the effective scalar potential of 4D $\mathcal{N}=2$
supergravity and the characterization of the BPS and non BPS 4D black holes.%
\newline
\textbf{(2)} the development of the \emph{"new attractor"} approach of the
4D $\mathcal{N}=2$\ supergravity and 4D $\mathcal{N}=1$\ supergravity with
fluxes \textrm{\cite{F1,F2}.}

Our purpose below is to build the analogue of the above relations for the
SHG of the 11D M-theory on K3. Using special features of the Hodge
decomposition of the second real cohomology of K3, we show that the analogue
of eq(\ref{db}) is, in some sense, given by (\ref{sa}) where $\Omega _{a}$
is an real isotriplet and $\Omega _{I}$ is a real 19-uplet.\newline
Because of the formal similarity with eqs(\ref{db}), we will sometimes refer
to the basis (\ref{sa}) as the \emph{Dalbeault like} basis for the second
real cohomology of K3. Nevertheless, one should note that there is a basic
difference between eqs(\ref{db}) and (\ref{sa}); the first one deals with
complex deformations of CY3 while the second deals with the combined Kahler
and complex deformations of K3.

\subsection{A special basis of $H_{2}\left( K3,R\right) $}

\qquad In this subsection, we derive the \emph{Dalbeault like} basis (\ref%
{sa}) by using special features of the underlying symmetries of the 7D $%
\mathcal{N}=2$ supergravity field theory; in particular:\newline
(\textbf{1}) the splitting of the fields content of 7D $\mathcal{N}=2$
supergravity in two irreducible supersymmetric representations,\newline
(\textbf{2}) the combination of the Kahler and complex deformations of the
metric of K3. This combination allows to group altogether the deformations
moduli into isotriplets. \newline
These two properties are not completely independent; they are in fact
different ways to state the implementation of the $SO\left( 3\right) \times
SO\left( 19\right) $ isotropy symmetry of the moduli space $\boldsymbol{M}%
_{7D}^{N=2}$ in the supergravity field theory.

\subsubsection{Supersymmetric representation constraints}

\qquad The 7D $\mathcal{N}=2$ supergravity embedded in 11D M-theory on K3
has several space time fields with different quantum numbers. For instance,
the \emph{22} abelian gauge fields $\mathcal{A}^{\underline{\Lambda }%
}=dx^{\mu }\mathcal{A}_{\mu }^{\underline{\Lambda }}$ with%
\begin{equation}
\mathcal{A}^{\underline{\Lambda }}=\int_{\Psi ^{\Lambda }}\mathcal{C}_{3}^{%
{\small 11D}},\qquad \Lambda =1,...,22,  \label{gfi}
\end{equation}%
belong to two different irreducible representations of 7D $\mathcal{N}=2$
supersymmetric algebra. These supersymmetric representations correspond to
the gravity multiplet%
\begin{equation}
\mathcal{G}_{7D,\mathcal{N}=2},
\end{equation}%
and the gauge (matter) supermultiplet 
\begin{equation}
\mathcal{V}_{7D,\mathcal{N}=2}.
\end{equation}%
From eqs(\ref{go},\ref{ga}), we see that the gauge fields $\mathcal{A}_{\mu
}^{\underline{\Lambda }}$ of eq(\ref{gfi}) split into two basic sets (\ref%
{dec}):\newline
(\textbf{i}) \emph{3} gauge fields $\mathcal{A}_{\mu }^{\underline{a}}$,
belonging to the gravity multiplet $\mathcal{G}_{7D,\mathcal{N}=2}$. \newline
(\textbf{ii}) \emph{19} gauge fields $\mathcal{A}_{\mu }^{I}$, belonging to
the gauge multiplets $\mathcal{V}_{7D,\mathcal{N}=2}^{I}$.\newline

\emph{Splitting the system }$\left\{ \mathcal{A}_{\mu }^{\underline{\Lambda }%
}\right\} $\newline
As noted before, the gauge fields $\mathcal{A}_{\mu }^{\underline{\Lambda }}$
and $\mathcal{F}_{\mu \nu }^{\underline{\Lambda }}$ are not exactly what it
seen by $\mathcal{N}=2$ supersymmetry in the 7D space time. What required by
the irreducible representations of the 7D $\mathcal{N}=2$ superalgebra are
precisely the gravi-photon isotriplet 
\begin{equation}
\mathcal{A}_{\mu }^{\underline{a}}=\mathcal{A}_{\mu }^{\underline{a}}\left(
x\right)
\end{equation}%
and the Maxwell gauge fields 
\begin{equation}
\mathcal{A}_{\mu }^{\underline{I}}=\mathcal{A}_{\mu }^{\underline{I}}\left(
x\right)
\end{equation}%
describing 19 "photons" in the gauge sector. \newline
This means that the "physical quantities"; in particular the gauge fields $%
\mathcal{A}_{\mu }^{\underline{a}}$ and $\mathcal{A}_{\mu }^{\underline{I}}$
as well as the corresponding field strengths $\mathcal{F}_{\mu \nu }^{%
\underline{a}}$ and $\mathcal{F}_{\mu \nu }^{\underline{I}}$ can be defined
as linear combinations of $\mathcal{A}_{\mu }^{\underline{\Lambda }}$ and $%
\mathcal{F}_{\mu \nu }^{\underline{\Lambda }}$ as follows%
\begin{equation}
\begin{tabular}{ll}
$\mathcal{A}_{\mu }^{\underline{a}}=\sum_{\Lambda =1}^{22}Q_{\underline{%
\Lambda }}^{\underline{a}}\mathcal{A}_{\mu }^{\underline{\Lambda }}$ & , \\ 
&  \\ 
$\mathcal{F}_{\mu \nu }^{\underline{a}}=\sum_{\Lambda =1}^{22}Q_{\underline{%
\Lambda }}^{\underline{a}}\mathcal{F}_{\mu \nu }^{\underline{\Lambda }}$ & .%
\end{tabular}
\label{61}
\end{equation}%
where the decomposition coefficients $Q_{\underline{\Lambda }}^{\underline{a}%
}=Q_{\underline{\Lambda }}^{\underline{a}}\left( \xi \right) $ are local
field tensors whose interpretation will be given in a moment. \newline
To fix the ideas, think about the $22\times 22$ matrix, which can be split
like 
\begin{equation}
Q_{\underline{\Lambda }}^{\underline{\Sigma }}=\left( Q_{\underline{\Lambda }%
}^{\underline{a}},Q_{\underline{\Lambda }}^{\underline{I}}\right) \text{ },
\end{equation}%
as an orthogonal matrix 
\begin{equation}
Q_{\underline{\Lambda }}^{\underline{\Sigma }}\in SO\left( 3,19\right) .
\label{60}
\end{equation}%
Similarly, the gauge fields $\mathcal{A}_{\mu }^{\underline{I}}$ and the
corresponding field strengths $\mathcal{F}_{\mu \nu }^{\underline{I}}$ may
be defined as well as linear combinations of $\mathcal{A}_{\mu }^{\underline{%
\Lambda }}$ and $\mathcal{F}_{\mu \nu }^{\underline{\Lambda }}$ like,%
\begin{equation}
\begin{tabular}{ll}
$\mathcal{A}_{\mu }^{\underline{I}}=\sum_{\Lambda =1}^{22}Q_{\underline{%
\Lambda }}^{\underline{I}}\mathcal{A}_{\mu }^{\underline{\Lambda }}$ & , \\ 
&  \\ 
$\mathcal{F}_{\mu \nu }^{\underline{I}}=\sum_{\Lambda =1}^{22}Q_{\underline{%
\Lambda }}^{\underline{I}}\mathcal{F}_{\mu \nu }^{\underline{\Lambda }}$ & $%
. $%
\end{tabular}
\label{62}
\end{equation}%
where $Q_{\underline{\Lambda }}^{\underline{I}}$ are as in eq(\ref{60}).
Moreover, inverting eqs(\ref{61}-\ref{62}) as follows,%
\begin{equation}
\begin{tabular}{ll}
$\mathcal{A}_{\mu }^{\underline{\Lambda }}=\sum_{a=1}^{3}L_{\underline{a}}^{%
\underline{\Lambda }}\mathcal{A}_{\mu }^{\underline{a}}+\sum_{I=1}^{19}L_{%
\underline{I}}^{\underline{\Lambda }}\mathcal{A}_{\mu }^{\underline{I}}$ & ,
\\ 
&  \\ 
$\mathcal{F}_{\mu \nu }^{\underline{\Lambda }}=\sum_{a=1}^{3}L_{\underline{a}%
}^{\underline{\Lambda }}\mathcal{F}_{\mu \nu }^{\underline{a}%
}+\sum_{I=1}^{19}L_{\underline{I}}^{\underline{\Lambda }}\mathcal{F}_{\mu
\nu }^{\underline{I}}$ & ,%
\end{tabular}
\label{63}
\end{equation}%
where the decomposition coefficients $L_{\underline{a}}^{\underline{\Lambda }%
}$ and $L_{\underline{I}}^{\underline{\Lambda }}$ are local fields, we can
get information on the matrices $Q_{\underline{\Lambda }}^{\underline{%
\Upsilon }}$ and $L_{\underline{\Upsilon }}^{\underline{\Sigma }}$. \newline
Substituting the decomposition (\ref{63}) back into (\ref{61}-\ref{62}), we
get the following relation%
\begin{equation}
\sum_{a=1}^{3}Q_{\underline{\Lambda }}^{\underline{a}}L_{\underline{a}}^{%
\underline{\Sigma }}+\sum_{I=1}^{19}Q_{\underline{\Lambda }}^{\underline{I}%
}L_{\underline{I}}^{\underline{\Sigma }}\equiv \sum_{\Upsilon =1}^{22}Q_{%
\underline{\Lambda }}^{\underline{\Upsilon }}L_{\underline{\Upsilon }}^{%
\underline{\Sigma }}=\delta _{\Lambda }^{\Sigma }.
\end{equation}%
Using the flat metric tensors $\eta _{\underline{a}\underline{b}}=+\delta _{%
\underline{a}\underline{b}}$ and $\eta _{\underline{I}\underline{J}}=-\delta
_{\underline{I}\underline{J}}$ of the inertial frame, we can put this
relation into the form%
\begin{equation}
Q_{\underline{\Lambda }}^{\underline{a}}\mathrm{\eta }_{\underline{a}%
\underline{b}}L_{\underline{\Sigma }}^{\underline{b}}+Q_{\underline{\Lambda }%
}^{\underline{I}}\mathrm{\eta }_{\underline{I}\underline{J}}L_{\underline{%
\Sigma }}^{\underline{J}}=\eta _{\underline{\Lambda }\underline{\Sigma }%
},\qquad \Leftrightarrow \qquad Q\mathrm{\eta }_{22\times 22}L=\mathrm{\eta }%
_{22\times 22}
\end{equation}%
which is precisely the $SO\left( 3,19\right) $ orthogonality relation we
have described in sections 2 and 3.

\subsubsection{The dual of $\left\{ \Omega _{a},\Omega _{I}\right\} $}

\qquad 7D $\mathcal{N}=2$ supersymmetry puts a strong constraint on the
underlying SHG of the 7D supergravity theory. It requires a particular real
2- cycle basis of $H_{2}\left( K3\right) $ 
\begin{equation}
\left\{ B^{a}\text{ },\text{ }B^{I}\right\} \text{ },
\end{equation}%
which allows to define the gauge fields $\mathcal{A}_{\mu }^{a}$ and $%
\mathcal{A}_{\mu }^{I}$ of the supergravity theory like 
\begin{equation}
\begin{tabular}{llll}
$\text{Gravity}:$ & $\mathcal{A}_{\mu }^{a}=\int_{B^{a}}\mathcal{C}_{3}^{%
{\small 11D}}$ & $\qquad ,\text{\qquad }$ & $\emph{3}\text{ gravi-photons}$
\\ 
&  & $\text{\qquad }$ & $\emph{\ }$ \\ 
$\text{Matter}:$ & $\mathcal{A}_{\mu }^{I}=\int_{B^{I}}\mathcal{C}_{3}^{%
{\small 11D}}$ & $\qquad $, & $\emph{19}\text{ abelian gauge fields}$%
\end{tabular}
\label{gm}
\end{equation}%
To get the relation between the new basis $\left\{ B^{a},B^{I}\right\} $
with $a=1,2,3$, $I=1,...,19$, and the old one 
\begin{equation}
\left\{ \Psi ^{\Lambda }\right\} ,\qquad \Lambda =1,...,22,  \label{psi}
\end{equation}%
considered previously, we proceed as follows:\newline
\textbf{(1)} start from the gauge 3-form $\mathcal{C}_{3}^{{\small 11D}}$ of
the 11D theory and compactify on K3. By using the $\left\{ \Psi ^{\Lambda
}\right\} $ 2- cycle basis, we get 
\begin{equation}
\mathcal{A}^{\underline{\Lambda }}=\int_{\Psi ^{\Lambda }}\mathcal{C}_{3}^{%
{\small 11D}},\qquad \mathcal{A}^{\underline{\Lambda }}=dx^{\mu }\mathcal{A}%
_{\mu }^{\underline{\Lambda }}.
\end{equation}%
If instead of (\ref{psi}), we use the 2- cycle basis $\left\{
B^{a},B^{I}\right\} $, we end with the relations (\ref{gm}). \newline
\textbf{(2)} compare the two expressions by using (\ref{61}-\ref{62}); we
obtain%
\begin{equation}
\begin{tabular}{ll}
$\mathcal{A}^{\underline{a}}=\sum_{\Lambda =1}^{22}Q_{\underline{\Lambda }}^{%
\underline{a}}\left( \int_{\Psi ^{\Lambda }}\mathcal{C}_{3}^{{\small 11D}%
}\right) $ & ,\quad \\ 
&  \\ 
$\mathcal{F}_{2}^{\underline{a}}=\sum_{\Lambda =1}^{22}Q_{\underline{\Lambda 
}}^{\underline{a}}\left( \int_{\Psi ^{\Lambda }}\mathcal{F}_{4}^{{\small 11D}%
}\right) $ & ,%
\end{tabular}
\label{fab}
\end{equation}%
and 
\begin{equation}
\begin{tabular}{ll}
$\mathcal{A}^{\underline{I}}=\sum_{\Lambda =1}^{22}Q_{\underline{\Lambda }}^{%
\underline{I}}\left( \int_{\Psi ^{\Lambda }}\mathcal{C}_{3}^{{\small 11D}%
}\right) $ & , \\ 
&  \\ 
$\mathcal{F}_{2}^{\underline{I}}=\sum_{\Lambda =1}^{22}Q_{\underline{\Lambda 
}}^{\underline{I}}\left( \int_{\Psi ^{\Lambda }}\mathcal{F}_{4}^{{\small 11D}%
}\right) $ & .%
\end{tabular}%
\end{equation}%
But, these relations read also as follows%
\begin{equation}
\begin{tabular}{llll}
$\mathcal{A}^{\underline{a}}=\int_{B^{a}}\mathcal{C}_{3}^{{\small 11D}}$ & $%
,\qquad $ & $\mathcal{F}_{2}^{\underline{a}}=\int_{B^{a}}\mathcal{F}_{4}^{%
{\small 11D}}$ & , \\ 
&  &  &  \\ 
$\mathcal{A}^{\underline{I}}=\int_{B^{I}}\mathcal{C}_{3}^{{\small 11D}}$ & $%
,\qquad $ & $\mathcal{F}_{2}^{\underline{I}}=\int_{B^{I}}\mathcal{F}_{4}^{%
{\small 11D}}$ & ,%
\end{tabular}
\label{fb}
\end{equation}%
with%
\begin{equation}
\begin{tabular}{ll}
$\left[ B^{a}\right] =\sum_{\Lambda =1}^{22}Q_{\underline{\Lambda }}^{%
\underline{a}}\left[ \Psi ^{\Lambda }\right] $ & $,\qquad $ \\ 
&  \\ 
$\left[ B^{I}\right] =\sum_{\Lambda =1}^{22}Q_{\underline{\Lambda }}^{%
\underline{I}}\left[ \Psi ^{\Lambda }\right] $ & ,%
\end{tabular}
\label{bb}
\end{equation}%
or equivalently 
\begin{equation}
\left[ \Psi ^{\underline{\Lambda }}\right] =\sum_{a=1}^{3}L_{\underline{a}}^{%
\underline{\Lambda }}\left[ B^{a}\right] +\sum_{I=1}^{19}L_{\underline{I}}^{%
\underline{\Lambda }}\left[ B^{I}\right] .
\end{equation}%
These similarity transformations show that the gauge fields $\left( \mathcal{%
A}_{\mu }^{\underline{a}},\mathcal{A}_{\mu }^{\underline{I}}\right) $ and
the 2-cycle basis $\left\{ B^{a},B^{I}\right\} $\ are related in same manner
as do the gauge fields $\mathcal{A}_{\mu }^{\underline{\Lambda }}$ with the
basis $\left\{ \Psi ^{\Lambda }\right\} $.\newline

\emph{Building the 2-cycle basis }$\left\{ B^{a}\text{ },\text{ }%
B^{I}\right\} $\newline
The "physical" $\emph{3}$ gravi-photons $\mathcal{A}_{\mu }^{\underline{a}}$
of the gravity multiplet and the $\emph{19}$ "physical" abelian gauge fields 
$\mathcal{A}_{\mu }^{\underline{I}}$ of the matter multiplets can be defined
in terms of the $\left\{ B^{a},B^{I}\right\} $ 2-cycle basis of $H_{2}(K3,R)$%
. This is a real 22 dimensional canonical basis 
\begin{equation}
\left\{ B^{a},B^{I}\right\} ,\qquad a=1,2,3,\qquad I=1,...,19,  \label{lan}
\end{equation}%
dual to $\left\{ \Omega _{a},\Omega _{I}\right\} $ and it is related to the
old basis $\left\{ \Psi ^{\Lambda }\right\} $ by eqs(\ref{bb}).\newline
Poincar\'{e} duality associates eq(\ref{lan}) and eq(\ref{sa}) through the
relation,%
\begin{equation}
\begin{tabular}{llllllll}
$\int_{B^{a}}\Omega _{c}$ & $\text{ \ }\sim \text{ \ }$ & $\lambda \delta
_{c}^{a}$ & $\qquad ,\qquad $ & $\int_{B^{a}}\Omega _{I}$ & $\text{ \ }\sim 
\text{ \ }$ & $\lambda \xi _{I}^{a}$ & , \\ 
$\int_{B^{I}}\Omega _{c}$ & $\text{ \ }\sim \text{ \ }$ & $\varrho \xi
_{c}^{I}$ & $\qquad ,\qquad $ & $\int_{B^{I}}\Omega _{J}$ & $\text{ \ }\sim 
\text{ \ }$ & $\varrho \delta _{J}^{I}$ & ,%
\end{tabular}
\label{ob}
\end{equation}%
where $\lambda =\sqrt{\frac{3+\xi ^{2}}{3}}$ and $\varrho =\sqrt{\frac{%
19+\xi ^{2}}{19}}$, with $\xi ^{2}=\sum \xi _{I}^{a}\xi _{a}^{I}$, are as in
eqs(\ref{lrr}). \newline
Thus, the physics of the 7D $\mathcal{N}=2$ supergravity theory teaches us
that eqs(\ref{sa}) (resp. (\ref{lan})) is the natural basis of the second
real cohomology of K3 (resp. $H_{2}\left( K3,R\right) $).\newline

\emph{Checking eqs}(\ref{sa}-\ref{lan})\emph{\ }\newline
To check the naturalness of eqs (\ref{sa}-\ref{lan}), we compute the
magnetic charges of the black hole and compare them with the results
obtained in section 3. \newline
With the gauge field strengths $\left( \mathcal{F}_{2}^{\underline{a}},%
\mathcal{F}_{2}^{\underline{I}}\right) $ defined as in eqs(\ref{fb}), the
"physical" magnetic charges are given by 
\begin{equation}
m^{\underline{a}}=\int_{S_{\infty }^{2}}\mathcal{F}_{2}^{\underline{a}}\quad
,\qquad m^{\underline{I}}=\int_{S_{\infty }^{2}}\mathcal{F}_{2}^{\underline{I%
}}\quad .  \label{dua}
\end{equation}%
Using eqs(\ref{61}-\ref{62}), we can put the above relations in the form
involving the field strength $\mathcal{F}_{2}^{\underline{\Lambda }}$ and
the field coordinates of the moduli space of the theory,%
\begin{equation}
\begin{tabular}{ll}
$\int_{S_{\infty }^{2}}\mathcal{F}_{2}^{\underline{a}}=\dsum\limits_{\Lambda
=1}^{22}Q_{\underline{\Lambda }}^{\underline{a}}\left( \int_{S^{2}}\mathcal{F%
}_{2}^{\underline{\Lambda }}\right) $ & $,\qquad $ \\ 
&  \\ 
$\int_{S_{\infty }^{2}}\mathcal{F}_{2}^{\underline{I}}=\dsum\limits_{\Lambda
=1}^{22}Q_{\underline{\Lambda }}^{\underline{I}}\left( \int_{S^{2}}\mathcal{F%
}_{2}^{\underline{\Lambda }}\right) $ & .%
\end{tabular}%
\end{equation}%
Then using the identity $\int_{S_{\infty }^{2}}\mathcal{F}_{2}^{\underline{%
\Lambda }}=p^{\underline{\Lambda }}$ considered in section 3, the above
relations can be reduced down to 
\begin{equation}
\begin{tabular}{lll}
$\int_{S^{2}}\mathcal{F}_{2}^{\underline{a}}$ & $=\sum_{\Lambda =1}^{22}p^{%
\underline{\Lambda }}Q_{\underline{\Lambda }}^{\underline{a}}$ & , \\ 
&  &  \\ 
$\int_{S^{2}}\mathcal{F}_{2}^{\underline{I}}$ & $=\sum_{\Lambda =1}^{22}p^{%
\underline{\Lambda }}Q_{\underline{\Lambda }}^{\underline{I}}$ & .%
\end{tabular}%
\end{equation}%
Comparing the expressions with \textrm{eqs(\ref{zl}),} we find that the
physical magnetic charges $m^{\underline{a}}$ and $m^{\underline{I}}$ are
precisely the dressed charges;%
\begin{equation}
m^{\underline{a}}=Z^{\underline{a}}\quad ,\qquad m^{\underline{I}}=Z^{%
\underline{I}}\quad ,  \label{duo}
\end{equation}%
involved in the supersymmetric transformations of the Fermi fields of the 7D 
$\mathcal{N}=2$ supergravity theory \textrm{\cite{FE0}-\cite{FE2}}.

\subsection{More on the basis $\left\{ \Omega _{a},\text{ }\Omega
_{I}\right\} $}

\qquad The real 22 dimensional 2-form basis $\left\{ \Omega _{a},\text{ }%
\Omega _{I}\right\} $ of $H^{2}\left( K3,R\right) $ has also an
interpretation in terms of the Hodge decomposition of the second real
cohomology group of K3,%
\begin{equation}
H^{2}\left( K3,R\right) =H^{\left( 2,0\right) }\oplus _{s}H^{\left(
1,1\right) }\oplus _{s}H^{\left( 0,2\right) }.  \label{da}
\end{equation}%
This Hodge decomposition has a particular property which we comment below:

\subsubsection{The isotriplet $\Omega _{a}$}

\qquad Compared with the Hodge decomposition of the half dimensional
cohomology $H^{n}\left( CYn,R\right) $ of generic complex n dimension
Calabi-Yau manifold, namely,%
\begin{equation}
H^{n}\left( CY,R\right) =H^{\left( n,0\right) }\oplus _{s}H^{\left(
n-1,1\right) }\oplus ...\oplus _{s}H^{\left( 1,n-1\right) }\oplus
_{s}H^{\left( 0,n\right) },
\end{equation}%
eq(\ref{da}) is particular and makes K3 a very special Calabi-Yau manifold.
The point is that for the particular case of complex $n=2$ Calabi-Yau
surfaces, it happens that the holomorphic and anti-holomorphic 2-forms 
\begin{equation}
\Omega ^{\left( n,0\right) }\qquad ,\qquad \Omega ^{\left( 0,n\right) },
\end{equation}%
as well as the Kahler 2-form 
\begin{equation}
\Omega ^{\left( 1,1\right) }
\end{equation}%
belong all of them to the same cohomology group. \newline
The property that $\Omega ^{\left( 2,0\right) }$, $\Omega ^{\left(
0,2\right) }$ and $\Omega ^{\left( 1,1\right) }$ are in the same second
cohomology of K3 allows us to combine altogether the complex moduli 
\begin{equation}
z_{i}=x_{i}+iy_{i}\equiv z_{i}^{+}\qquad ,\qquad \ \overline{z}%
_{i}=x_{i}-iy_{i}\equiv z_{i}^{-},
\end{equation}%
and the Kahler ones 
\begin{equation}
t_{i}\equiv z_{i}^{0}\ ,
\end{equation}%
to form $SO\left( 3\right) $ isotriplets%
\begin{equation}
\xi _{i}^{a}=\left( t_{i},x_{i},y_{i}\right) \qquad \leftrightarrow \qquad
\xi _{i}^{a}=\left( z_{i}^{0},z_{i}^{+},z_{i}^{-}\right) .
\end{equation}%
Recall that these moduli are given by the following integrals%
\begin{equation}
\begin{tabular}{llll}
$z_{i}^{+}=\int_{C_{i}}\Omega ^{+}$ & $\qquad ,\qquad $ & $x_{i}=\int_{C_{i}}%
\func{Re}\Omega ^{+}$ & , \\ 
$z_{i}^{-}=\int_{C_{i}}\Omega ^{-}$ & $\qquad ,$ & $y_{i}=\int_{C_{i}}\func{%
Im}\Omega ^{+}$ & , \\ 
$z_{i}^{0}=\int_{C_{i}}\Omega ^{0}$ & $\qquad ,$ & $t_{i}=\int_{C_{i}}\Omega
^{0}$ & ,%
\end{tabular}
\label{zi}
\end{equation}%
where $\left\{ C_{i}\right\} $ is a generic basis of real 2-cycles of K3 and
where we have set,%
\begin{equation}
\begin{tabular}{llll}
$\Omega ^{+}=\Omega ^{\left( 2,0\right) }$ & $\qquad ,\qquad $ & $\overline{%
\Omega ^{+}}=\Omega ^{-}$ & , \\ 
$\Omega ^{-}=\Omega ^{\left( 0,2\right) }$ & $\qquad ,$ & $\overline{\Omega
^{-}}=\Omega ^{+}$ & , \\ 
$\Omega ^{0}=\Omega ^{\left( 1,1\right) }$ & $\qquad ,$ & $\overline{\Omega
^{0}}=\Omega ^{0}$ & .%
\end{tabular}%
\end{equation}%
For later use, we also set 
\begin{equation}
\func{Re}\Omega ^{+}\equiv \Omega ^{1}\qquad ,\qquad \func{Im}\Omega
^{-}\equiv \Omega ^{2}.
\end{equation}%
and, 
\begin{equation}
\left\langle F,G\right\rangle =\int_{K3}F\wedge G,\qquad F,\text{ }G\in
H^{2}\left( K3\right) .
\end{equation}%
The above inner product is bilinear and symmetric%
\begin{equation}
\begin{tabular}{llll}
$\left\langle aF+bF^{\prime },G\right\rangle $ & $=$ & $a\left\langle
F,G\right\rangle +b\left\langle F^{\prime },G\right\rangle $ & , \\ 
$\left\langle F,G\right\rangle $ & $=$ & $\left\langle G,F\right\rangle $ & .%
\end{tabular}%
\end{equation}%
Using the orthogonality relations,%
\begin{equation}
\begin{tabular}{llll}
$\left\langle \Omega ^{+},\Omega ^{+}\right\rangle $ & $=$ & $0$ & , \\ 
$\left\langle \Omega ^{\pm },\Omega ^{0}\right\rangle $ & $=$ & $0$ & , \\ 
$\left\langle \Omega ^{-},\Omega ^{-}\right\rangle $ & $=$ & $0$ & ,%
\end{tabular}
\label{71}
\end{equation}%
and the identity 
\begin{equation}
\left\langle \Omega ^{-},\Omega ^{+}\right\rangle =2\left\langle \Omega
^{0},\Omega ^{0}\right\rangle \text{ },  \label{72}
\end{equation}%
required by $SO\left( 3\right) $ symmetry, it is not difficult to see that
we also have the orthogonality relations%
\begin{equation}
\left\langle \Omega ^{1},\Omega ^{2}\right\rangle =\left\langle \Omega
^{1},\Omega ^{0}\right\rangle =\left\langle \Omega ^{2},\Omega
^{0}\right\rangle =0,  \label{73}
\end{equation}%
together with%
\begin{equation}
\left\langle \Omega ^{1},\Omega ^{1}\right\rangle =\left\langle \Omega
^{2},\Omega ^{2}\right\rangle =\left\langle \Omega ^{0},\Omega
^{0}\right\rangle .  \label{74}
\end{equation}%
Eqs (\ref{71}-\ref{74}) can be put altogether in the following relation%
\begin{equation}
\left\langle \Omega ^{a},\Omega ^{b}\right\rangle =\lambda \delta
^{ab},\qquad  \label{75}
\end{equation}%
where the real number can be determined by computing $\lambda =\frac{1}{3}%
\delta _{ab}\left\langle \Omega ^{a},\Omega ^{b}\right\rangle $.

\subsubsection{The 19- uplet $\Omega _{I}$}

\qquad The metric of the complex surface K3 has two kinds of deformations: 
\newline
(\textbf{i}) Complex deformations$\left( z_{I}^{+},z_{I}^{-}\right) $
captured by the periods of the holomorphic $\Omega ^{+}$ and antiholomorphic 
$\Omega ^{-}$ 2-forms. \newline
(\textbf{ii}) Kahler deformations $t_{I}$ captured by the periods of the
Kahler 2-form $\Omega ^{0}$. \newline
Here we want to show that the real 2-form $\Omega _{I}$ is given by the
following $SO\left( 3\right) $ invariant%
\begin{equation}
\Omega _{I}=D_{+I}\Omega ^{+}+D_{-I}\Omega ^{-}+D_{0I}\Omega ^{0},
\label{oi}
\end{equation}%
where $D_{0,\pm I}$ are covariant derivatives to be defined later on.\newline

\textbf{(1)}\emph{\ Complex holomorphic deformations}\newline
The complex holomorphic deformations (\ref{zi}) with moduli $z^{+I}$ are
generated by the typical complex $\left( 1,1\right) $- form $\Omega
_{+I}^{+} $ following from the complex variation $\delta \Omega ^{+}$ of the
holomorphic 2-form $\Omega _{+}$,%
\begin{equation}
\delta \Omega ^{+}=\sum_{I=1}^{19}\left( \Omega _{+I}^{+}\right) \delta
z^{+I},\qquad \Omega _{+I}^{+}=D_{+I}\Omega ^{+}.
\end{equation}%
The gauge covariant derivative $D_{+I}\Omega ^{+}$ is defined in term of the
gauge field $A_{+I}$, associated with the coordinate transformations of the
moduli space of complex deformations, as follows 
\begin{equation}
D_{+I}\Omega ^{+}=\left( \frac{\partial }{\partial z^{+I}}-A_{+I}\right)
\Omega ^{+}.
\end{equation}%
Under a Kahler gauge transformation with holomorphic gauge parameter $%
\mathrm{f}\left( z\right) $ 
\begin{equation}
\Omega ^{+}\qquad \rightarrow \qquad e^{\mathrm{f}\left( z\right) }\Omega
^{+}\text{ },
\end{equation}%
the covariant derivative $D_{+I}\Omega ^{+}$ should transform in same
manner; i.e%
\begin{equation}
\left( D_{+I}\Omega ^{+}\right) \qquad \rightarrow \qquad e^{\mathrm{f}%
\left( z\right) }\left( D_{+I}\Omega ^{+}\right) .
\end{equation}%
So we should also have%
\begin{equation}
A_{+I}\qquad \rightarrow \qquad A_{+I}+\frac{\partial \mathrm{f}\left(
z\right) }{\partial z^{+I}}\text{ },\qquad
\end{equation}%
Notice also that the complex moduli $\left\{ z^{+I}\right\} $ parameterize
the complex $19$ dimension manifold 
\begin{equation}
\frac{SO\left( 2,19\right) }{SO\left( 2\right) \times SO\left( 19\right) }%
\text{ },  \label{sum}
\end{equation}%
which is a submanifold of the moduli space (\ref{md}). \newline
Notice moreover that we also have the following trivial variations 
\begin{equation}
\Omega _{-I}^{+}=\left( \frac{\delta \Omega ^{+}}{\delta z^{-I}}\right)
=0\qquad ,\qquad \Omega _{0I}^{+}=\left( \frac{\delta \Omega ^{+}}{\delta
t^{I}}\right) =0.  \label{cco}
\end{equation}%
Using the real notations,%
\begin{equation}
\begin{tabular}{llll}
$\Omega ^{\pm }$ & $=\Omega ^{1}\pm i\Omega ^{2}$ & , &  \\ 
$z^{\pm I}$ & $=x^{I}\pm iy^{I}$ & , & 
\end{tabular}
\label{deco}
\end{equation}%
and the parametrization 
\begin{equation}
\begin{tabular}{llll}
$\mathrm{f}\left( z\right) $ & $=$ & $r\left( x,y\right) +i\theta \left(
x,y\right) $ & , \\ 
$\frac{\partial \mathrm{\theta }\left( x,y\right) }{\partial x}$ & $=$ & $-%
\frac{\partial \mathrm{r}\left( x,y\right) }{\partial y}$ & , \\ 
$\frac{\partial \mathrm{r}\left( x,y\right) }{\partial x}$ & $=$ & $\frac{%
\partial \mathrm{\theta }\left( x,y\right) }{\partial y}$ & , \\ 
$e^{\mathrm{f}\left( z\right) }$ & $=$ & $e^{\mathrm{r}\left( x,y\right)
}e^{i\mathrm{\theta }\left( x,y\right) }$ & ,%
\end{tabular}%
\end{equation}%
the Kahler gauge transformation of real 2-forms $\Omega ^{1}$ and $\Omega
^{2}$ read as follows 
\begin{equation}
\left( 
\begin{array}{c}
\Omega ^{1} \\ 
\Omega ^{2}%
\end{array}%
\right) \quad \rightarrow \quad e^{r\left( x,y\right) }\left( 
\begin{array}{c}
\Omega ^{1}\cos \theta +\Omega ^{2}\sin \theta \\ 
-\Omega ^{1}\sin \theta +\Omega ^{2}\cos \theta%
\end{array}%
\right) ,  \label{soo}
\end{equation}%
where $r\left( x,y\right) $ and $\theta \left( x,y\right) $ are respectively
the local scale and local $SO\left( 2\right) $ transformations.\newline

\textbf{(2)}\emph{\ Complex antiholomorphic deformations}\newline
Along with the $z^{+I}$ complex moduli, we have also the antiholomorphic
moduli $z^{-I}$ (\ref{zi}). They correspond to the variations, 
\begin{equation}
\Omega _{-I}^{-}=D_{-I}\Omega ^{-}=\overline{\left( \Omega _{+I}^{+}\right) }%
.
\end{equation}%
We also have%
\begin{equation}
\begin{tabular}{llll}
$\Omega _{+I}^{-}$ & $=\left( \frac{\delta \Omega ^{-}}{\delta z^{+I}}%
\right) $ & $=0$ & , \\ 
$\Omega _{0I}^{-}$ & $=\left( \frac{\delta \Omega ^{-}}{\delta t^{I}}\right) 
$ & $=0$ & ,%
\end{tabular}
\label{cca}
\end{equation}%
which are just the complex conjugation of eqs(\ref{cco}).\newline
With the above relations, one can define the complex deformation tensor as 
\begin{equation}
\Omega _{aI}^{b}=\left( 
\begin{array}{cc}
\Omega _{+I}^{+} & \Omega _{+I}^{-} \\ 
\Omega _{-I}^{+} & \Omega _{-I}^{-}%
\end{array}%
\right) =\left( 
\begin{array}{cc}
\Omega _{+I}^{+} & 0 \\ 
0 & \Omega _{-I}^{-}%
\end{array}%
\right) ,\qquad a,b=+,-,
\end{equation}%
The trace of this deformation tensor is $SO\left( 2\right) $ invariant and
reads as follows 
\begin{equation}
Tr_{\text{{\small SO}}\left( {\small 2}\right) }\left( \Omega
_{aI}^{b}\right) =\left( \sum_{a=\pm }\Omega _{aI}^{a}\right) \equiv \Omega
_{I}.
\end{equation}%
Moreover, using the decomposition (\ref{deco}), we have the following
identities%
\begin{eqnarray}
\frac{\delta \Omega ^{+}}{\delta z^{+I}} &=&\frac{1}{2}\left( \frac{\delta
\Omega ^{1}}{\delta x^{I}}+\frac{\delta \Omega ^{2}}{\delta y^{I}}\right) +%
\frac{i}{2}\left( \frac{\delta \Omega ^{2}}{\delta x^{I}}-\frac{\delta
\Omega ^{1}}{\delta y^{I}}\right) \text{ },  \notag \\
&&  \label{iid} \\
\frac{\delta \Omega ^{-}}{\delta z^{-I}} &=&\frac{1}{2}\left( \frac{\delta
\Omega ^{1}}{\delta x^{I}}+\frac{\delta \Omega ^{2}}{\delta y^{I}}\right) -%
\frac{i}{2}\left( \frac{\delta \Omega ^{2}}{\delta x^{I}}-\frac{\delta
\Omega ^{1}}{\delta y^{I}}\right) \text{ },  \notag
\end{eqnarray}%
and%
\begin{eqnarray}
\frac{\delta \Omega ^{+}}{\delta z^{-I}} &=&\frac{1}{2}\left( \frac{\delta
\Omega ^{1}}{\delta x^{I}}-\frac{\delta \Omega ^{2}}{\delta y^{I}}\right) +%
\frac{i}{2}\left( \frac{\delta \Omega ^{2}}{\delta x^{I}}+\frac{\delta
\Omega ^{1}}{\delta y^{I}}\right) \text{ },  \notag \\
&& \\
\frac{\delta \Omega ^{-}}{\delta z^{+I}} &=&\frac{1}{2}\left( \frac{\delta
\Omega ^{1}}{\delta x^{I}}-\frac{\delta \Omega ^{2}}{\delta y^{I}}\right) -%
\frac{i}{2}\left( \frac{\delta \Omega ^{2}}{\delta x^{I}}+\frac{\delta
\Omega ^{1}}{\delta y^{I}}\right) \text{ },  \notag
\end{eqnarray}%
from which we read%
\begin{equation}
\frac{\delta \Omega ^{1}}{\delta x^{i}}=\frac{\delta \Omega ^{2}}{\delta
y^{i}}\qquad ,\qquad \frac{\delta \Omega ^{2}}{\delta x^{i}}=-\frac{\delta
\Omega ^{1}}{\delta y^{i}}.  \label{idd}
\end{equation}%
Using the identities (\ref{iid}-\ref{idd}), we can rewrite the deformation
tensor $\Omega _{aI}^{b}$ in the real coordinate frame as follows%
\begin{equation}
\Omega _{aI}^{b}=\left( 
\begin{array}{cc}
\Omega _{1I}^{1} & \Omega _{1I}^{2} \\ 
\Omega _{2I}^{1} & \Omega _{2I}^{2}%
\end{array}%
\right) ,\qquad a,b=1,2.
\end{equation}%
The trace is 
\begin{equation}
\Omega _{I}=\left( \frac{\delta \Omega ^{+}}{\delta z^{+I}}+\frac{\delta
\Omega ^{-}}{\delta z^{-I}}\right) =\left( \frac{\delta \Omega ^{1}}{\delta
x^{I}}+\frac{\delta \Omega ^{2}}{\delta y^{I}}\right) .
\end{equation}

\textbf{(2)}\emph{\ Kahler deformations}\newline
The Kahler deformations (\ref{zi}) of the metric of K3 captured by the real
moduli $\sigma $ and $t^{I}$ are generated by the variation $\delta \Omega
^{0}$ of the Kahler 2-form,%
\begin{equation}
\delta \Omega ^{0}=\left( \sum_{I=1}^{19}\Omega _{I}^{0}\delta t^{I}\right)
+\Omega _{\sigma }^{0}\delta \sigma ,\qquad \Omega _{\sigma }^{0}=\left( 
\frac{\partial \Omega ^{0}}{\partial \sigma }\right) .
\end{equation}%
By setting $t^{I}=z^{0I}$, we can put the above relation into the form,%
\begin{equation}
\delta \Omega ^{0}=\left( \sum_{I=1}^{19}\Omega _{0I}^{0}\delta
z^{0I}\right) +\left( \Omega _{\sigma }^{0}\delta \sigma \right) ,
\end{equation}%
with%
\begin{equation}
\begin{tabular}{llll}
$\Omega _{0I}^{0}$ & $=$ & $\left( D_{0I}\Omega ^{0}\right) $ & , \\ 
$D_{0I}\Omega ^{0}$ & $=$ & $\left( \frac{\partial }{\partial z^{0I}}%
-A_{0I}\ \right) \Omega ^{0}$ & ,%
\end{tabular}%
\end{equation}%
where $A_{0I}\left( t\right) $ is the gauge field capturing the local scale
transformation 
\begin{equation}
\Omega ^{0}\rightarrow e^{\tau \left( t\right) }\Omega ^{0},\qquad
A_{0I}\rightarrow A_{0I}+\frac{\partial \tau \left( t\right) }{\partial t^{I}%
}.  \label{soa}
\end{equation}%
The real deformations $\left\{ t^{I}\right\} $ and $\sigma $ parameterize
the real $20$ dimension manifold%
\begin{equation}
\frac{SO\left( 1,19\right) }{SO\left( 19\right) }\times SO\left( 1,1\right) .
\end{equation}%
This is a submanifold of the moduli space (\ref{md}) and can be thought of
as the transverse space to the space $\frac{SO\left( 2,19\right) }{SO\left(
2\right) \times SO\left( 19\right) }$ eq(\ref{sum}) in the full moduli space 
$\frac{SO\left( 3,19\right) }{SO\left( 3\right) \times SO\left( 19\right) }%
\times SO\left( 1,1\right) $.\newline
We also have the analogue of eqs(\ref{cco}),%
\begin{equation}
\begin{tabular}{llll}
$\Omega _{+I}^{0}$ & $=\left( \frac{\delta \Omega ^{0}}{\delta z^{+I}}%
\right) $ & $=0$ & , \\ 
$\Omega _{-I}^{0}$ & $=\left( \frac{\delta \Omega ^{0}}{\delta z^{-I}}%
\right) $ & $=0$ & .%
\end{tabular}
\label{ccb}
\end{equation}

\subsubsection{Deformation tensor $\Omega _{aI}^{b}$}

\qquad From the above analysis, we learn the two following remarkable
properties:\newline
\textbf{(1)} the moduli $\left\{ \sigma ,t^{I},x^{I},y^{I}\right\} $
describing Kahler and complex deformations of the metric of K3 parameterize
the space 
\begin{equation}
SO\left( 1,1\right) \times \left( \frac{SO\left( 1,19\right) }{SO\left(
19\right) }\right) \times \left( \frac{SO\left( 2,19\right) }{SO\left(
2\right) \times SO\left( 19\right) }\right) ,
\end{equation}%
with isotropy symmetry $SO\left( 2\right) \times SO\left( 19\right) $. as
mentioned earlier. This is a sub-manifold of eq(\ref{md}).\newline
\textbf{(2)} the $3\times 3$ deformation matrix $\left( \Omega
_{aI}^{b}\right) $, capturing both the Kahler and complex deformations of
the metric of K3, is generally given by%
\begin{equation}
\left( \Omega _{aI}^{b}\right) =\left( 
\begin{array}{ccc}
\Omega _{0I}^{0} & \Omega _{0I}^{+} & \Omega _{0I}^{-} \\ 
\Omega _{+I}^{0} & \Omega _{+I}^{+} & \Omega _{+I}^{-} \\ 
\Omega _{-I}^{0} & \Omega _{-I}^{+} & \Omega _{-I}^{-}%
\end{array}%
\right) .  \label{def}
\end{equation}%
However, because of eqs(\ref{cco},\ref{cca},\ref{ccb}), this matrix reduces
to the diagonal form%
\begin{equation}
\Omega _{aI}^{b}=\left( 
\begin{array}{ccc}
\Omega _{+I}^{+} & 0 & 0 \\ 
0 & \Omega _{0I}^{0} & 0 \\ 
0 & 0 & \Omega _{-I}^{-}%
\end{array}%
\right) .  \label{omm}
\end{equation}%
Eq(\ref{omm}) captures the $1+57$ deformations of the metric of K3; the
dilaton can be exhibited by factorizing it as follows: 
\begin{equation}
\Omega _{aI}^{b}=e^{-\sigma }\varpi _{aI}^{b}.
\end{equation}%
However, seen that $\dim H^{2}\left( K3,\mathbb{R}\right) =22$, and seen
that three of the vector basis of $H^{2}\left( K3,\mathbb{R}\right) $ namely 
$\Omega _{a}$ have been already identified, it follows that the remaining
nineteen 2-forms vector basis are given by 
\begin{equation}
\Omega _{I}=\Omega _{+I}^{+}+\Omega _{+I}^{+}+\Omega _{+I}^{+}.
\end{equation}%
This trace is precisely eq(\ref{oi}); and it reads, in the real notations,
as follows 
\begin{equation}
\Omega _{I}=\sum_{a=0}^{2}\left( D_{aI}\Omega ^{a}\right) =\sum_{a=0,\pm
}\left( D_{aI}\Omega ^{a}\right) .
\end{equation}%
Notice that gauge transformations 
\begin{equation}
\begin{tabular}{llll}
$\Omega ^{0}\left( t\right) $ & $\qquad \rightarrow \qquad $ & e$^{\mathrm{%
\tau }\left( t\right) }\Omega ^{0}\left( t\right) $ & , \\ 
$\Omega ^{+}\left( z\right) $ & $\qquad \rightarrow \qquad $ & e$^{\mathrm{f}%
\left( z\right) }\Omega ^{+}\left( z\right) $ & , \\ 
$\Omega ^{-}\left( \overline{z}\right) $ & $\qquad \rightarrow \qquad $ & e$%
^{\overline{\mathrm{f}}\left( \overline{z}\right) }\Omega ^{-}\left( 
\overline{z}\right) $ & ,%
\end{tabular}%
\qquad  \label{rx}
\end{equation}%
as well eqs(\ref{soo}) and (\ref{soa}), are not the most general one. The
most general gauge change for the isotriplet 2-form $\Omega ^{a}=\Omega
^{a}\left( \phi \right) $ should be like%
\begin{equation}
\begin{tabular}{llll}
$\Omega ^{a}$ & $\rightarrow $ & $e^{\lambda }\left( U_{b}^{a}\Omega
^{b}\right) $ & ,%
\end{tabular}%
\end{equation}%
where $U_{b}^{a}=U_{b}^{a}\left( \phi \right) $ is a local $SO\left(
3\right) $ gauge transformation and $\exp \left[ \lambda \left( \phi \right) %
\right] $ being a local scale factor.

\section{SHG: the basic relations}

\qquad The special hyperKahler geometry (SHG) of the moduli space of the 11D
M-theory on K3 can be nicely described by specifying: \newline
(\textbf{1}) the usual Hodge 2- form basis $\left\{ \mathbf{\alpha }_{%
{\small \Lambda }},\text{ \ }{\small \Lambda =1,...,22}\right\} $ and its
dual 2-cycle basis $\left\{ \Psi ^{\Lambda }\right\} $ of $H_{2}\left(
K3,R\right) $ satisfying 
\begin{equation}
\int_{\Psi ^{\Lambda }}\mathbf{\alpha }_{{\small \Sigma }}\sim \delta
_{\Sigma }^{\Lambda }.
\end{equation}%
(\textbf{2}) the new basis $\left\{ \Omega _{a},\Omega _{I}\right\} $ with $%
\Omega _{I}=D_{aI}\Omega ^{a}$ and its dual 2-cycle basis $\left\{
B^{a},B^{I}\right\} $ considered in previous section. \newline
The "old" real 2-forms basis $\left\{ \mathbf{\alpha }_{\Lambda }\right\} $
and the "new" $\left\{ \Omega _{a},\Omega _{I}\right\} $ one are globally
defined on K3; they generate the second real cohomology group $H^{2}\left(
K3,R\right) $. The passage from the old Hodge basis $\mathbf{\alpha }%
_{\Lambda }$ to the new basis $\left\{ \Omega _{a},\Omega _{I}\right\} $ is
given, at each point $\varphi ^{m}=\left( \sigma ,\phi ^{aI}\right) $ of the
moduli space, by the similarity transformations%
\begin{equation}
\begin{tabular}{lll}
$\Omega _{a}\ $ & $=\sum \mathbf{\alpha }_{\Lambda }X_{a}^{\Lambda }\left(
\varphi \right) $ & , \\ 
$\Omega _{I}$ & $=\sum \mathbf{\alpha }_{\Lambda }X_{I}^{\Lambda }\left(
\varphi \right) $ & .%
\end{tabular}%
\qquad  \label{ex}
\end{equation}%
The expansion modes $X_{a}^{\Lambda }\left( \varphi \right) $ and $%
X_{I}^{\Lambda }\left( \varphi \right) $ are local fields on the moduli
space and can be interpreted as the periods of $\Omega _{a}$ and $\Omega
_{I} $ over the 2-cycles $\Psi ^{\Lambda }$ as shown below,%
\begin{equation}
\begin{tabular}{lll}
$X_{a}^{\Lambda }\left( \varphi \right) =$ & $\int_{\Psi ^{\Lambda }}\Omega
_{a}$ & , \\ 
$X_{I}^{\Lambda }\left( \varphi \right) =$ & $\int_{\Psi ^{\Lambda }}\Omega
_{I}$ & .%
\end{tabular}%
\end{equation}%
The 2-forms $\Omega _{a}$ and $\Omega _{I}$ are defined up to a local $%
SO\left( 3\right) \times SO\left( 19\right) $ gauge transformations,%
\begin{equation}
\begin{tabular}{llllll}
$\Omega _{a}\left( \varphi \right) \equiv $ & $U_{a}^{b}\left( \varphi
\right) \Omega _{b}\left( \varphi \right) $ & \qquad ,\qquad & $\Omega
_{I}\left( \varphi \right) \equiv $ & $V_{I}^{J}\left( \varphi \right)
\Omega _{J}\left( \varphi \right) $ & , \\ 
$X_{a}^{\Lambda }\left( \varphi \right) \equiv $ & $U_{a}^{b}\left( \varphi
\right) X_{b}^{\Lambda }\left( \varphi \right) $ & \qquad ,\qquad & $%
X_{I}^{\Lambda }\left( \varphi \right) \equiv $ & $V_{I}^{J}\left( \varphi
\right) X_{J}^{\Lambda }\left( \varphi \right) $ & ,%
\end{tabular}
\label{ch}
\end{equation}%
with%
\begin{equation}
\begin{tabular}{lll}
$U_{a}^{c}\left( \varphi \right) U_{c}^{b}\left( \varphi \right) $ & $%
=\delta _{a}^{b}$ & , \\ 
$V_{I}^{K}\left( \varphi \right) V_{K}^{J}\left( \varphi \right) $ & $%
=\delta _{I}^{J}$ & ,%
\end{tabular}%
\end{equation}%
where $\varphi $ parameterizes a generic local point on $\boldsymbol{M}%
_{7D}^{N=2}$.

\subsection{Fundamental relations}

\qquad The constraint eqs(\ref{71}-\ref{75}) describing the Kahler and
complex deformations of the metric of K3 can be reformulated in an $SO\left(
3\right) \times SO\left( 19\right) $ covariant manner by using the basis $%
\left\{ \Omega _{a},\Omega _{I}\right\} $ and the symmetric inner products $%
\left\langle \Omega _{a},\Omega _{b}\right\rangle $, $\left\langle \Omega
_{I},\Omega _{J}\right\rangle $ and so on. Notice that the inner product $%
\left\langle F,G\right\rangle $ of generic local 2-forms $F,G\in H^{2}\left(
K3,\mathbb{R}\right) $ is defined as 
\begin{equation}
\left\langle F,G\right\rangle =\int_{K3}F\wedge G.
\end{equation}%
It is bilinear and symmetric.

\subsubsection{Gauge invariant constraint eqs}

\qquad Because of their local nature and because of their symmetries, the
constraint eqs(\ref{71}-\ref{74}) can be rewritten as follows:%
\begin{equation}
\begin{tabular}{llll}
$\left\langle \Omega ^{a}\left( \sigma ,\phi \right) ,\Omega _{b}\left(
\sigma ,\phi \right) \right\rangle $ & $=$ & $\delta _{b}^{a}$ & , \\ 
$\left\langle \Omega ^{a}\left( \sigma ,\phi \right) ,\Omega _{I}\left(
\sigma ,\phi \right) \right\rangle $ & $=$ & $0$ & , \\ 
$\left\langle \Omega ^{I}\left( \sigma ,\phi \right) ,\Omega _{J}\left(
\sigma ,\phi \right) \right\rangle $ & $=$ & $\delta _{J}^{I}$ & .%
\end{tabular}
\label{fo}
\end{equation}%
These relations are gauge invariant under the $SO\left( 3\right) \times
SO\left( 19\right) $ local gauge transformations (\ref{ch}); thanks to the
local orthogonality relations%
\begin{equation}
\begin{tabular}{lllll}
$\delta _{a}^{b}=$ & $U_{a}^{c}\left( \varphi \right) U_{d}^{b}\left(
\varphi \right) \delta _{c}^{d}$ & $\qquad ,\qquad $ & $U\in SO\left(
3\right) $ & , \\ 
$\delta _{J}^{I}=$ & $V_{J}^{L}\left( \varphi \right) V_{K}^{I}\left(
\varphi \right) \delta _{L}^{K}$ & $\qquad ,\qquad $ & $V\in SO\left(
19\right) $ & .%
\end{tabular}%
\end{equation}%
Now, think about the $\delta _{b}^{a}$ and $\delta _{J}^{I}$ invariants as
the products of the local field matrix $K_{ab}$ (resp. $K_{LJ}$) and its $%
K^{ac}$ inverse (resp. $K^{IL}$);%
\begin{equation}
\begin{tabular}{lllll}
$\delta _{b}^{a}=$ & $\mathcal{K}^{ac}\left( \varphi \right) \mathcal{K}%
_{cb}\left( \varphi \right) $ & $=$ & $K^{ac}\left( \phi \right)
K_{cb}\left( \phi \right) $ & , \\ 
$\delta _{J}^{I}=$ & $\mathcal{K}^{IL}\left( \varphi \right) \mathcal{K}%
_{LJ}\left( \varphi \right) $ & $=$ & $K^{IL}\left( \phi \right)
K_{LJ}\left( \phi \right) $ & ,%
\end{tabular}%
\end{equation}%
with the field matrices $\mathcal{K}_{ab}$ and $\mathcal{K}_{IJ}$ factorized
like, 
\begin{equation}
\begin{tabular}{llllll}
$\mathcal{K}^{ac}\left( \sigma ,\phi \right) =$ & $e^{+\sigma }K^{ac}\left(
\phi \right) $ & $\qquad ,\qquad $ & $\mathcal{K}_{cb}\left( \sigma ,\phi
\right) =$ & $e^{-\sigma }K_{cb}\left( \phi \right) $ & , \\ 
$\mathcal{K}^{IL}\left( \sigma ,\phi \right) =$ & $e^{+\sigma }K^{IL}\left(
\phi \right) $ & $\qquad ,\qquad $ & $\mathcal{K}_{LJ}\left( \sigma ,\phi
\right) =$ & $e^{-\sigma }K_{LJ}\left( \phi \right) $ & .%
\end{tabular}%
\end{equation}%
Then put back into eqs(\ref{fo}), we can bring it to the following covariant
form%
\begin{equation}
\begin{tabular}{llll}
$\left\langle \Omega _{a},\Omega _{b}\right\rangle =e^{-2\sigma }K_{ab}$ & $%
\qquad ,\qquad $ & $\left\langle \Omega ^{a},\Omega ^{b}\right\rangle
=e^{+2\sigma }K^{ab}$ & , \\ 
$\left\langle \Omega _{a},\Omega _{I}\right\rangle =0$ & $\qquad ,\qquad $ & 
$\left\langle \Omega ^{a},\Omega ^{I}\right\rangle =0$ & , \\ 
$\left\langle \Omega _{I},\Omega _{J}\right\rangle =e^{-2\sigma }K_{IJ}$ & $%
\qquad ,\qquad $ & $\left\langle \Omega ^{I},\Omega ^{J}\right\rangle
=e^{+2\sigma }K^{IJ}$ & .%
\end{tabular}
\label{sik}
\end{equation}%
Moreover, setting $\Omega _{a}=\Omega _{a}\left( \sigma ,\phi \right) $ and $%
\Omega _{I}=\Omega _{I}\left( \sigma ,\phi \right) $ as 
\begin{equation}
\Omega _{a}=e^{-\sigma }\varpi _{a}\left( \phi \right) \qquad ,\qquad \Omega
_{I}=e^{-\sigma }\varpi _{I}\left( \phi \right) \text{ },
\end{equation}%
the above eqs reduce further down to 
\begin{equation}
\begin{tabular}{llll}
$\left\langle \varpi _{a},\varpi _{b}\right\rangle =K_{ab}$ & $\qquad
,\qquad $ & $\left\langle \varpi ^{a},\varpi ^{b}\right\rangle =K^{ab}$ & ,
\\ 
$\left\langle \varpi _{a},\varpi _{I}\right\rangle =0$ & $\qquad $, & $%
\left\langle \varpi ^{a},\varpi ^{I}\right\rangle =0$ & , \\ 
$\left\langle \varpi _{I},\varpi _{J}\right\rangle =K_{IJ}$ & $\qquad
,\qquad $ & $\left\langle \varpi ^{I},\varpi ^{J}\right\rangle =K^{IJ}$ & ,%
\end{tabular}
\label{sid}
\end{equation}%
where now the dependence into the dilaton field $\sigma $ has been
completely factorized out. \newline
Besides locality, we learn from the above fundamental relations, a set of
special features; in particular the following. \newline

\emph{Metric tensors and potentials}\newline
First notice that because of the following symmetry properties 
\begin{equation}
\begin{tabular}{llll}
$\Omega _{a}\wedge \Omega _{b}$ & $=$ & $\Omega _{b}\wedge \Omega _{a}$ & ,
\\ 
$\Omega _{I}\wedge \Omega _{J}$ & $=$ & $\Omega _{J}\wedge \Omega _{I}$ & ,%
\end{tabular}%
\end{equation}%
the local field matrices $K_{ab}$ and $K_{IJ}$ are real and symmetric%
\begin{equation}
K_{ab}=K_{ba}\qquad ,\qquad K_{IJ}=K_{JI}.
\end{equation}%
These rank two tensor fields play also the role of metric tensors that can
be used to rise and lower the $SO\left( 3\right) $ and $SO\left( 19\right) $
indices as shown below:%
\begin{equation}
\Omega _{a}=K_{ab}\Omega ^{b},\qquad \Omega _{I}=K_{IJ}\Omega ^{J}.
\end{equation}%
Under the $SO\left( 3\right) $ gauge transformations $\Omega _{a}^{\prime
}=U_{a}^{b}\Omega _{b}$, the matrix $K_{IJ}$ is invariant while $K_{ab}$
transforms like%
\begin{equation}
K_{ab}\qquad \rightarrow \qquad K_{ab}^{\prime }=U_{a}^{c}K_{cd}U_{b}^{d}.
\label{gs}
\end{equation}%
Eq(\ref{gs}) shows that $K_{cd}$ captures three physical degrees of freedom
(a 3-vector potential $\kappa _{a}$) since one can usually perform an
appropriate $SO\left( 3\right) $ gauge transformation to put $K_{ab}$ in a
diagonal form 
\begin{equation}
K_{ab}^{\prime }=\kappa _{a}\delta _{ab}.
\end{equation}

\begin{proposition}
\ \ \newline
(\textbf{i}) The 2-form isotriplet $\varpi _{a}$ and the matrix potential $%
K_{ab}\left( \phi \right) $ are defined up to the $SO\left( 3\right) $ gauge
symmetry eq(\ref{gs}).\newline
(\textbf{ii}) The geometry of the moduli space of the 11D M-theory on K3 is
characterized by a 3-vector potential $\left( \kappa _{0},\kappa _{1},\kappa
_{2}\right) $. These potentials reflects the hyperKahler structure that
lives on K3. They could be thought of as the analogue of the Kahler
potential of the special Kahler geometry of type IIB superstring on
Calabi-Yau threefolds. \newline
(\textbf{iii}) The real 3-vector potential $\kappa _{a}$ describes the
"physical" degrees of freedom captured by the local field metric $%
K_{ab}\left( \phi \right) $ defining the intersections $\left\langle \varpi
_{a},\varpi _{b}\right\rangle $. SHG is then specified by the isovector $%
\left( \kappa _{0},\kappa _{1},\kappa _{2}\right) .$
\end{proposition}

\emph{Volume of K3}\newline
The $SO\left( 3\right) $ invariant real volume of K3 reads as 
\begin{equation}
\mathcal{V}\left( K3\right) =\frac{1}{3}K^{ab}\left\langle \Omega
_{a},\Omega _{b}\right\rangle .  \label{sig}
\end{equation}%
We can write this volume in different, but equivalent, ways: \newline
First by using eq(\ref{ex}), we have, up on integrating over K3, the
following result%
\begin{equation}
\mathcal{V}\left( K3\right) =\frac{1}{3}K^{ab}\left( \varphi \right)
X_{a}^{\Lambda }\left( \varphi \right) J_{\Lambda \Sigma }\left( \varphi
\right) X_{b}^{\Sigma }\left( \varphi \right) ,
\end{equation}%
where $J_{\Lambda \Sigma }\left( \varphi \right) $ will be defined below and 
$X_{a}^{\Lambda }\left( \varphi \right) $ as before. \newline
Moreover, by using the first relation of eqs(\ref{sik}), we find that $%
\mathcal{V}\left( K3\right) $ is given by the exponential of the dilaton
field%
\begin{equation}
\mathcal{V}\left( K3\right) =e^{-2\sigma }.
\end{equation}%
Notice that $\mathcal{V}\left( K3\right) $ is non degenerate$\footnote{$%
\mathcal{V}\left( K3\right) =e^{-2\sigma }\rightarrow 0$ for $\sigma
\rightarrow \infty $ and to infinity for $\sigma \rightarrow -\infty $.}$
only for finite values of $\sigma $, see also \emph{footnote 5}.\newline
Furthermore, using the third relation of eqs(\ref{sik}), the volume $%
\mathcal{V}\left( K3\right) $ is also invariant under $SO\left( 19\right) $
and can be expressed as well like,%
\begin{equation}
e^{-2\sigma }=\frac{1}{19}K^{IJ}\left\langle \Omega _{I},\Omega
_{J}\right\rangle =\frac{1}{19}K^{IJ}X_{I}^{\Lambda }\left( \varphi \right)
J_{\Lambda \Sigma }X_{J}^{\Sigma }\left( \varphi \right) .  \label{sim}
\end{equation}%
Comparing eq(\ref{sig}) and eq(\ref{sim}), we end with the identity%
\begin{equation}
\frac{1}{19}K^{IJ}\left\langle \Omega _{I},\Omega _{J}\right\rangle =\frac{1%
}{3}K^{ab}\left\langle \Omega _{a},\Omega _{b}\right\rangle .
\end{equation}%
By substituting $\Omega _{I}=D_{aI}\Omega ^{a}$ into the third relation of
eqs(\ref{sik}), we get%
\begin{equation}
\begin{tabular}{llll}
$K_{IJ}$ & $=$ & $e^{+2\sigma }\left\langle D_{aI}\Omega ^{a},D_{bJ}\Omega
^{b}\right\rangle $ & , \\ 
& $=$ & $e^{+2\sigma }K^{ac}K^{bd}\left( D_{aI}X_{c}^{\Lambda }\right)
J_{\Lambda \Sigma }\left( D_{bJ}X_{d}^{\Sigma }\right) $ & .%
\end{tabular}%
\end{equation}%
Notice that we cannot pull out the covariant derivatives $D_{aI}$ and $%
D_{bJ} $ outside the inner product $\left\langle D_{aI}\Omega
^{a},D_{bJ}\Omega ^{b}\right\rangle $. As such the relation between the
matrices $K_{IJ}$ and $K_{ab}$ is not trivial as in SKG eqs(\ref{G}-\ref{H}%
). It will be considered later on by using the vielbeins $e_{a}^{\underline{b%
}}$ and their derivatives.\newline

\emph{SHG\ using old basis }\newline
The constraint relations (\ref{sik}) have been formulated in terms of the
inner product of 2-forms $\Omega _{a}$ and $\Omega _{I}$. We can also
rewrite these constraint eqs by using the Hodge basis $\left\{ \mathbf{%
\alpha }_{\Lambda }\right\} $ as follows: 
\begin{equation}
\begin{tabular}{lll}
$\left\langle \mathbf{\alpha }_{\Lambda },\mathbf{\alpha }_{\Sigma
}\right\rangle $ & $=e^{-2\sigma }J_{\Lambda \Sigma }$ & , \\ 
$\left\langle \mathbf{\alpha }^{\Lambda },\mathbf{\alpha }^{\Sigma
}\right\rangle $ & $=e^{+2\sigma }J^{\Lambda \Sigma }$ & , \\ 
$\left\langle \mathbf{\alpha }_{\Lambda },\mathbf{\alpha }^{\Sigma
}\right\rangle $ & $=\delta _{\Lambda }^{\Sigma }$ & ,%
\end{tabular}
\label{sil}
\end{equation}%
with%
\begin{equation}
J_{\Lambda \Upsilon }J^{\Upsilon \Sigma }=\delta _{\Lambda }^{\Sigma
},\qquad J_{\Lambda \Sigma }=J_{\Sigma \Lambda }.
\end{equation}%
The field matrix $J_{\Lambda \Upsilon }$ can be interpreted as the metric
tensor to rise and lower the indices $\Lambda $ of the $SO\left( 3,19\right) 
$ vectors as 
\begin{equation}
\mathbf{\alpha }_{\Lambda }=\sum_{\Upsilon =1}^{22}J_{\Lambda \Upsilon }%
\mathbf{\alpha }^{\Upsilon }.
\end{equation}%
Eqs(\ref{sil}) are invariant under the local $SO\left( 3,19\right) $ gauge
transformations,%
\begin{equation}
\mathbf{\alpha }_{\Lambda }\equiv \mathbf{\alpha }_{\Sigma }P_{\Lambda
}^{\Sigma }\left( \varphi \right) ,\qquad P_{\Lambda }^{\Upsilon }\left(
\varphi \right) P_{\Upsilon }^{\Sigma }\left( \varphi \right) =\delta
_{\Lambda }^{\Sigma }.
\end{equation}%
Using the expansions (\ref{ex}) and their inverse, which we write as%
\begin{equation}
\mathbf{\alpha }_{\Lambda }=\sum_{a=1}^{3}\Omega _{a}T_{\Lambda }^{a}\left(
\varphi \right) +\sum_{I=1}^{19}\Omega _{I}T_{\Lambda }^{I}\left( \varphi
\right) ,  \label{alm}
\end{equation}%
we can work out the relations between the field moduli $X_{a}^{\Lambda
}\left( \varphi \right) $, $X_{I}^{\Lambda }\left( \varphi \right) ,$ $%
T_{\Lambda }^{a}\left( \varphi \right) ,$ $T_{\Lambda }^{I}\left( \varphi
\right) $ and the matrices $K_{ab}$, $K_{IJ}$ and $J_{\Lambda \Sigma }$. 
\newline
First we have 
\begin{equation}
\begin{tabular}{lll}
$X_{a}^{\Lambda }\left( \varphi \right) J_{\Lambda \Sigma }X_{b}^{\Sigma
}\left( \varphi \right) $ & $=K_{ab}\left( \varphi \right) $ & , \\ 
$X_{a}^{\Lambda }\left( \varphi \right) J_{\Lambda \Sigma }X_{J}^{\Sigma
}\left( \varphi \right) $ & $0$ & , \\ 
$X_{I}^{\Lambda }\left( \varphi \right) J_{\Lambda \Sigma }X_{J}^{\Sigma
}\left( \varphi \right) $ & $K_{IJ}\left( \varphi \right) .$ & .%
\end{tabular}
\label{xj}
\end{equation}%
Similarly,%
\begin{equation}
K_{ab}\left( \varphi \right) T_{\Lambda }^{a}\left( \varphi \right)
T_{\Sigma }^{b}\left( \varphi \right) +K_{IJ}\left( \varphi \right)
T_{\Lambda }^{I}\left( \varphi \right) T_{\Sigma }^{J}\left( \varphi \right)
=J_{\Lambda \Sigma }\left( \varphi \right) .
\end{equation}%
By integrating eq(\ref{alm}) over the 2-cycle $\Psi ^{\Sigma }$, we also have%
\begin{equation}
X_{a}^{\Sigma }\left( \varphi \right) T_{\Lambda }^{a}\left( \varphi \right)
+X_{I}^{\Sigma }\left( \varphi \right) T_{\Lambda }^{I}\left( \varphi
\right) =\delta _{\Lambda }^{\Sigma },  \label{xl}
\end{equation}%
showing that the matrix $\left( T_{\Lambda }^{a},T_{\Lambda }^{I}\right) $
is just the inverse of $\left( X_{a}^{\Lambda },X_{I}^{\Lambda }\right) $.

\subsubsection{Inertial coordinate frame}

\qquad To get more insight into eqs(\ref{fo}-\ref{sik}-\ref{sil}-\ref{xj})
and also to make contact with the analysis of section 2, it is useful to
rewrite the above gauge invariant constraint eqs in the inertial coordinate
frame $\left\{ \xi \right\} $. \newline

\emph{Field matrix potentials}\newline
Using the vielbeins $e_{a}^{\underline{c}}$, $e_{I}^{\underline{K}}$ and
their inverses $e_{\underline{c}}^{a}$, $e_{\underline{K}}^{I}$, we can
rewrite the field matrices $K_{ab}\left( \varphi \right) $ and $K_{IJ}\left(
\varphi \right) $\ as 
\begin{equation}
\begin{tabular}{lllll}
$K_{ab}\left( \varphi \right) =$ & $\left( e_{a}^{\underline{c}}e_{b}^{%
\underline{d}}\right) \eta _{\underline{c}\underline{d}}\left( \xi \right) $
& $\qquad ,\qquad $ & $e_{a}^{\underline{c}}=e_{a}^{\underline{c}}\left(
\varphi ,\xi \right) $ & , \\ 
$K_{IJ}\left( \varphi \right) =$ & $\left( e_{I}^{\underline{K}}e_{J}^{%
\underline{L}}\right) \eta _{\underline{K}\underline{L}}\left( \xi \right) $
& $\qquad ,\qquad $ & $e_{I}^{\underline{K}}=e_{I}^{\underline{K}}\left(
\varphi ,\xi \right) $ & ,%
\end{tabular}%
\end{equation}%
where $\eta _{\underline{a}\underline{b}}\left( \xi \right) =+\delta _{%
\underline{a}\underline{b}}$ and $\eta _{\underline{I}\underline{J}}\left(
\xi \right) =-\delta _{\underline{I}\underline{J}}$. \newline
Similar factorizations may be done for the real 2-forms $\Omega _{a}$ and $%
\Omega _{I}=D_{aI}\Omega ^{a}$. We have%
\begin{equation}
\begin{tabular}{llll}
$\Omega _{a}\left( \varphi \right) $ & $=$ & $e_{a}^{\underline{c}}\Omega _{%
\underline{c}}\left( \xi \right) $ & , \\ 
$\Omega _{I}\left( \varphi \right) $ & $=$ & $e_{I}^{\underline{L}}\Omega _{%
\underline{L}}\left( \xi \right) $ & , \\ 
$D_{aI}$ & $=$ & $e_{a}^{\underline{c}}e_{I}^{\underline{L}}D_{\underline{c}%
\underline{L}}$ & , \\ 
$\frac{\partial }{\partial \phi ^{aI}}$ & $=$ & $e_{a}^{\underline{c}}e_{I}^{%
\underline{L}}\frac{\partial }{\partial \xi ^{\underline{c}\underline{L}}}$
& , \\ 
$A_{aI}\left( \varphi \right) $ & $=$ & $e_{a}^{\underline{c}}e_{I}^{%
\underline{L}}A_{\underline{c}\underline{L}}\left( \xi \right) $ & .%
\end{tabular}%
\end{equation}%
Using these relations, the gauge invariant constraint eqs read in the
inertial coordinate frame $\left\{ \xi \right\} $ as follows:%
\begin{equation}
\begin{tabular}{llll}
$\left\langle \Omega _{\underline{a}}\left( \varphi ,\xi \right) ,\Omega _{%
\underline{b}}\left( \sigma ,\xi \right) \right\rangle =e^{-2\sigma }\eta _{%
\underline{a}\underline{b}}$ & $\quad ,\quad $ & $\left\langle \Omega ^{%
\underline{a}}\left( \sigma ,\xi \right) ,\Omega ^{\underline{b}}\left(
\sigma ,\xi \right) \right\rangle =e^{+2\sigma }\eta ^{\underline{a}%
\underline{b}}$ & , \\ 
$\left\langle \Omega _{\underline{a}}\left( \sigma ,\xi \right) ,\Omega _{%
\underline{I}}\left( \sigma ,\xi \right) \right\rangle =0$ & $\quad ,\quad $
& $\left\langle \Omega ^{\underline{a}}\left( \sigma ,\xi \right) ,\Omega ^{%
\underline{I}}\left( \sigma ,\xi \right) \right\rangle =0$ & , \\ 
$\left\langle \Omega _{\underline{I}}\left( \sigma ,\xi \right) ,\Omega _{%
\underline{J}}\left( \sigma ,\xi \right) \right\rangle =e^{-2\sigma }\eta _{%
\underline{I}\underline{J}}$ & $\quad ,\quad $ & $\left\langle \Omega ^{%
\underline{I}}\left( \sigma ,\xi \right) ,\Omega ^{\underline{J}}\left(
\sigma ,\xi \right) \right\rangle =e^{+2\sigma }\eta ^{\underline{I}%
\underline{J}}$ & .%
\end{tabular}%
\end{equation}%
Setting 
\begin{equation}
\begin{tabular}{llll}
$\Omega _{\underline{a}}$ & $=$ & $e^{-\sigma }\varpi _{\underline{a}}$ & ,
\\ 
$\Omega _{\underline{I}}$ & $=$ & $e^{-\sigma }\varpi _{\underline{I}}$ & ,%
\end{tabular}%
\end{equation}%
we can reduce the above relations down to%
\begin{equation}
\begin{tabular}{llll}
$\left\langle \varpi _{\underline{a}}\left( \xi \right) ,\varpi _{\underline{%
b}}\left( \xi \right) \right\rangle =\eta _{\underline{a}\underline{b}}$ & $%
\qquad ,\qquad $ & $\left\langle \varpi ^{\underline{a}}\left( \xi \right)
,\varpi ^{\underline{b}}\left( \xi \right) \right\rangle =\eta ^{\underline{a%
}\underline{b}}$ & , \\ 
$\left\langle \varpi _{\underline{a}}\left( \xi \right) ,\varpi _{\underline{%
I}}\left( \sigma ,\xi \right) \right\rangle =0$ & $\qquad ,$ & $\left\langle
\varpi ^{\underline{a}}\left( \xi \right) ,\varpi ^{\underline{I}}\left( \xi
\right) \right\rangle =0$ & , \\ 
$\left\langle \varpi _{\underline{I}}\left( \xi \right) ,\varpi _{\underline{%
J}}\left( \xi \right) \right\rangle =\eta _{\underline{I}\underline{J}}$ & $%
\qquad ,\qquad $ & $\left\langle \varpi ^{\underline{I}}\left( \xi \right)
,\varpi ^{\underline{J}}\left( \xi \right) \right\rangle =\eta ^{\underline{I%
}\underline{J}}$ & .%
\end{tabular}
\label{aw}
\end{equation}%
These relations are invariant under the transformations%
\begin{equation}
\begin{tabular}{lllll}
$\varpi _{\underline{a}}\left( \xi \right) \equiv $ & $U_{\underline{a}}^{%
\underline{b}}\left( \xi \right) \varpi _{\underline{b}}\left( \xi \right) $
& \qquad $,$\qquad & $U_{\underline{a}}^{\underline{b}}\left( \xi \right)
\in SO\left( 3\right) $ & , \\ 
$\varpi _{\underline{I}}\left( \xi \right) \equiv $ & $V_{\underline{I}}^{%
\underline{J}}\left( \xi \right) \varpi _{\underline{J}}\left( \xi \right) $
& \qquad $,$\qquad & $V_{\underline{I}}^{\underline{J}}\left( \xi \right)
\in SO\left( 19\right) $ & .%
\end{tabular}%
\end{equation}%
Below, we give explicit computations in the frame $\left\{ \xi \right\} $.%
\newline

\emph{Isopin gauge connection }$A_{\underline{a}\underline{I}}\left( \xi
\right) $\newline
The spin gauge connection on the moduli space is explicitly computed by help
of the constraint eq 
\begin{equation}
\left\langle \Omega _{\underline{a}}\left( \sigma ,\xi \right) ,\Omega _{%
\underline{I}}\left( \sigma ,\xi \right) \right\rangle =0.
\end{equation}%
Substituting 
\begin{equation}
D_{\underline{a}\underline{I}}\Omega ^{\underline{a}}=\partial _{\underline{a%
}\underline{I}}\Omega ^{\underline{a}}-A_{\underline{a}\underline{I}}\Omega
^{\underline{a}},
\end{equation}%
we first obtain%
\begin{equation}
\left\langle \Omega _{\underline{b}}A_{\underline{a}\underline{I}}\Omega ^{%
\underline{a}}\right\rangle =\left\langle \Omega _{\underline{b}}\partial _{%
\underline{a}\underline{I}}\Omega ^{\underline{a}}\right\rangle .
\end{equation}%
More explicit expressions can be written down by using the following $%
SO\left( 3\right) $ group parametrization%
\begin{equation}
\begin{tabular}{lll}
$U\left[ \lambda \left( \xi \right) \right] $ & $=\exp \lambda \left( \xi
\right) $ & , \\ 
$\lambda \left( \xi \right) $ & $=\sum_{m=1}^{3}T_{\underline{m}}\lambda ^{%
\underline{m}}\left( \xi \right) $ & , \\ 
$A_{\underline{a}\underline{I}}\left( \xi \right) $ & $=\sum_{m=1}^{3}T_{%
\underline{m}}A_{\underline{a}\underline{I}}^{\underline{m}}\left( \xi
\right) $ & ,%
\end{tabular}%
\end{equation}%
with $\lambda ^{\underline{m}}\left( \xi \right) $ and $T_{\underline{m}}$ ($%
T_{\underline{m}}^{t}=-T_{\underline{m}}$) are respectively the gauge group
parameters and the corresponding $so\left( 3\right) $ Lie algebra
generators. We have%
\begin{equation}
\left( A_{\underline{a}\underline{I}}\right) _{\underline{b}}^{\underline{a}%
}=\sum_{m,a=1}^{3}A_{\underline{a}\underline{I}}^{\underline{m}}\left( \xi
\right) \left( T_{\underline{m}}\right) _{\underline{b}}^{\underline{a}%
}=\left\langle \Omega _{\underline{b}}\partial _{\underline{a}\underline{I}%
}\Omega ^{\underline{a}}\right\rangle .
\end{equation}%
Using the vielbeins, this relation can be as well expressed as follows:%
\begin{equation}
\left( A_{\underline{a}\underline{I}}\right) _{\underline{b}}^{\underline{a}%
}=e_{\underline{b}}^{c}\left( \frac{\partial e_{c}^{\underline{a}}}{\partial
\xi ^{\underline{a}\underline{I}}}\right) =-e_{c}^{\underline{a}}\left( 
\frac{\partial e_{\underline{b}}^{c}}{\partial \xi ^{\underline{a}\underline{%
I}}}\right) .  \label{ae}
\end{equation}%
We can also compute the infinitesimal variation of the gauge field $A_{%
\underline{a}\underline{I}}\left( \xi \right) $. We have%
\begin{equation}
\begin{tabular}{llll}
$\delta A_{\underline{a}\underline{I}}\left( \xi \right) $ & $=$ & $D_{%
\underline{a}\underline{I}}\lambda \left( \xi \right) $ & , \\ 
$\delta A_{\underline{a}\underline{I}}^{\underline{m}}\left( \xi \right) $ & 
$=$ & $D_{\underline{a}\underline{I}}\lambda ^{\underline{m}}\left( \xi
\right) $ & ,%
\end{tabular}%
\end{equation}%
with%
\begin{equation}
\begin{tabular}{llll}
$D_{\underline{a}\underline{I}}\lambda $ & $=$ & $\partial _{\underline{a}%
\underline{I}}\lambda -\left[ A_{\underline{a}\underline{I}},\lambda \right] 
$ & , \\ 
$D_{\underline{a}\underline{I}}\lambda ^{\underline{m}}$ & $=$ & $\frac{%
\partial \lambda ^{\underline{m}}}{\partial \xi ^{\underline{a}\underline{I}}%
}-\mathrm{f}_{\underline{n}\underline{k}}^{\underline{m}}A_{\underline{a}%
\underline{I}}^{\underline{k}}\lambda ^{\underline{n}}$ & ,%
\end{tabular}%
\end{equation}%
where $\mathrm{f}_{\underline{n}\underline{k}}^{\underline{m}}=-\mathrm{f}_{%
\underline{k}\underline{n}}^{\underline{m}}$ are the usual $so\left(
3\right) $ structure constants.\newline

\emph{Relation between }$K_{IJ}$\emph{\ and }$K_{ab}$ \emph{via the vielbeins%
} \newline
Starting from the identity 
\begin{equation}
K_{IJ}=\left\langle D_{aI}\Omega ^{a},D_{bJ}\Omega ^{b}\right\rangle ,
\end{equation}%
and substituting 
\begin{equation}
D_{aI}\Omega ^{a}=\Omega ^{\underline{c}}\left( D_{aI}e_{\underline{c}%
}^{a}\right) ,
\end{equation}%
we first get%
\begin{equation}
\begin{tabular}{llll}
$K_{IJ}$ & $=$ & $\eta ^{\underline{c}\underline{d}}\left( D_{aI}e_{%
\underline{c}}^{a}\right) \left( D_{bJ}e_{\underline{d}}^{b}\right) $ & .%
\end{tabular}%
\end{equation}%
By replacing $\eta ^{\underline{c}\underline{d}}=K^{gh}e_{g}^{\underline{c}%
}e_{h}^{\underline{d}}$, we can also put $K_{IJ}$ in the form 
\begin{equation}
K_{IJ}=K^{gh}\left( e_{g}^{\underline{c}}D_{aI}e_{\underline{c}}^{a}\right)
\left( e_{h}^{\underline{d}}D_{bJ}e_{\underline{d}}^{b}\right) .  \label{ijk}
\end{equation}%
Now using the identities 
\begin{equation}
e_{\underline{c}}^{a}D_{aI}e_{g}^{\underline{c}}=-e_{g}^{\underline{c}%
}D_{aI}e_{\underline{c}}^{a}\qquad ,\qquad e_{h}^{\underline{d}}D_{bJ}e_{%
\underline{d}}^{b}=-e_{\underline{d}}^{b}D_{bJ}e_{h}^{\underline{d}},
\end{equation}%
following from the variation of $\left\langle \Omega ^{a},\Omega
_{g}\right\rangle =\delta _{g}^{a}$, we can bring eq(\ref{ijk}) to the form%
\begin{equation}
K_{IJ}=K^{gh}\left( e_{\underline{c}}^{a}D_{aI}e_{g}^{\underline{c}}\right)
\left( e_{\underline{d}}^{b}D_{bJ}e_{h}^{\underline{d}}\right) .
\end{equation}%
Then using 
\begin{equation}
D_{\underline{c}I}=e_{\underline{c}}^{a}D_{aI}\qquad ,\qquad D_{\underline{d}%
I}=e_{\underline{d}}^{b}D_{bJ},
\end{equation}%
the above relation reads as follows%
\begin{equation}
K_{IJ}=K^{gh}\left( D_{\underline{c}I}e_{g}^{\underline{c}}\right) \left( D_{%
\underline{d}J}e_{h}^{\underline{d}}\right) ,  \label{kac}
\end{equation}%
or equivalently%
\begin{equation}
K_{IJ}=\eta ^{\underline{a}\underline{b}}\left( e_{\underline{a}}^{g}D_{%
\underline{c}I}e_{g}^{\underline{c}}\right) \left( e_{\underline{b}}^{h}D_{%
\underline{d}J}e_{h}^{\underline{d}}\right)  \label{kij}
\end{equation}

\emph{Deriving the constraint eqs on the moduli }$L_{\underline{a}}^{%
\underline{\Lambda }}$\newline
To get the constraint eqs in the inertial coordinate frame $\left\{ \xi
\right\} $, we begin by giving some useful results%
\begin{equation}
\begin{tabular}{lll}
$\mathbf{\alpha }_{\underline{\Lambda }}$ & $=\mathcal{E}_{\underline{%
\Lambda }}^{\Upsilon }\mathbf{\alpha }_{\Upsilon }$ & , \\ 
$\mathbf{\alpha }_{\Lambda }$ & $=\mathcal{E}_{\Lambda }^{\underline{%
\Upsilon }}\mathbf{\alpha }_{\underline{\Upsilon }}$ & , \\ 
$\delta _{\underline{\Lambda }}^{\underline{\Sigma }}$ & $=\mathcal{E}_{%
\underline{\Lambda }}^{\Upsilon }\mathcal{E}_{\Upsilon }^{\underline{\Sigma }%
}$ & , \\ 
$\mathcal{E}_{\underline{\Lambda }}^{\Upsilon }$ & $=\mathcal{E}_{\underline{%
\Lambda }}^{\Upsilon }\left( \varphi ,\xi \right) $ & ,%
\end{tabular}%
\end{equation}%
where $\mathcal{E}_{\underline{\Lambda }}^{\Upsilon }$ and $\mathcal{E}%
_{\Lambda }^{\underline{\Upsilon }}$ are vielbeins. The metric tensor $%
J_{\Lambda \Sigma }\left( \varphi \right) $ is mapped to 
\begin{equation}
\begin{tabular}{llll}
$J_{\Lambda \Sigma }\left( \varphi \right) =$ & $\left( \mathcal{E}_{\Lambda
}^{\underline{\Upsilon }}\mathcal{E}_{\Sigma }^{\underline{\Gamma }}\right)
\eta _{\underline{\Upsilon }\underline{\Gamma }}\left( \xi \right) $ & $,$ & 
\end{tabular}%
\end{equation}%
and the constraint eqs becomes 
\begin{equation}
\begin{tabular}{lll}
$\left\langle \mathbf{\alpha }_{\underline{\Lambda }},\mathbf{\alpha }%
\underline{_{\Sigma }}\right\rangle $ & $=e^{-2\sigma }\eta _{\underline{%
\Lambda }\underline{\Sigma }}$ & $,$ \\ 
$\left\langle \mathbf{\alpha }^{\underline{\Lambda }},\mathbf{\alpha }^{%
\underline{\Sigma }}\right\rangle $ & $=e^{+2\sigma }\eta ^{\underline{%
\Lambda }\underline{\Sigma }}$ & , \\ 
$\left\langle \mathbf{\alpha }_{\underline{\Lambda }},\mathbf{\alpha }^{%
\underline{\Sigma }}\right\rangle $ & $=\delta _{\underline{\Lambda }}^{%
\underline{\Sigma }}$ & .%
\end{tabular}%
\end{equation}%
Expanding the 2-forms $\varpi _{\underline{a}}\left( \xi \right) $ and $%
\varpi _{\underline{I}}\left( \xi \right) $ in the 2- form basis $\left\{ 
\mathbf{\alpha }_{\underline{\Lambda }}\right\} $ as follows,%
\begin{equation}
\begin{tabular}{llll}
$\varpi _{\underline{a}}\ $ & $=$ & $\sum_{\Lambda }\mathbf{\alpha }_{%
\underline{\Lambda }}L_{\underline{a}}^{\underline{\Lambda }}\left( \xi
\right) $ & $,$ \\ 
$\varpi _{\underline{I}}$ & $=$ & $\sum \mathbf{\alpha }_{\underline{\Lambda 
}}L_{\underline{I}}^{\underline{\Lambda }}\left( \xi \right) $ & $,$%
\end{tabular}%
\end{equation}%
and integrating over the 2- cycles $\left\{ \Psi ^{\Lambda }\right\} $, we
get%
\begin{equation}
\begin{tabular}{llll}
$L_{\underline{a}}^{\underline{\Lambda }}\left( \xi \right) $ & $=$ & $%
\int_{\Psi ^{\Lambda }}\varpi _{\underline{a}}$ & , \\ 
$L_{\underline{I}}^{\underline{\Lambda }}\left( \xi \right) $ & $=$ & $%
\int_{\Psi ^{\Lambda }}\varpi _{\underline{I}}$ & .%
\end{tabular}%
\end{equation}%
Substituting these expansions back into (\ref{aw}), we obtain%
\begin{equation}
\begin{tabular}{llll}
$L_{\underline{a}}^{\underline{\Lambda }}\left( \xi \right) \eta _{%
\underline{\Lambda }\underline{\Sigma }}L_{\underline{b}}^{\underline{%
\Lambda }}\left( \xi \right) $ & $=$ & $\eta _{\underline{a}\underline{b}}$
& $,$ \\ 
$L_{\underline{a}}^{\underline{\Lambda }}\left( \xi \right) \eta _{%
\underline{\Lambda }\underline{\Sigma }}L_{\underline{I}}^{\underline{%
\Lambda }}\left( \xi \right) $ & $=$ & $0$ & , \\ 
$L_{\underline{I}}^{\underline{\Lambda }}\left( \xi \right) \eta _{%
\underline{\Lambda }\underline{\Sigma }}L_{\underline{J}}^{\underline{%
\Lambda }}\left( \xi \right) $ & $=$ & $\eta _{\underline{I}\underline{J}}$
& $.$%
\end{tabular}%
\end{equation}%
These relations, which are invariant under $SO\left( 3\right) \times
SO\left( 19\right) $ gauge change, are precisely the defining constraint
equations of the moduli space of metric deformations of K3.

\subsection{Metric of the moduli space}

\qquad We first give the expression of the metric $g_{aIbJ}$ in terms of the
matrix potentials $K_{ab}$ and $K_{IJ}$. Then we give the expression of $%
g_{aIbJ}$ in terms of the vielbeins $e_{a}^{\underline{c}}$ and their
covariant derivatives.

\subsubsection{Factorization of the metric $g_{aIbJ}$}

\qquad We begin by recalling that the complex and Kahler deformations of the
metric of K3 are captured by the deformation tensor $\Omega _{aI}^{c}$ (\ref%
{def}). In terms of this deformation tensor, the metric $g_{IJ}^{ab}$ reads
in the curved coordinate frame as follows%
\begin{equation}
g_{aIbJ}=\gamma \sum_{c,d=0}^{2}K_{cd}\left\langle \Omega _{aI}^{c},\Omega
_{bJ}^{d}\right\rangle ,
\end{equation}%
where $\gamma $ is a normalization constant number which can be chosen as $%
\gamma =\gamma _{1}\gamma _{2}$; with $\gamma _{1}$ for the $SO\left(
3\right) $ sector and $\gamma _{2}$ for $SO\left( 19\right) $. For
simplicity, we set $\gamma =1.$\newline
Using the relation $\Omega _{aI}^{c}=D_{aI}\Omega ^{c}$, we can also define
the metric in terms of the inner product of the covariant derivatives of the
isotriplet form like,%
\begin{equation}
g_{aIbJ}=\sum_{c,d=0}^{2}K_{cd}\left\langle D_{aI}\Omega ^{c},D_{aI}\Omega
^{d}\right\rangle .
\end{equation}%
However, since in the case of 11D M-theory on K3, the deformation tensor $%
\Omega _{aI}^{c}$ has only non zero diagonal terms (\ref{omm}), 
\begin{equation}
\Omega _{aI}^{c}=\delta _{a}^{c}\Omega _{I}\qquad ,\qquad \Omega
_{I}=D_{aI}\Omega ^{a},
\end{equation}%
the metric $g_{IJ}^{ab}$ gets reduced down to%
\begin{equation}
g_{aIbJ}=\left( \sum_{c,d=1}^{3}K_{cd}\delta _{a}^{c}\delta _{b}^{d}\right)
\left\langle \Omega _{I},\Omega _{J}\right\rangle .
\end{equation}%
Moreover, using the identity $K_{IJ}=\left\langle \Omega _{I},\Omega
_{J}\right\rangle $, we get the remarkable factorization%
\begin{equation}
g_{aIbJ}=K_{ab}K_{IJ}.  \label{kak}
\end{equation}%
The metric $g_{aIbJ}$ of the special hyperKahler geometry of 11D M-theory on
K3 is given by the product of $K_{IJ}$ and $K_{ab}$. In the inertial frame $%
\left\{ \xi \right\} $, the vielbeins $e_{a}^{\underline{c}}$ and $e_{I}^{%
\underline{L}}$ reduce to Kroneker symbols ($e_{a}^{\underline{c}%
}\rightarrow \delta _{a}^{\underline{c}},$ $e_{I}^{\underline{L}}\rightarrow
\delta _{I}^{\underline{L}}$) and the metric $g_{aIbJ}\rightarrow \eta _{%
\underline{a}\underline{b}}\eta _{\underline{I}\underline{J}}$.

\subsubsection{Expression of $g_{aIbJ}$ in terms of the vielbeins}

\qquad The relation (\ref{kak}) can be rewritten in different, but
equivalent, manners. First, we can use the metrics $K_{ab}$ and $K_{IJ}$ to
write the metric like 
\begin{equation}
\begin{tabular}{lll}
$g_{IJ}^{ab}=K_{IJ}K^{ab}$ &  & , \\ 
$g_{aIJ}^{b}=K_{IJ}K_{a}^{b}$ & $=K_{IJ}\delta _{a}^{b}$ & , \\ 
$g_{a}^{bIJ}=K^{IJ}K_{a}^{b}$ & $=K^{IJ}\delta _{a}^{b}$ & , \\ 
$g_{aI}^{bJ}=K_{I}^{J}K_{a}^{b}$ & $=\delta _{I}^{J}\delta _{a}^{b}$ & , \\ 
$g_{I}^{abJ}=K_{I}^{J}K^{ab}$ & $=\delta _{I}^{J}K^{ab}$ & , \\ 
$g_{abI}^{J}=K_{I}^{J}K_{ab}$ & $=\delta _{I}^{J}K_{ab}$ & , \\ 
$g_{ab}^{IJ}=K^{IJ}K_{ab}$ &  & .%
\end{tabular}%
\end{equation}%
We can also use this relation to express $K^{IJ}$ (resp. $K_{ab}$) in terms
of $g_{ab}^{IJ}$ and $K^{ab}$ (resp. $K_{IJ}$),%
\begin{equation}
\begin{tabular}{ll}
$K^{IJ}=g_{ab}^{IJ}K^{ab}$ & , \\ 
$K_{ab}=g_{ab}^{IJ}K_{IJ}$ & .%
\end{tabular}%
\end{equation}%
In these relations, the metric $g_{IJ}^{ab}$ can be interpreted as the
bridge from $K_{IJ}$ to $K_{ab}$ and vice versa. Eq(\ref{kak}) tells us
moreover that the vielbeins $E_{aI}^{\underline{c}\underline{K}}$,
introduced in section 2 to factorize the metric like 
\begin{equation}
g_{aIbJ}=E_{aI}^{\underline{c}\underline{K}}E_{bJ}^{\underline{d}\underline{L%
}}\mathrm{\eta }_{\underline{c}\underline{d}}\mathrm{\eta }_{\underline{K}%
\underline{L}}\text{ },  \label{kab}
\end{equation}%
get themselves factorized as shown below,%
\begin{equation}
E_{aI}^{\underline{c}\underline{K}}=e_{a}^{\underline{c}}e_{I}^{\underline{K}%
}.
\end{equation}%
By substituting back in the previous relations, we get%
\begin{equation}
g_{aIbJ}=\left( e_{a}^{\underline{c}}e_{b}^{\underline{d}}\mathrm{\eta }_{%
\underline{c}\underline{d}}\right) \left( e_{I}^{\underline{K}}e_{J}^{%
\underline{L}}\mathrm{\eta }_{\underline{K}\underline{L}}\right) ,
\label{gab}
\end{equation}%
which is an equivalent way to state (\ref{kak}). Moreover using eq(\ref{kij}%
), we can also put the metric in the equivalent form 
\begin{equation}
g_{aIbJ}=e_{a}^{\underline{c}}e_{b}^{\underline{d}}\left( D_{\mathrm{m}I}e_{%
\underline{g}}^{\mathrm{m}}\right) \left( D_{\mathrm{n}J}e_{\underline{h}}^{%
\mathrm{n}}\right) \eta _{\underline{c}\underline{d}}\eta ^{\underline{g}%
\underline{h}}.  \label{gac}
\end{equation}%
Eqs (\ref{kak}), (\ref{kab}), (\ref{gab}) and (\ref{gac}) are obviously
equivalent.

\section{New attractor approach in 7D}

\qquad The effective potential of the 7D black hole and black 3-brane have
been considered in\emph{\ section 3} by using the criticality method. In
this section, we complete this study by developing the extension of the new
attractor approach to 7D space time. We recall that new attractor approach
has been first introduced by Kallosh \textrm{\cite{BNB3}} in the framework
of 4 dimensional black hole physics and it is remarkably useful in dealing
with fluxes\textrm{.\cite{BNB6,F2}}

\subsection{Further on criticality method}

\qquad The effective scalar potential $\mathcal{V}_{eff}=\mathcal{V}%
_{eff}\left( \varphi \right) $ of the 7D black attractors is given by the
Weinhold relation \textrm{\cite{FE0,FE1}}. This is a gauge invariant
quadratic relation (\ref{vef}) in the \emph{dressed} charges,%
\begin{equation}
\begin{tabular}{lllll}
$\mathcal{Z}_{a}$ & $=$ & $e^{-\sigma }Z_{a}$ & $,\qquad a=1,2,3$ & , \\ 
$\mathcal{Z}_{I}$ & $=$ & $e^{-\sigma }Z_{I}$ & $,\qquad I=1,...,19$ & $.$%
\end{tabular}
\label{cec}
\end{equation}%
The charge $\mathcal{Z}_{a}$ and $\mathcal{Z}_{I}$ are the physical charges (%
\ref{duo}); they appear in the supersymmetric transformations of the
gravitinos $\left\{ \mathrm{\psi }_{\mu }^{1},\mathrm{\psi }_{\mu
}^{2}\right\} $, the gravi-photinos $\left\{ \mathrm{\chi }_{\mu }^{1},%
\mathrm{\chi }_{\mu }^{2}\right\} $ and the photinos $\left\{ \mathrm{%
\lambda }^{I}\right\} $ of the 7D $\mathcal{N}=2$ supergravity theory; eqs(%
\ref{go}-\ref{ga}). They induce a matrix mass to the fermionic fields and
play a crucial role in the attractor mechanism of the 7D black objects.%
\newline
The idea of the attractor mechanism is that, at the event horizon of the 7D
black objects, the attractor potential $\mathcal{V}_{eff}$ reaches its
minimum and the real field moduli $\varphi ^{m}$, which parameterize $\frac{%
{\small SO}\left( {\small 1,1}\right) {\small \times SO}\left( {\small 3,19}%
\right) }{{\small SO}\left( {\small 3}\right) {\small \times SO}\left( 
{\small 19}\right) }$, get fixed by the magnetic (electric) bare charges $%
\mathrm{p}^{\Lambda }$ ( $\mathrm{q}^{\Lambda }$) of the \emph{22} abelian
gauge fields strengths $\mathcal{F}_{2}^{\Lambda }$ (dual dual $\mathcal{G}%
_{5}^{\Lambda }$). The gauge invariants fields $\mathcal{F}_{2}^{\Lambda }$
and $\mathcal{G}_{5}^{\Lambda }$ follow from the compactification of the 11D
M- theory on K3%
\begin{equation}
\begin{tabular}{llll}
$\mathcal{F}_{2}^{\Lambda }$ & $=$ & $\int_{\Psi ^{\Lambda }}\mathcal{F}_{4}$
& , \\ 
$\mathcal{G}_{5}^{\Lambda }$ & $=$ & $\int_{\Psi ^{\Lambda }}\mathcal{G}_{5}$
& ,%
\end{tabular}%
\end{equation}%
with fluxes as%
\begin{equation}
\begin{tabular}{llll}
$\mathrm{p}^{\underline{\Lambda }}$ & $=$ & $\int_{S_{\infty }^{2}}\mathcal{F%
}_{2}^{\underline{\Lambda }}$ & , \\ 
$\mathrm{q}^{\underline{\Lambda }}$ & $=$ & $\int_{S_{\infty }^{5}}\mathcal{G%
}_{5}^{\underline{\Lambda }}$ & , \\ 
$\mathrm{q}_{\Lambda }\mathrm{p}^{\Sigma }$ & $=$ & $2\pi k_{\Lambda }\delta
_{\Lambda }^{\Sigma }$ & ,%
\end{tabular}%
\end{equation}%
where the $k_{\Lambda }$'s are non zero integers ; $k_{\Lambda }\in \mathbb{N%
}^{\ast }$.\newline
Notice that $\mathrm{p}^{\underline{\Lambda }}$ and $\mathrm{q}^{\underline{%
\Lambda }}$ are bare (undressed) charges; the physical ones are given by the
dressed $Z_{a}$ and $Z_{I}$ which coincide exactly with magnetic $\left( 
\mathrm{m}^{a},\mathrm{m}^{I}\right) $ and physical electric $\left( \mathrm{%
e}^{a},\mathrm{e}^{I}\right) $. The latter are given by the fluxes of the $%
\left( 3+19\right) $ abelian gauge field strengths $\left( \mathcal{F}%
_{2}^{a},\mathcal{F}_{2}^{I}\right) $ and $\left( \mathcal{G}_{5}^{a},%
\mathcal{G}_{5}^{I}\right) $ of the 7D $\mathcal{N}=2$ supergravity theory.
Using the relations 
\begin{equation}
\begin{tabular}{lll}
$\mathcal{G}_{5}^{a}$ & $=\text{ }\left( ^{\ast }\mathcal{F}_{2}^{a}\right) $
& , \\ 
$\mathcal{G}_{5}^{I}$ & $=\text{ }\left( ^{\ast }\mathcal{F}_{2}^{I}\right) $
& , \\ 
$\mathcal{G}_{5}^{\Lambda }$ & $=\text{ }\left( ^{\ast }\mathcal{F}%
_{2}^{\Lambda }\right) $ & ,%
\end{tabular}%
\end{equation}%
we have, 
\begin{equation}
\begin{tabular}{llll}
$\mathrm{m}^{a}=\int_{S_{\infty }^{2}}\mathcal{F}_{2}^{a}$ & \qquad ,\qquad
& $\mathrm{m}^{I}=\int_{S_{\infty }^{2}}\mathcal{F}_{2}^{I}$ & , \\ 
$\mathrm{e}^{a}=\int_{S_{\infty }^{5}}\mathcal{G}_{5}^{a}$ & \qquad ,\qquad
& $\mathrm{e}^{I}=\int_{S_{\infty }^{2}}\mathcal{G}_{5}^{I}$ & ,%
\end{tabular}%
\end{equation}%
obeying the electric/magnetic quantization condition%
\begin{equation}
\begin{tabular}{lll}
$\mathrm{m}^{a}\mathrm{e}_{b}$ & $=2\pi k_{a}\delta _{b}^{a}$ & , \\ 
$\mathrm{m}^{I}\mathrm{e}_{J}$ & $=2\pi k_{I}\delta _{J}^{I}$ & ,%
\end{tabular}%
\end{equation}%
where the $k_{a}$'s and the $k_{I}$'s are non zero integers. \newline
Recall that the relation between $\left( \mathcal{F}_{2}^{a},\mathcal{F}%
_{2}^{I}\right) $ and $\mathcal{F}_{2}^{\Lambda }$ (resp. $\mathcal{G}%
_{5}^{a},$ $\mathcal{G}_{5}^{I}$ and $\mathcal{G}_{5}^{\Lambda }$) are
related as follows, 
\begin{equation}
\begin{tabular}{llll}
$\mathcal{F}_{2}^{a}=\dsum\limits_{\Lambda =1}^{22}X_{\Lambda }^{a}\left(
\varphi \right) \mathcal{F}_{2}^{\Lambda }$ & \qquad ,\qquad & $\mathcal{G}%
_{5}^{a}=\dsum\limits_{\Lambda =1}^{22}\widetilde{X}_{\Lambda }^{a}\left(
\varphi \right) \mathcal{G}_{5}^{\Lambda }$ & , \\ 
$\mathcal{F}_{2}^{I}=\dsum\limits_{\Lambda =1}^{22}X_{\Lambda }^{I}\left(
\varphi \right) \mathcal{F}_{2}^{\Lambda }$ & \qquad ,\qquad & $\mathcal{G}%
_{5}^{I}=\dsum\limits_{\Lambda =1}^{22}\widetilde{X}_{\Lambda }^{I}\left(
\varphi \right) \mathcal{G}_{5}^{\Lambda }$ & ,%
\end{tabular}%
\end{equation}%
where $X_{\Lambda }^{a}\left( \varphi \right) $\ and $X_{\Lambda }^{I}\left(
\varphi \right) $ (resp $\widetilde{X}_{\Lambda }^{a}$ and $\widetilde{X}%
_{\Lambda }^{I}$ ) are as in eqs(%
\index{xj}-\ref{xl}).\newline
The attractor equations of the 7D black attractors can be obtained by
extremizing the effective potential $\mathcal{V}_{eff}$. This potential has
a set of symmetries; in particular it is invariant under general coordinate
transformations $\varphi ^{m}\rightarrow \xi ^{m}\left( \varphi \right) $ in
the moduli space $%
\frac{{\small SO}\left( {\small 3,19}\right) }{{\small SO}\left( {\small 3}%
\right) {\small \times SO}\left( {\small 19}\right) }$. Under the coordinate
change ,%
\begin{equation}
\begin{tabular}{llll}
$\sigma $ & \qquad $\rightarrow \qquad $ & $\zeta ^{0}=\zeta ^{0}\left(
\sigma ,\phi \right) $ & , \\ 
$\phi ^{aI}$ & \qquad $\rightarrow \qquad $ & $\zeta ^{\underline{a}%
\underline{I}}=\xi ^{\underline{a}\underline{I}}\left( \sigma ,\phi \right) $
& ,%
\end{tabular}%
\end{equation}%
with the convenient choice $\zeta ^{0}=\sigma $, we have%
\begin{equation}
\begin{tabular}{llll}
$\mathcal{V}_{eff}\left( \varphi \right) $ & $=$ & $\mathcal{V}_{eff}\left(
\zeta \right) $ & .%
\end{tabular}%
\end{equation}%
The attractor eqs can be stated in two different, but equivalent, ways.
Either in the generic curved coordinate frame $\left\{ \varphi \right\} $ as 
\begin{equation}
\begin{tabular}{lll}
$\frac{\partial \mathcal{V}_{eff}\left( \sigma ,\phi \right) }{\partial
\sigma }$ & $=0$ & , \\ 
$\frac{\partial \mathcal{V}_{eff}\left( \sigma ,\phi \right) }{\partial \phi
^{aI}}$ & $=0$ & ,%
\end{tabular}%
\end{equation}%
or in the inertial coordinate frame $\left\{ \sigma ,\xi \right\} $ like, 
\begin{equation}
\begin{tabular}{lll}
$\frac{\partial \mathcal{V}_{eff}\left( \sigma ,\xi \right) }{\partial
\sigma }$ & $=0$ & , \\ 
$e_{a}^{\underline{b}}\times e_{I}^{\underline{J}}\times \frac{\partial 
\mathcal{V}_{eff}\left( \xi \right) }{\partial \xi ^{\underline{b}\underline{%
J}}}$ & $=0$ & .%
\end{tabular}%
\end{equation}%
For non singular $e_{a}^{\underline{b}}\left( \varphi ,\xi \right) $ and $%
e_{I}^{\underline{J}}\left( \varphi ,\xi \right) $, the last relation can be
reduced down to 
\begin{equation}
\frac{\partial \mathcal{V}_{eff}\left( \xi \right) }{\partial \xi ^{%
\underline{b}\underline{J}}}=0.  \label{vvv}
\end{equation}%
Leaving aside the condition\footnote{%
Notice that $\frac{\partial \mathcal{V}_{eff}\left( \sigma ,\xi \right) }{%
\partial \sigma }=0$ requires $-2e^{-2\sigma }\mathcal{V}_{BH}\left( \xi
\right) =0$ which is solved either by $\sigma \rightarrow \infty $ whatever $%
\mathcal{V}_{BH}\left( \xi \right) $ is; or by $\sigma =\sigma _{0}$ finite
and $\mathcal{V}_{BH}\left( \xi \right) =0$. These two cases are singular
and so disregarded; see also footnotes 1 and 5.} $\partial \mathcal{V}%
_{eff}/\partial \sigma =0$, (\emph{see also footnotes 3,5 and 10}), the
solutions of eqs (\ref{vvv}) fix the field moduli in terms of the bare
charges $p_{\underline{\Lambda }}$. For the case of the 7D black hole, we
have:%
\begin{equation}
\begin{tabular}{lll}
$\left( \varphi \right) _{\text{horizon}}$ & $=f\left( p_{\underline{a}},p_{%
\underline{I}}\right) $ & ,%
\end{tabular}%
\end{equation}%
or equivalently in the inertial coordinate frame $\left\{ \xi \right\} $ like%
\begin{equation}
\begin{tabular}{lll}
$\left( \xi \right) _{\text{horizon}}$ & $=g\left( p_{\underline{a}},p_{%
\underline{I}}\right) $ & .%
\end{tabular}%
\end{equation}

(\textbf{1}) \emph{Potential in the inertial frame}\newline
In the inertial coordinates frame $\left\{ \xi \right\} $, the 7D black hole
potential $\mathcal{V}_{BH}^{7D,\mathcal{N}=2}\left( \sigma ,\xi \right) $
has a simple expression in terms of the geometric and matter charges $Z_{%
\underline{a}}\left( \xi \right) $ and $Z_{\underline{I}}\left( \xi \right) $
and can be factorized as follows,%
\begin{equation}
\mathcal{V}_{BH}^{7D,\mathcal{N}=2}\left( \sigma ,\xi \right) =e^{-2\sigma }%
\mathcal{V}_{BH}\left( \xi \right) ,
\end{equation}%
with 
\begin{equation}
\mathcal{V}_{BH}\left( \xi \right) =\sum_{\underline{a},\underline{b}}\delta
^{\underline{a}\underline{b}}Z_{\underline{a}}\left( \xi \right) Z_{%
\underline{b}}\left( \xi \right) +\sum_{\underline{I},\underline{J}}\delta ^{%
\underline{I}\underline{J}}Z_{\underline{I}}\left( \xi \right) Z_{\underline{%
J}}\left( \xi \right) .  \label{51}
\end{equation}%
Since $\delta ^{\underline{a}\underline{b}}=\eta ^{\underline{a}\underline{b}%
}$ and $\delta ^{\underline{I}\underline{J}}=-\eta ^{\underline{I}\underline{%
J}}$, we also have 
\begin{equation}
\mathcal{V}_{BH}\left( \xi \right) =\sum_{\underline{a},\underline{b}}\eta ^{%
\underline{a}\underline{b}}Z_{\underline{a}}\left( \xi \right) Z_{\underline{%
b}}\left( \xi \right) -\sum_{\underline{I},\underline{J}}\eta ^{\underline{I}%
\underline{J}}Z_{\underline{I}}\left( \xi \right) Z_{\underline{J}}\left(
\xi \right) .
\end{equation}%
Using the identity $Z_{\underline{I}}=D_{\underline{c}\underline{I}}Z^{%
\underline{c}}$ where $D_{\underline{c}\underline{I}}$ is the covariant
derivative in the inertial coordinate frame, we can rewrite the black hole
potential like%
\begin{equation}
\mathcal{V}_{BH}^{7D,\mathcal{N}=2}=e^{-2\sigma }\left( \sum_{\underline{a},%
\underline{b}}\delta ^{\underline{a}\underline{b}}Z_{\underline{a}}Z_{%
\underline{b}}+\sum_{\underline{I},\underline{J}}\delta ^{\underline{I}%
\underline{J}}D_{\underline{c}\underline{I}}Z^{\underline{c}}D_{\underline{d}%
\underline{J}}Z^{\underline{d}}\right) .
\end{equation}%
The criticality conditions of eq(\ref{vvv}) has been studied in section 3;
see eqs(\ref{sd}-\ref{cz}). There, it was shown the existence of three non
trivial sectors: One of them describes a $\frac{1}{2}$BPS state and the two
others describe non BPS states referred to as type 1 and type 2. Below, we
give a classification of these states by using the sign the semi-norm 
\begin{equation}
p^{2}=\left( p_{\underline{a}}\delta ^{\underline{a}\underline{b}}p_{%
\underline{b}}-p_{\underline{I}}\delta ^{\underline{I}\underline{J}}p_{%
\underline{J}}\right)  \label{za}
\end{equation}%
of the bare charge vector $p_{\underline{\Lambda }}$.\newline
Notice that because of the $SO\left( 3\right) \times SO\left( 19\right) $
isotropy symmetry, we can usually perform a particular special
transformations to simplify the above relations. Instead of dealing with the 
$3+19$ magnetic charges $p_{\underline{a}}$ and $p_{\underline{I}}$, one can
focus on two of them, 
\begin{equation}
\begin{tabular}{llll}
$\left( p_{\underline{1}},p_{\underline{2}},p_{\underline{3}}\right) $ & 
\qquad $\rightarrow \qquad $ & $\left( \mathrm{r},0,0\right) $ & , \\ 
$\left( p_{\underline{1}},...,p_{\underline{19}}\right) $ & \qquad $%
\rightarrow \qquad $ & $\left( \mathrm{s},0,...,0\right) $ & .%
\end{tabular}%
\end{equation}%
The $SO\left( 3\right) \times SO\left( 19\right) $ invariance ensures that
the results obtained by using the charges \textrm{r} and \textrm{s} are also
valid for all others. \newline
Besides the singular state associated with $p^{2}=0$ and the degenerate case
where the dressed charges are equal zero, $Z_{\underline{a}}=0$ et $Z_{%
\underline{I}}=0$, we the following classification according to the values
of the couple $\left( \mathrm{r},\mathrm{s}\right) $:\newline
(\textbf{a}) $\frac{1}{2}$ \emph{BPS state} with $\left( \mathrm{r,s}\right)
=\left( \mathrm{r,0}\right) $; $\mathrm{rs=0}$.\newline
This state has $p^{2}>0$ and corresponds to $Z_{\underline{a}}\neq 0$ et $Z_{%
\underline{I}}=0$. Entropy $\mathcal{S}_{\emph{BPS}}^{{\small entropy}}$ is
proportional to $p^{2}$, 
\begin{equation}
\mathcal{S}_{\emph{BPS}}^{{\small entropy}}\sim +p^{2}.
\end{equation}%
(\textbf{b}) non \emph{BPS state} \emph{type 1}$\ $with $\left( \mathrm{r,s}%
\right) =\left( \mathrm{0,s}\right) $; $\mathrm{rs=0.}$ \newline
This non supersymmetric state has $p^{2}<0$ and corresponds to $Z_{%
\underline{a}}=0$ and $Z_{\underline{I}}\neq 0.$ Entropy $\mathcal{S}%
_{\left( N\emph{BPS}\right) _{1}}^{{\small entropy}}$ is proportional to $%
\left( -p^{2}\right) $; 
\begin{equation}
\mathcal{S}_{\left( N\emph{BPS}\right) _{1}}^{{\small entropy}}\sim -p^{2}.
\end{equation}%
(\textbf{c}) non \emph{BPS state} \emph{type 2} with $\left( \mathrm{r,s}%
\right) $ and $\mathrm{rs\neq 0}$. \newline
This non supersymmetric state is characterized by$\ p^{2}$ which has an
indefinite sign. It corresponds to, 
\begin{equation}
\begin{tabular}{lllll}
$Z_{\underline{a}}$ & $\neq $ & $0$ & $a\in \mathcal{J}\subset \mathcal{I}%
_{3}=\left\{ 1,2,3\right\} $ & , \\ 
$Z_{\underline{a}}$ & $=$ & $0$ & $a\in \left( \mathcal{I}_{3}/\mathcal{J}%
\right) $ & , \\ 
$Z_{\underline{I}}$ & $\neq $ & $0$ & $I\in \mathcal{J}^{\prime }\subset 
\mathcal{I}_{19}=\left\{ 1,...,19\right\} $ & , \\ 
$Z_{\underline{I}}$ & $\neq $ & $0$ & $I\in \left( \mathcal{I}_{19}/\mathcal{%
J}^{\prime }\right) $ & .%
\end{tabular}
\label{zz}
\end{equation}%
The entropy $\mathcal{S}_{\left( N\emph{BPS}\right) _{2}}^{{\small entropy}}$
is proportional to $\left\vert p^{2}\right\vert $. \newline

(\textbf{2}) \emph{Potential in curved coordinate frame}\newline
To get the form of the potential in the curved coordinate frame, we use the
vielbeins $e_{\underline{a}}^{c}$ and $e_{\underline{I}}^{K}$ to rewrite $Z_{%
\underline{a}}$ and $Z_{\underline{I}}$ as%
\begin{equation}
\begin{tabular}{llll}
$Z_{\underline{a}}$ & $=e_{\underline{a}}^{c}Y_{c}$ &  & $,$ \\ 
$Z_{\underline{I}}$ & $=e_{\underline{I}}^{K}Y_{K}$ & $=e_{\underline{I}}^{K}%
\mathcal{D}_{cK}Y^{c}$ & ,%
\end{tabular}%
\end{equation}%
\ where $Y_{c}=Y_{c}\left( \varphi \right) $ and $Y_{K}=Y_{K}\left( \varphi
\right) $ are the dressed charges in the curved frame. By putting these
relations back into $\mathcal{V}_{BH}^{7D,\mathcal{N}=2}$, we obtain $%
\mathcal{V}_{BH}^{7D,\mathcal{N}=2}=e^{-2\sigma }\mathcal{V}_{BH}\left( \phi
\right) $ with%
\begin{equation}
\begin{tabular}{llll}
$\mathcal{V}_{BH}\left( \phi \right) $ & $=$ & $\delta ^{\underline{a}%
\underline{b}}e_{\underline{a}}^{c}e_{\underline{b}}^{d}Y_{c}Y_{d}+\delta ^{%
\underline{I}\underline{J}}e_{\underline{I}}^{K}e_{\underline{J}}^{L}\left( 
\mathcal{D}_{cK}Y^{c}\right) \left( \mathcal{D}_{dK}Y^{d}\right) $ & .%
\end{tabular}%
\end{equation}%
Now, using the identities 
\begin{equation}
\begin{tabular}{llll}
$K^{cd}$ & $=$ & $+\delta ^{\underline{a}\underline{b}}e_{\underline{a}%
}^{c}e_{\underline{b}}^{d}$ & , \\ 
$K^{KL}$ & $=$ & $-\delta ^{\underline{I}\underline{J}}e_{\underline{I}%
}^{K}e_{\underline{J}}^{L}$ & , \\ 
& $=$ & $+\delta ^{\underline{a}\underline{b}}\left( e_{\underline{a}}^{g}%
\mathcal{D}_{\underline{c}I}e_{g}^{\underline{c}}\right) \left( e_{%
\underline{b}}^{h}\mathcal{D}_{\underline{d}J}e_{h}^{\underline{d}}\right) $
& ,%
\end{tabular}%
\end{equation}%
we can rewrite the black hole potential as follows:%
\begin{equation}
\mathcal{V}_{BH}\left( \phi \right) =K^{cd}Y_{c}Y_{d}-K^{KL}\left( \mathcal{D%
}_{cK}Y^{c}\right) \left( \mathcal{D}_{dK}Y^{d}\right) .
\end{equation}%
Furthermore, using the relation 
\begin{equation}
K^{KL}=\frac{1}{3}K^{cd}g_{cd}^{KL},
\end{equation}%
where $g_{cd}^{KL}$ is the metric of the moduli space, we end with the
following form of the potential%
\begin{equation}
\mathcal{V}_{BH}\left( \phi \right) =\sum_{a,b=1}^{3}K^{ab}\left( Y_{a}Y_{b}-%
\frac{1}{3}\sum_{I,J=1}^{19}g_{ab}^{KL}\left( \mathcal{D}_{cK}Y^{c}\right)
\left( \mathcal{D}_{dK}Y^{d}\right) \right) .  \label{vds}
\end{equation}%
Notice that relaxing the the sums $\sum_{a,b=1}^{3}$ and $\sum_{I,J=1}^{19}$
respectively as $\sum_{a,b=1}^{r}$ and $\sum_{I,J=1}^{n}$ where $r$ and $n$
are positive definite integers, the above equation appears as a particular
relation of a general relation associated with the target space manifold%
\begin{equation}
\frac{SO\left( r,n\right) }{SO\left( r\right) \times SO\left( n\right) }.
\end{equation}%
However the above geometric interpretation cease to be valid since $K_{ab}$
and $K_{IJ}$ can no longer be defined as intersection matrices and are not
necessary symmetric. Nevertheless, it is interesting to note that for the
case $r=2$ (resp $r=4$), eq(\ref{vds}) could be related to the usual
expression of the black hole potential in $4D$ (resp. $6D$) $\mathcal{N}=2$
supergravity.

\subsection{7D attractor eqs}

\qquad We begin by recalling that in 4D $\mathcal{N}=2$ supergravity
embedded in type IIB superstrings on CY3, one generally uses two different,
but equivalent, approaches \textrm{\cite{F2}} to determining the black hole
attractor eqs. These two methods are: \newline
(\textbf{1}) the critically conditions approach based on computing the
critical points of the black hole potential $\delta \mathcal{V}_{BH}^{4D,%
\mathcal{N}=2}=0$.\newline
(\textbf{2}) the so called \emph{new attractor} approach using projections
along the "geometric" and "matter" directions of the Dalbeault basis of the
third cohomology of the CY3. \newline
The first method has been systematically used to deal with black objects in
higher dimensional supergravity theories; in particular in the 5D and 6D
space times. \newline
In 7D $\mathcal{N}=2$ supergravity we are interested in here, assuming non
degeneracy condition, 
\begin{equation}
\left( \mathcal{V}_{BH}^{7D,\mathcal{N}=2}\right) |_{_{\partial \mathcal{V}_{%
{\small BH}}=0}}>0,
\end{equation}%
the critically conditions of the black hole potential reads as 
\begin{equation}
\begin{tabular}{llll}
$\delta \mathcal{V}_{BH}^{7D,\mathcal{N}=2}$ & $=$ & $2\delta ^{\underline{a}%
\underline{b}}\left( \delta Z_{\underline{a}}\right) Z_{\underline{b}%
}+2\delta ^{\underline{I}\underline{J}}Z_{\underline{J}}\delta \left( Z_{%
\underline{I}}\right) =0$ & , \\ 
$\delta Z_{\underline{a}}$ & $=$ & $\left( \frac{\partial Z_{\underline{a}}}{%
\partial \sigma }\right) \delta \sigma +\left( \frac{\partial Z_{\underline{a%
}}}{\partial \phi ^{cI}}\right) \delta \phi ^{cI}=0$ & , \\ 
$\delta Z_{\underline{I}}$ & $=$ & $\left( \frac{\partial Z_{\underline{I}}}{%
\partial \sigma }\right) \delta \sigma +\left( \frac{\partial Z_{\underline{I%
}}}{\partial \phi ^{cI}}\right) \delta \phi ^{cI}=0$ & ,%
\end{tabular}%
\end{equation}%
and leads to the critical\textrm{\ }solutions (\ref{cr}-\ref{cz}) studied in
section 3 and previous subsection. \newline
Below, we develop the \emph{new attractor} approach of Kallosh to the 7D
black attractors.

\subsubsection{Extending the new attractor approach to 7D}

\qquad Here, we study the attractor eqs for the extremal 7D black hole in
the framework of the new attractor approach. The latter is given by
extending the idea of \textrm{\cite{BNB3} }dealing with black holes in type
IIB on CY3-folds\textrm{\ }to the case of black attractors in 11D M-theory
on K3. \newline
The attractor eqs are obtained by evaluating the Hodge decomposition
identity (\ref{da}) along the constraint eqs determining the various classes
of critical points (\ref{cr}-\ref{cz}) of the potential. To get these eqs,
we proceed as follows:\newline
First, we consider from the field strength $\mathcal{F}_{4}=dC_{3}$ in 11D
M-theory compactified on K3 and compute its fluxes as in eq(\ref{flu})
namely,%
\begin{equation}
\begin{tabular}{llll}
$p^{\Lambda }$ & $=$ & $\int_{S_{\infty }^{2}\times \Psi ^{\Lambda }}%
\mathcal{F}_{4}$ & ,%
\end{tabular}%
\end{equation}%
where $p^{\Lambda }$ are integers. This relation can be decomposed in two
equivalent ways; either as 
\begin{equation}
\begin{tabular}{llllll}
$p^{\Lambda }$ & $=$ & $\int_{S_{\infty }^{2}}\left( \int_{\Psi ^{\Lambda }}%
\mathcal{F}_{4}\right) $ & $=$ & $\int_{S_{\infty }^{2}}\mathcal{F}%
_{2}^{\Lambda }$ & ,%
\end{tabular}%
\end{equation}%
or like 
\begin{equation}
\begin{tabular}{llllll}
$p^{\Lambda }$ & $=$ & $\int_{\Psi ^{\Lambda }}\left( \int_{S_{\infty }^{2}}%
\mathcal{F}_{4}\right) $ & $\equiv $ & $\int_{\Psi ^{\Lambda }}\mathcal{H}%
_{2}$ & ,%
\end{tabular}%
\end{equation}%
where we have set%
\begin{equation}
\begin{tabular}{llll}
$\mathcal{F}_{2}^{\Lambda }$ & $=$ & $\int_{\Psi ^{\Lambda }}\mathcal{F}_{4}$
& , \\ 
$\mathcal{H}_{2}$ & $=$ & $\int_{S_{\infty }^{2}}\mathcal{F}_{4}$ & .%
\end{tabular}%
\end{equation}%
Since $\mathcal{H}_{2}\in H^{2}\left( K3,R\right) $, we also have the
decomposition with respect to the basis $\mathbf{\alpha }_{\Lambda }$,%
\begin{equation}
\mathcal{H}_{2}=\sum_{\Lambda =1}^{22}p^{\Lambda }\mathbf{\alpha }_{\Lambda
},\qquad p^{\Lambda }=\int_{S_{\infty }^{2}}\mathcal{F}_{2}^{\Lambda }.
\label{ff}
\end{equation}%
The next step is to Hodge decompose the real gauge invariant 2- form field
strength $\mathcal{H}_{2}$ on the $\left\{ \Omega _{a},\Omega _{I}\right\} $
2-form basis as 
\begin{equation}
\mathcal{H}_{2}=\sum \mathcal{H}^{a}\Omega _{a}+\sum \mathcal{H}^{I}\Omega
_{I},
\end{equation}%
or equivalently like,%
\begin{equation}
\mathcal{H}_{2}=\mathrm{\varsigma }K^{ab}\left( \int_{K3}\mathcal{H}%
_{2}\wedge \Omega _{a}\right) \Omega _{b}+\mathrm{\varkappa }K^{IJ}\left(
\int_{K3}\mathcal{H}_{2}\wedge \Omega _{I}\right) \Omega _{J},  \label{fh}
\end{equation}%
where $\mathrm{\varsigma }$\ and $\mathrm{\varkappa }$\ are numbers which
will be determined below. \newline
Putting $\mathcal{H}_{2}=\sum_{\Lambda =1}^{22}p^{\Lambda }\mathbf{\alpha }%
_{\Lambda }$ back into the right hand side of the above relation and using
the following expressions, 
\begin{equation}
\begin{tabular}{llll}
$X_{a}^{\Lambda }$ & $=$ & $\int_{K3}\mathbf{\alpha }^{\Lambda }\wedge
\Omega _{a}$ & , \\ 
$X_{I}^{\Lambda }$ & $=$ & $\int_{K3}\mathbf{\alpha }^{\Lambda }\wedge
\Omega _{I}$ & ,%
\end{tabular}%
\qquad
\end{equation}%
we can rewrite $\mathcal{H}_{2}$ like,%
\begin{equation}
\mathcal{H}_{2}=\mathrm{\varsigma }K^{ab}\left( \sum_{\Lambda }p_{\Lambda
}X_{a}^{\Lambda }\right) \Omega _{b}+\mathrm{\varkappa }K^{IJ}\left(
\sum_{\Lambda }p_{\Lambda }X_{I}^{\Lambda }\right) \Omega _{J}.
\end{equation}%
The coefficients $\mathrm{\varsigma }$ and $\mathrm{\varkappa }$\ can be
determined by computing 
\begin{equation}
\int_{K3}\mathcal{H}_{2}\wedge \Omega _{a}\text{\qquad },\qquad \int_{%
{\small K3}}\mathcal{H}_{2}\wedge \Omega _{I},
\end{equation}%
in two ways and compare the results. On one hand, we have%
\begin{equation}
\begin{tabular}{llll}
$\int_{K3}\mathcal{H}_{2}\wedge \Omega _{c}$ & $=$ & $\dsum\limits_{\Lambda
}p_{\Lambda }X_{c}^{\Lambda }$ & , \\ 
$\int_{K3}\mathcal{F}_{2}\wedge \Omega _{L}$ & $=$ & $\dsum\limits_{\Lambda
}p_{\Lambda }X_{L}^{\Lambda }$ & ,%
\end{tabular}%
\qquad
\end{equation}%
and on the other hand%
\begin{equation}
\begin{tabular}{llll}
$\int_{K3}\mathcal{H}_{2}\wedge \Omega _{c}$ & $=$ & $\mathrm{\varsigma }%
e^{-2\sigma }\dsum\limits_{\Lambda }p_{\Lambda }X_{c}^{\Lambda }$ & , \\ 
$\int_{K3}\mathcal{H}_{2}\wedge \Omega _{L}$ & $=$ & $\mathrm{\varkappa }%
e^{-2\sigma }\dsum\limits_{\Lambda }p_{\Lambda }X_{L}^{\Lambda }$ & .%
\end{tabular}%
\qquad
\end{equation}%
The identification of the two relations give,%
\begin{equation}
\mathrm{\varsigma }=\mathrm{\varkappa }=e^{2\sigma }.
\end{equation}%
Now using the dressed charges%
\begin{equation}
\begin{tabular}{llllll}
$Y_{a}$ & $=$ & $\int_{K3}\mathcal{H}_{2}\wedge \Omega _{a}$ & $=$ & $%
\dsum\limits_{\Lambda }p_{\Lambda }X_{a}^{\Lambda }$ & , \\ 
$Y_{I}$ & $=$ & $\int_{K3}\mathcal{H}_{2}\wedge \Omega _{I}$ & $=$ & $%
\dsum\limits_{\Lambda }p_{\Lambda }X_{I}^{\Lambda }$ & ,%
\end{tabular}%
\end{equation}%
with $Y_{I}=K^{ab}\mathcal{D}_{aI}Y_{b}$, we can put the Hodge decomposition
into the real 2-form as follows,%
\begin{equation}
\mathcal{H}_{2}=e^{2\sigma }K^{ab}Y_{a}\Omega _{b}+e^{2\sigma
}K^{IJ}Y_{I}\Omega _{J}.  \label{AE}
\end{equation}%
Finally, integrating both sides of (\ref{AE}) over the $\left\{ \Psi
^{\Lambda }\right\} $ basis, we get the 7D black hole attractor eqs%
\begin{equation}
p^{\Lambda }=K^{ab}Y_{a}X_{b}^{\Lambda }+K^{IJ}Y_{I}X_{J}^{\Lambda }.
\label{ate}
\end{equation}%
Notice that this equation can be put in other forms as given below. \newline
First by substituting $K^{ab}=e_{\underline{c}}^{a}e_{\underline{d}}^{b}%
\mathrm{\eta }^{\underline{c}\underline{d}}$, $K^{IJ}=e_{\underline{K}%
}^{I}e_{\underline{L}}^{J}\mathrm{\eta }^{\underline{K}\underline{L}}$ and
using $Z_{\underline{c}}=e_{\underline{c}}^{a}Y_{a}$, $Z_{\underline{K}}=e_{%
\underline{K}}^{I}Y_{I}$, eq(\ref{ate}) becomes%
\begin{equation}
p^{\underline{\Lambda }}=\mathrm{\eta }^{\underline{c}\underline{d}}Z_{%
\underline{c}}L_{\underline{d}}^{\underline{\Lambda }}+\mathrm{\eta }^{%
\underline{K}\underline{L}}Z_{\underline{K}}L_{\underline{L}}^{\underline{%
\Lambda }},  \label{bte}
\end{equation}%
where $\left( L_{\underline{d}}^{\underline{\Lambda }},L_{\underline{L}}^{%
\underline{\Lambda }}\right) $ are as in eq(\ref{bte}).\newline
Second, multiplying eq(\ref{bte}) $p_{\underline{\Lambda }}$ and summing
over $\Lambda $, we rediscover the relations (\ref{inf},\ref{nf}) that we
have used in section 3,%
\begin{equation}
p^{2}=\mathrm{\eta }^{\underline{a}\underline{b}}Z_{\underline{a}}Z_{%
\underline{b}}+\mathrm{\eta }^{\underline{I}\underline{J}}Z_{\underline{I}%
}Z_{\underline{J}},  \label{cte}
\end{equation}%
with $p^{2}=p_{\underline{\Lambda }}p^{\underline{\Lambda }}$.

\subsubsection{Solving the attractor eqs}

\qquad Here we evaluate the fundamental SHG identities along the constraints
determining the various classes of critical points of the black hole (black
3-brane) potential in the moduli space. We show that the supersymmetry
breaking at the horizon of the static, spherically symmetric extremal black
hole (3-brane) solution, can be traced back to the non-vanishing
intersections between the field strength $\mathcal{H}_{2}$ and the
components of the basis $\left\{ \Omega _{\underline{a}}\Omega _{\underline{I%
}}\right\} $. We have:\newline

(1) \emph{Supersymmetric }$\frac{1}{2}$\emph{\ BPS} \newline
This supersymmetric 7D attractor corresponds to the critical point $Z_{%
\underline{a}}\neq \left( 0,0,0\right) $ and$\ Z_{\underline{I}}=\left(
0,...,0\right) $. Putting $Z_{\underline{I}}=0$ $\forall $ $I$ $\in \mathcal{%
I}$ $=$ $\left\{ {\small 1,...,19}\right\} $ back in eq(\ref{AE}), we find
that the real 2- form $\mathcal{H}_{2}$ of M-theory on K3 has vanishing
components along the second cohomologies $H^{\left( 1,1\right) }\left(
K3\right) $ generated by $\Omega _{I}=\mathcal{D}_{aI}\Omega ^{a}$. As such
the 2- form $\left( \mathcal{H}_{2}\right) _{\frac{{\small 1}}{{\small 2}}%
{\small BPS}}$ reduces down to,%
\begin{equation}
\begin{tabular}{llll}
$\left( \mathcal{H}_{2}\right) _{\frac{{\small 1}}{{\small 2}}{\small BPS}}$
& $=$ & $\left( e^{2\sigma }K^{ab}Y_{a}\Omega _{b}\right) _{\frac{{\small 1}%
}{{\small 2}}{\small BPS}}$ & , \\ 
& $=$ & $\left( e^{2\sigma }\mathrm{\eta }^{\underline{c}\underline{d}}Z_{%
\underline{c}}\Omega _{\underline{d}}\right) _{\frac{{\small 1}}{{\small 2}}%
{\small BPS}}$ & .%
\end{tabular}
\label{ab}
\end{equation}%
The BPS non degeneracy condition $\left( Z_{\underline{a}}\right) _{\frac{%
{\small 1}}{{\small 2}}{\small BPS}}\neq 0$ corresponds therefore to a
condition of \emph{non orthogonality} between $\mathcal{H}_{2}\ $and $\Omega
_{\underline{a}}$,%
\begin{equation}
\begin{tabular}{llll}
$\int_{K3}\mathcal{H}_{2}\wedge \Omega _{a}$ & $\neq 0$ & ,$\qquad $at least
for one of the $a$'s & , \\ 
$\int_{K3}\mathcal{H}_{2}\wedge \Omega _{I}$ & $=0$ & ,$\qquad \forall $ $%
I=1,\ldots ,19$ & .%
\end{tabular}%
\end{equation}

(2) \emph{Non BPS type 1}\newline
This non supersymmetric attractor corresponds to the critical point\emph{\ }$%
Z_{a}=\left( 0,0,0\right) $; but $Z_{I}\neq \left( 0,...,0\right) $. \newline
The real flux 2-form $\mathcal{H}_{2}$ of M-theory on K3 has non zero
components along $\Omega _{I}$; but no component along $\Omega _{a}$, 
\begin{equation}
\begin{tabular}{llll}
$\int_{K3}\mathcal{H}_{2}\wedge \Omega _{a}$ & $=0$ & ,$\qquad \forall $ $%
a=1,2,3,$ & , \\ 
$\int_{K3}\mathcal{H}_{2}\wedge \Omega _{I}$ & $\neq 0$ & ,$\qquad $\ at
least for one of the $I$'s & .%
\end{tabular}%
\end{equation}%
Then, we have%
\begin{equation}
\begin{tabular}{llll}
$\left( \mathcal{H}_{2}\right) _{\left( N{\small BPS}\right) _{1}}$ & $=$ & $%
\left( e^{2\sigma }K^{IJ}Y_{I}\Omega _{J}\right) _{\left( N{\small BPS}%
\right) _{1}}$ & , \\ 
& $=$ & $\left( e^{2\sigma }\mathrm{\eta }^{\underline{I}\underline{J}}Z_{%
\underline{I}}\Omega _{\underline{J}}\right) _{\left( N{\small BPS}\right)
_{1}}$ & .%
\end{tabular}%
\end{equation}

(\textbf{3}) \emph{Non BPS type 2}\newline
This is a non supersymmetric attractor corresponding to the critical point $%
Z_{a}\neq \left( 0,0,0\right) $ and $Z_{I}\neq \left( 0,...,0\right) $.%
\newline
The real flux 2-form $\mathcal{H}_{2}$ of M-theory on K3 has at least one
non zero component along $\Omega _{a}$ and at least one non zero component
along $\Omega _{I}$, 
\begin{equation}
\begin{tabular}{llll}
$\int_{K3}\mathcal{H}_{2}\wedge \Omega _{a}$ & $=0$ & ,$\qquad $\ at least
for one of the $a$'s\  & , \\ 
$\int_{K3}\mathcal{H}_{2}\wedge \Omega _{I}$ & $\neq 0$ & ,$\qquad $\ at
least for one of the $I$'s\  & .%
\end{tabular}
\label{ac}
\end{equation}

\section{Conclusion and discussion}

\qquad In this paper we have studied the extremal BPS and non BPS black
attractors in the seven dimensional $\mathcal{N}=2$ supergravity embedded in
11D M- theory on K3. The attractor eqs and their solutions have been treated
by using both the criticality condition of the attractor potential (black
hole and the dual black 3-brane) as well as by extending the 4D attractor
approach of Kallosh to $\mathcal{N}=2$ supergravity in 7D space time.

After having given some useful tools on ways to deal with the moduli space
of the theory,%
\begin{equation}
\boldsymbol{M}_{7D}^{N=2}=\frac{SO\left( 1,1\right) \times SO\left(
3,19\right) }{SO\left( 3\right) \times SO\left( 19\right) }\text{ },
\end{equation}%
we have described the brane realizations of the 7D black objects in terms of
M2 and M5 branes wrapping 2-cycles of K3. Then, we have studied explicitly
the corresponding attractor mechanism: First, by using the critically
condition method, in both inertial and curved frames $\left\{ \xi ^{%
\underline{m}}\left( x\right) \right\} $ and $\left\{ \varphi ^{m}\left(
x\right) \right\} $ of the moduli space (sections 3 and 7). Second, by
extending the so called "new attractor approach" of Kallosh (section 7).

Moreover, using specific properties of the quantum numbers of the fields of
the 7D\ theory, we have derived the 2-form basis eq(\ref{sa}) for the second
real cohomology of K3,%
\begin{equation}
\left\{ \Omega _{a},\Omega _{I}\right\} _{I=1,...,19}^{a=1,2,3}\text{ \ \ .}
\end{equation}%
This basis, refereed to as the new basis of $H^{2}\left( K3,R\right) $,
exhibits manifestly the $SO\left( 3\right) \times SO\left( 19\right) $
isotropy symmetry of the moduli space and plays an important role in the
study the underlying special hyperKahler geometry of 11D M-theory on K3. The
new basis, which could be also motivated by using properties of the Picard
group of complex curves in K3 \textrm{\cite{AS,GO}}, has been derived here
from the two following physical arguments:\newline
(\textbf{i}) the $7D$ $\mathcal{N}=2$ supergravity field theory has two
kinds of irreducible supersymmetric fields representations, namely the
supergravity multiplet $\mathcal{G}_{7D}^{N=2}$ eq\textrm{(\ref{go})} and
the Maxwell-matter supermultiplet $\mathcal{V}_{7D}^{N=2}$ eq\textrm{(\ref%
{ga})}. Each one of these two representations contains its own abelian
Maxwell gauge fields: $\mathcal{G}_{7D}^{N=2}$ has \emph{three} 7D space
time gauge fields 
\begin{equation}
\mathcal{A}_{\mu }^{a}\left( x\right) ,\qquad a=1,2,3,
\end{equation}%
while the gauge-matter sector with the set $\left\{ \left( \mathcal{V}%
_{7D}^{N=2}\right) _{I}\right\} $\ has \emph{nineteen} 
\begin{equation}
\mathcal{A}_{\mu }^{I}\left( x\right) ,\qquad I=1,...,19,
\end{equation}%
constituting altogether the \emph{twenty} \emph{two} gauge fields of the
underlying $U^{22}\left( 1\right) $ gauge invariance. This splitting allows
to classify the field strengths of the 7D supergravity theory into two kinds
namely $\mathcal{F}_{\mu \nu }^{a}$ and $\mathcal{F}_{\mu \nu }^{I}$; and
leads then to two types of physical gauge invariant (magnetic) charges 
\begin{equation}
m^{a}=\left( \int_{S^{2}}\mathcal{F}^{a}\right) \qquad ,\qquad m^{I}=\left(
\int_{S^{2}}\mathcal{F}^{a}\right) .
\end{equation}%
These magnetic black hole charges are precisely the dressed charges $Z^{%
\underline{a}}$ and $Z^{\underline{I}}$ of the extended brane version of the
7D $\mathcal{N}=2$ superalgebra \textrm{\cite{FE0,07,03,08,09}}.\newline
(\textbf{ii}) the compactification of 11D M- theory on K3, together with the
Calabi-Yau condition preventing 1-cycles, lead to the possibility to combine
both the Kahler moduli%
\begin{equation*}
t^{I}\equiv z^{0I}
\end{equation*}%
and the complex deformations 
\begin{equation*}
\left( z^{I},\overline{z}^{I}\right) \equiv \left( z^{+I},z^{-I}\right)
\end{equation*}%
of the metric of K3 into \emph{nineteen} isotriplets 
\begin{equation}
\xi ^{aI}=\left( z^{0I},z^{+I},z^{-I}\right) ,\text{\qquad }I=1,...,19,
\end{equation}%
which are nothing but the \emph{fifty seven} scalars of the \emph{nineteen}
Maxwell-matter gauge multiplets of the gauge sector of the supergravity
theory. This combination is a very special property of the K3 surface; which
reflects in some sense its hyperKahler nature; it has no analogue in higher
dimensional Calabi-Yau manifolds.

Furthermore, using the new basis $\left\{ \Omega _{a},\Omega _{I}\right\} $
of $H^{2}\left( K3,R\right) $ and the deformation tensor $\Omega _{aI}^{b}$
eqs(\ref{def}-\ref{omm}) of the metric of K3 as well as the symmetric inner
product $\left\langle F,G\right\rangle =\int_{K3}F\wedge G$, we have derived
the fundamental relations (\ref{kkg}-\ref{kkk}) of the SHG geometry of the
moduli space $\frac{SO\left( 1,1\right) \times SO\left( 3,19\right) }{%
SO\left( 3\right) SO\left( 19\right) }$; see also eqs(\ref{sik}-\ref{sid}). 
\newline
By decomposing $\Omega _{a}$ and $\Omega _{I}$ with respect to the standard (%
\emph{old})\emph{\ basis} Hodge of $H^{2}\left( K3,R\right) $, \ 
\begin{equation}
\left\{ \mathbf{\alpha }_{\Lambda }\right\} _{\Lambda =1,...,22}
\end{equation}%
we recover all usual constraint eqs of the 7D theory given in \textrm{\cite%
{FE0}}; especially the canonical coordinates eqs(\ref{ql}-\ref{qq}), the
dressed charges eqs(\ref{dua}-\ref{duo}) and the constraint eqs(\ref{xj}-\ref%
{xl}) described in section 2.\newline
It is remarkable that the physical field strength $\mathcal{F}_{\mu \nu
}^{a} $ of the gravity multiplet and the field strength $\mathcal{F}_{\mu
\nu }^{I} $ of the Maxwell-matter multiplet are given by the \emph{linear
combinations} (\ref{61}-\ref{63}),%
\begin{equation}
\mathcal{F}_{\mu \nu }^{\underline{a}}=\sum_{\Lambda =1}^{22}L_{\underline{%
\Lambda }}^{\underline{a}}\mathcal{F}_{\mu \nu }^{\underline{\Lambda }%
}\qquad ,\qquad \mathcal{F}_{\mu \nu }^{\underline{I}}=\sum_{\Lambda
=1}^{22}L_{\underline{\Lambda }}^{\underline{I}}\mathcal{F}_{\mu \nu }^{%
\underline{\Lambda }},
\end{equation}%
where $\mathcal{F}_{\mu \nu }^{\underline{\Lambda }}$ is the compactified
4-form of the 11D M-theory on the 2-cycles basis $\Psi ^{\Lambda }\in
H_{2}\left( K3,R\right) $ 
\begin{equation}
\mathcal{F}_{2}^{\Lambda }=\int_{\Psi ^{\Lambda }}\mathcal{F}_{4}\qquad
,\qquad \int_{\Psi ^{\Lambda }}\mathbf{\alpha }_{\Sigma }=\delta _{\Sigma
}^{\Lambda }.
\end{equation}%
The decomposition coefficients $L_{\underline{\Sigma }}^{\underline{\Delta }%
}=\left( L_{\underline{a}}^{\underline{\Delta }},L_{\underline{I}}^{%
\underline{\Delta }}\right) $ are given by 
\begin{equation}
\int_{\Psi ^{\underline{\Lambda }}}\Omega _{\underline{a}}=L_{\underline{a}%
}^{\underline{\Delta }}\qquad ,\qquad \int_{\Psi ^{\underline{\Lambda }%
}}\Omega _{\underline{I}}=L_{\underline{I}}^{\underline{\Delta }},
\end{equation}%
and form precisely the $SO\left( 3,19\right) $ orthogonal field matrix $L_{%
\underline{\Sigma }}^{\underline{\Delta }}$ considered in section 2, eqs(\ref%
{pn}-\ref{ln}).

With the $\left\{ \Omega _{a},\Omega _{I}\right\} $ basis at hand, we have
also extended the Kallosh attractor approach to the case of 7D $\mathcal{N}%
=2 $ supergravity. Then we have used this "extended new approach" to
rederive the 7D black hole (7D black 3-brane) attractor eqs(\ref{AE}-\ref%
{bte}) and their solutions (\ref{ab}-\ref{ac}) which have been also
classified in terms of the sign of $p^{2}$; see eqs(\ref{za}-\ref{zz}).%
\newline

In the end, we would like to add that the compactification of the 7D $%
\mathcal{N}=2$ supergravity theory on a circle leads to 6D $\mathcal{N}=2$
non chiral supergravity. This is also equivalent to compactifying 10D type
IIA superstring on K3 \textrm{\cite{5D4} }or the heterotic string on the
3-torus. Then, one can think about the analysis given in this paper as the
uplifting of 6D $\mathcal{N}=2$ supergravity theory to the 7D; in analogy
with the uplifting of 4D $\mathcal{N}=2$ supergravity theory to the 5D with 
\emph{real} cubic prepotential \textrm{\cite{FC,FCA, FCB,5D41}}. \newline
This property allows us to ask whether results concerning 4D/5D
correspondence with cubic prepotential could be generalized to the 6D/7D
case where we have a quadratic prepotential. Below, we give an heuristic
exploration of this issue.

\subsection{6D/7D correspondence}

An interesting field theoretical way to study the link between the 6D/7D BPS
and non BPS attractors is to follow the analysis of Ceresole, Ferrara and
Marrani (CFM) \textrm{\cite{FC} }concerning\textrm{\ }the 4D/5D
correspondence and explore how it could be extended to get the 6D/7D
correspondence for the black attractor potentials and their critical points. 
\newline
In the CFM field theory set up, the extension 
\begin{equation}
\begin{tabular}{llll}
4D/5D {\small correspondence} & $\qquad \rightarrow \qquad $ & 6D/7D {\small %
correspondence} & ,%
\end{tabular}%
\end{equation}%
could, \`{a} priori, be done by first working out a dictionary regarding the
links between the moduli spaces of the 4D, 5D, 6D and 7D supergravity
theories. \newline
Second, determine the various effective potentials from which we may read
the critical points and their relations.\newline

\textbf{(1) }\emph{Dictionary}\newline
A first step in the way to 6D/7D correspondence can be made by working out
the relation between the geometries of the underlying moduli spaces of 4D
(resp. 5D) and 6D (resp. 7D) $\mathcal{N}=2$ supergravity theories. We have
the following picture,

\begin{equation}
\begin{tabular}{lll}
{\small 4D}$:${\small \ \ SK Geometry} & $\qquad \mathbf{\longleftrightarrow 
}\qquad $ & {\small 6D}$:${\small \ \ SQ Geometry} \\ 
$\qquad \updownarrow $ &  & $\qquad \updownarrow $ \\ 
{\small 5D}$:${\small \ \ SR Geometry} & $\qquad \mathbf{\longleftrightarrow 
}\qquad $ & {\small 7D}$:${\small \ \ SH Geometry}%
\end{tabular}
\label{ssss}
\end{equation}

\ \ \newline
where SQG and and SHG stands for special quaternionic and special
hyperkahler geometries respectively. \newline
Much about the 4D/5D $\leftrightarrow $\ 6D/7D dictionary can be also learnt
from the isotropy symmetries of the underlying $\mathcal{N}=2$ supergravity
theories and from the way the fields have been generated from the 10D
superstrings and M-theory compactifications. In the type IIA set up, we have

\begin{equation}
\begin{tabular}{lll}
10D\ {\small Type IIA/CY3} & $\qquad \mathbf{\longleftrightarrow }\qquad $ & 
10D\ {\small Type IIA/K3} \\ 
$\qquad \downarrow $ &  & $\qquad \downarrow $ \\ 
Uplift to 5D & $\qquad \mathbf{\longleftrightarrow }\qquad $ & Uplift to 7D%
\end{tabular}
\label{m5}
\end{equation}

\ \ \newline
These correspondences can be translated in the language of 2-forms on the
corresponding moduli space as follows

\begin{equation}
\begin{tabular}{lll}
$B^{{\small NS}}+iJ$ & $\qquad \mathbf{\longleftrightarrow }\qquad $ & $B^{%
{\small NS}}+\sigma ^{a}\Omega _{a}$ \\ 
$\qquad \updownarrow $ &  & $\qquad \updownarrow $ \\ 
$\qquad J$ & $\qquad \mathbf{\longleftrightarrow }\qquad $ & $\qquad \Omega
_{a}$%
\end{tabular}%
\end{equation}

\ \ \newline
Here $B^{{\small NS}}+iJ$ is the complexified Kahler form with $B^{{\small NS%
}}$ standing for the {\small NS-NS} B-field of type II superstrings and give
axions $\chi ^{i}$ up on integration over the 2-cycles $C_{2}^{i}$ of the
compact spaces,%
\begin{equation}
\chi ^{i}=\int_{C_{2}^{i}}B^{{\small NS}}.  \label{ax}
\end{equation}%
Notice by the way that the table (\ref{m5}) can be also stated by starting
from 11D M-theory on CY3 and on K3; then compactifying on a circle.%
\begin{equation}
\begin{tabular}{lll}
{\small down lift to 4D} & $\qquad \mathbf{\longleftrightarrow }\qquad $ & 
{\small down lift to 6D} \\ 
$\qquad \uparrow $ &  & $\qquad \uparrow $ \\ 
{\small M-theory on CY3} & $\qquad \mathbf{\longleftrightarrow }\qquad $ & 
{\small M-theory on K3}%
\end{tabular}%
\end{equation}
Using results of \textrm{\cite{FC}} and the analysis given in \textrm{\cite%
{5D4}}; although more explicit and handleable expressions are still needed,
we learnt that the CFM method could be applied to the 6D/7D case provided we
can have the explicit expressions of the potentials in the special
coordinate.\newline

\textbf{(2) }\emph{Potentials}\newline
With the relations (\ref{ssss}-\ref{ax}) in mind, the second step to 6D/7D
correspondence is to mimic the CFM analysis of ref.\textrm{\cite{FC}. }%
There, the 5D black hole potential $\mathcal{V}_{BH}^{5D,N=2}$ is determined
by using the known expression of $\mathcal{V}_{BH}^{4D,N=2}$ and putting
constraints on the axions $\chi ^{i}$ (\ref{ax}) and the volume of the CY3. 
\newline
The extension of the CFM field theoretical method towards a 6D/7D
correspondence can be done in a similar manner. For this purpose, we need to
know the effective potential of 6D black attractors $\mathcal{V}%
_{BH}^{6D,N=2}$ in the special quaternionic coordinates on which we put
constraints on the axions $\chi ^{i}$ (mainly $\chi ^{i}\rightarrow 0,$ $%
i=1,...22$) and on the volume of K3. In the language of the moduli space
group symmetries, the uplifting from 6D to 7D corresponds to the symmetry
breaking%
\begin{equation}
\begin{tabular}{llll}
$SO\left( 4,20\right) $ & $\qquad \rightarrow \qquad $ & $SO\left(
3,19\right) $ & , \\ 
$SO\left( 4\right) $ & $\qquad \rightarrow \qquad $ & $SO\left( 3\right) $ & 
, \\ 
$SO\left( 20\right) $ & $\qquad \rightarrow \qquad $ & $SO\left( 19\right) $
& .%
\end{tabular}%
\end{equation}%
At the level of the scalar field manifolds, the 6D$\rightarrow $7D uplifting
is accompanied by the breaking $\frac{SO\left( 4,20\right) }{SO\left(
4\right) \times SO\left( 20\right) }\rightarrow \frac{SO\left( 3,19\right) }{%
SO\left( 3\right) \times SO\left( 19\right) }$ reducing the dimension from
real 80 dimension down to the real dimension 57 sub-manifold. This reduction
corresponds then to fixing 23 real moduli and these are precisely given by
the constraints on the axions, $\chi ^{i}\rightarrow 0$, $i=1,...22$; and by
fixing the volume of K3.\newline
However, the knowledge of the explicit expression $\mathcal{V}_{BH}^{6D,N=2}$
in the special quaternionic coordinates is some how problematic; since it
requires the knowledge of the explicit expression of the quaternionic metric 
$G_{mn}^{\text{quaternion}}$ of the moduli space$\footnote{%
the dilaton $\sigma $, captured by the $SO\left( 1,1\right) $ subgroup
factor, is freezed in (\ref{m6d})}$ of 6D $\mathcal{N}=2$ supergravity, 
\begin{equation}
\boldsymbol{M}_{6D}^{N=2}=SO\left( 1,1\right) \times \frac{SO\left(
4,20\right) }{SO\left( 4\right) \times SO\left( 20\right) }.  \label{m6d}
\end{equation}%
To our knowledge, the explicit expression of $G_{mn}^{\text{quaternion}}$ is
still missing although it is suspected to be a \emph{real 80 dimensional}
generalization of the Taub-NUT metric of 4D Euclidean gravity. Thought
lengthy and technical, the explicit expression of $G_{mn}^{\text{quaternion}%
} $ could be however derived by using harmonic superspace method \textrm{%
\cite{HS}-\cite{HS2}}. The explicit expression of $G_{mn}^{{\scriptsize %
quaternion}}$ will be considered in a future occasion.\newline
Nevertheless, partial results can be still given by using the Weinhold
potential (\ref{vef}) and the constrained matrix representation of
sub-section 2.2. The 7D black hole potential $\mathcal{V}_{BH}^{7D,N=2}$ can
be put in a form quite similar to the $\mathcal{V}_{BH}^{5D,N=2}$
corresponding one. Up on solving underlying constraints, $\mathcal{V}%
_{BH}^{7D,N=2}$ can be expressed in terms of the special coordinates $\xi ^{%
\underline{a}\underline{I}}$ eq(\ref{ob}) and the magnetic bare charges $p^{%
\underline{a}}$ and $p^{\underline{I}}$. \newline
To see how this can be done, we start from $\mathcal{V}_{BH}^{7D,N=2}$ in
terms of the dressed central charges $\mathcal{Z}^{\underline{a}}$ and $%
\mathcal{Z}^{\underline{I}}$ eq(\ref{vef}). Then, we put this potential in
the quadratic form,%
\begin{equation}
\mathcal{V}_{BH}^{7D,N=2}=\frac{1}{2}\left( \mathcal{M}_{\underline{a}%
\underline{b}}p^{\underline{a}}p^{\underline{b}}+\mathcal{M}_{\underline{a}%
\underline{J}}p^{\underline{a}}p^{\underline{J}}+\mathcal{M}_{\underline{I}%
\underline{b}}p^{\underline{I}}p^{\underline{a}}+\mathcal{M}_{\underline{I}%
\underline{J}}p^{\underline{I}}p^{\underline{J}}\right) ,
\end{equation}%
or equivalently like%
\begin{equation}
\mathcal{V}_{BH}^{7D,N=2}=\frac{1}{2}\left( p^{\underline{a}},p^{\underline{I%
}}\right) \left( 
\begin{array}{cc}
\mathcal{M}_{\underline{a}\underline{b}} & \mathcal{M}_{\underline{a}%
\underline{J}} \\ 
\mathcal{M}_{\underline{I}\underline{b}} & \mathcal{M}_{\underline{I}%
\underline{J}}%
\end{array}%
\right) \left( 
\begin{array}{c}
p^{\underline{b}} \\ 
p^{\underline{J}}%
\end{array}%
\right) ,
\end{equation}%
where the $22\times 22$ matrix $\mathcal{M}_{\underline{\Lambda }\underline{%
\Sigma }}$ is given by 
\begin{equation}
\mathcal{M}_{\underline{\Lambda }\underline{\Sigma }}=2\left(
\sum_{c,d=1}^{3}\mathcal{L}_{\underline{\Lambda }}^{\underline{c}}\delta _{%
\underline{c}\underline{d}}\mathcal{L}_{\underline{\Sigma }}^{\underline{d}%
}\right) +2\left( \sum_{K,L=1}^{19}\mathcal{L}_{\underline{\Lambda }}^{%
\underline{K}}\delta _{\underline{K}\underline{L}}\mathcal{L}_{\underline{%
\Sigma }}^{\underline{L}}\right) ,
\end{equation}%
with $\mathcal{L}_{\underline{\Lambda }}^{\underline{c}}$ and $\mathcal{L}_{%
\underline{\Lambda }}^{\underline{K}}$ as in eqs(\ref{LL})\textrm{.} \newline
Next, using the constraint eq(\ref{qq}), we can also rewrite the matrix $%
\mathcal{M}_{\underline{\Lambda }\underline{\Sigma }}$ as, 
\begin{equation}
\mathcal{M}_{\underline{\Lambda }\underline{\Sigma }}=2e^{-2\sigma }\left[
\eta _{\underline{\Lambda }\underline{\Sigma }}+2\left( \sum_{I,J=1}^{19}L_{%
\underline{\Lambda }}^{\underline{I}}\delta _{\underline{I}\underline{J}}L_{%
\underline{\Sigma }}^{\underline{J}}\right) \right] ,  \label{mab}
\end{equation}%
where the dependence into the dilaton has been factorized. This expression
can be simplified further by replacing $L_{\underline{\Lambda }}^{\underline{%
I}}$ as in eq(\ref{lrr},\ref{ob}), which we rewrite as follows,%
\begin{equation}
L_{\underline{\Lambda }}^{\underline{\Sigma }}=\left( 
\begin{array}{cc}
\sqrt{\frac{3+\xi ^{2}}{3}}\delta _{\underline{a}}^{\underline{b}} & \sqrt{%
\frac{19+\xi ^{2}}{19}}\xi _{\underline{a}}^{\underline{J}} \\ 
\sqrt{\frac{3+\xi ^{2}}{3}}\xi _{\underline{I}}^{\underline{b}} & \sqrt{%
\frac{19+\xi ^{2}}{19}}\delta _{\underline{I}}^{\underline{J}}%
\end{array}%
\right) ,
\end{equation}%
where $\xi _{\underline{I}}^{\underline{b}}=\left( \xi _{\underline{b}}^{%
\underline{I}}\right) ^{t}=\eta _{\underline{I}\underline{J}}\eta ^{%
\underline{a}\underline{b}}\xi _{\underline{a}}^{\underline{J}}$ and $\xi
^{2}=\sum \xi _{\underline{a}}^{\underline{I}}\xi _{\underline{I}}^{%
\underline{a}}=\sum \xi ^{\underline{a}\underline{I}}\xi _{\underline{a}%
\underline{I}}$. \newline
Putting these relations back into (\ref{mab}), we get the explicit
expression of the black hole potential in terms of the special coordinates $%
\xi $. \newline
The next step is to do the same thing for the potential of the 6D black hole 
$\mathcal{V}_{BH}^{6D,N=2}$. Then, try to figure out the 6D/7D
correspondence by following the method of Ceresole, Ferrara and Marrani.
Progress in this direction will be reported elsewhere.

\section{Appendix}

In this appendix, we describe some useful relations regarding SKG in curved
and the inertial frames. These relations complete the analysis of
sub-section 5.1 and allows to make formal analogies with the analysis given
in section 6 regarding the fundamental relations of SHG. \newline
4D $\mathcal{N}=2$ supergravity has been extensively studied in literature,
it can be realized as the effective field theory of 10D superstring II on
Calabi-Yau threefolds. We first review the fundamentals of the SKG geometry
underlying its scalar manifold $\boldsymbol{M}_{4D}^{N=2}$, with $\dim _{C}%
\boldsymbol{M}_{4D}^{N=2}=n$ in curved frame. Then, we consider the same
relations; but now in the inertial frame set up.\newline

\textbf{(1)} \emph{SKG in curved frame}\newline
To fix the ideas, consider 10D superstring$\footnote{%
In type IIA set up, the complex variables $z^{i}$ are given by the moduli of
the complexified Kahler 2- form $B^{{\small NS}}+iJ$ over the the 2- cycles $%
C_{2}^{i}$ of $H_{2}\left( CY3\right) $.}$ IIB on Calabi-Yau threefolds and
let $\left( z^{+i},z^{-i}\right) _{i=1,...,n}$ be the local (special)
coordinates of the $\boldsymbol{M}_{4D}^{N=2}$ with $n$ being the number of
abelian vector supermultiplets that couple the supergravity multiplet. The
metric $g_{i\overline{j}}$ of this Kahler manifold which, for convenience,
we rewrite it as $g_{-i+j}$, is given by.%
\begin{equation}
\begin{tabular}{llll}
$g_{-i+j}$ & $=$ & $\partial _{-i}\partial _{+j}\mathcal{K}$ & , \\ 
$\partial _{\mp i}$ & $=$ & $\frac{\partial }{\partial z^{\pm i}}$ & , \\ 
$\overline{\left( z^{+i}\right) }$ & $=$ & $z_{i}^{-}$ & .%
\end{tabular}%
\end{equation}%
In this relation, $\mathcal{K}=\mathcal{K}\left( z^{+},z^{-}\right) $ is the
Kahler potential with the usual gauge transformation%
\begin{equation}
\mathcal{K\qquad }\rightarrow \qquad \mathcal{K}+\mathrm{f}\left(
z^{+}\right) +\overline{\mathrm{f}}\left( z^{-}\right) ,  \label{kt}
\end{equation}%
where $\mathrm{f}\left( z^{+}\right) $ is an arbitrary holomorphic function.
The abelian gauge transformation (\ref{kt}) leaves the metric $g_{-i+j}$
invariant since the variation $\partial _{-i}\partial _{+j}\mathrm{f}\left(
z^{+}\right) =0$. \newline
Let also 
\begin{equation}
\begin{tabular}{lllllll}
{\small Hodge}:\qquad & $\qquad \mathbf{\alpha }_{\Lambda }$ & , & $\qquad 
\mathbf{\beta }^{\Lambda }$ & , & $\Lambda ={\small 0,...,n}$ & , \\ 
{\small Dalbeault:\qquad } & $\Omega _{+}$ $\ ,$ $\ \Omega _{-i+}$ & , & $%
\Omega _{-}$ $\ ,$ $\ \Omega _{+i-}$ & , & $i={\small 1,...,n}$ & ,%
\end{tabular}%
\end{equation}%
be respectively the Hodge and Dalbeault basis of 3-forms of $H^{3}\left(
CY3\right) $ with 
\begin{equation}
\begin{tabular}{llll}
$\Omega _{-}$ & $=$ & $\overline{\left( \Omega _{+}\right) }$ & , \\ 
$\Omega _{+i-}$ & $=$ & $\overline{\left( \Omega _{-i+}\right) }$ & , \\ 
$n$ & $=$ & $h^{2,1}\left( CY3\right) $ & ,%
\end{tabular}%
\end{equation}%
and $\left\{ A^{\Lambda },B_{\Lambda }\right\} $ being the usual symplectic
basis of real 3-cycles given by eqs(\ref{alb}). \newline
Since both Hodge and Dalbeault 3-forms are two independent basises of the
third real cohomology of CY3, we have the following relation%
\begin{equation}
\begin{tabular}{llll}
$\Omega _{\pm }$ & $=$ & $\mathbf{\alpha }_{\Lambda }X_{\pm }^{\Lambda }-%
\mathbf{\beta }^{\Lambda }F_{\Lambda \pm }$ & , \\ 
$\Omega _{-i+}$ & $=$ & $\mathbf{\alpha }_{\Lambda }X_{-i+}^{\Lambda }-%
\mathbf{\beta }^{\Lambda }F_{\Lambda -i+}$ & , \\ 
$\Omega _{+i-}$ & $=$ & $\mathbf{\alpha }_{\Lambda }X_{+i-}^{\Lambda }-%
\mathbf{\beta }^{\Lambda }F_{\Lambda +i-}$ & ,%
\end{tabular}%
\end{equation}%
with%
\begin{equation}
\begin{tabular}{llllllll}
$X_{\pm }^{\Lambda }$ & $=$ & $\int_{A^{\Lambda }}\Omega _{\pm }$ & \qquad
,\qquad & $F_{\Lambda \pm }$ & $=$ & $\int_{B_{\Lambda }}\Omega _{\pm }$ & ,
\\ 
$X_{-i+}^{\Lambda }$ & $=$ & $\int_{A^{\Lambda }}\Omega _{-i+}$ & \qquad
,\qquad & $F_{\Lambda -i+}$ & $=$ & $\int_{B_{\Lambda }}\Omega _{-i+}$ & ,
\\ 
$X_{+i-}^{\Lambda }$ & $=$ & $\int_{A^{\Lambda }}\Omega _{+i-}$ & \qquad
,\qquad & $F_{\Lambda +i-}$ & $=$ & $\int_{B_{\Lambda }}\Omega _{+i-}$ & ,%
\end{tabular}
\label{xf}
\end{equation}%
and 
\begin{equation}
\begin{tabular}{llllllll}
$X_{+}^{\Lambda }$ & $=$ & $X_{+}^{\Lambda }\left( z^{+}\right) $ & \qquad
,\qquad & $X_{-}^{\Lambda }$ & $=$ & $X_{-}^{\Lambda }\left( z^{-}\right) $
& , \\ 
$F_{\Lambda +}$ & $=$ & $F_{\Lambda +}\left( z^{+}\right) $ & \qquad ,\qquad
& $F_{\Lambda -}$ & $=$ & $F_{\Lambda -}\left( z^{-}\right) $ & ,%
\end{tabular}
\label{xg}
\end{equation}%
Using these 3-forms, we can define the fundamental relations of the SKG in
curved frame:\newline

\textbf{(a)} \emph{the} \emph{Kahler potential}\newline
It is defined by computing the volume $\left( 3,3\right) $- form on the
moduli space and reads as%
\begin{equation}
\begin{tabular}{llll}
$\int_{CY3}\Omega _{+}\wedge \Omega _{-}$ & $=$ & $ie^{-\mathcal{K}}$ & , \\ 
$\int_{CY3}\Omega _{+}\wedge \Omega _{+}$ & $=$ & $0$ & , \\ 
$\int_{CY3}\Omega _{-}\wedge \Omega _{-}$ & $=$ & $0$ & ,%
\end{tabular}
\label{pm}
\end{equation}%
where $\mathcal{K}$ is the Kahler potential. The number $i$ is required by
the reality condition and antisymmetry $\Omega _{+}\wedge \Omega
_{-}=-\Omega _{-}\wedge \Omega _{+}$. \newline
Notice that setting 
\begin{equation}
\begin{tabular}{llll}
$z^{\pm j}$ & $=$ & $x^{j}\pm iy^{j}$ & , \\ 
$\partial _{\pm j}$ & $=$ & $\frac{\partial }{2\partial x^{j}}\mp i\frac{%
\partial }{2\partial y^{j}}$ & , \\ 
$\Omega _{\pm }$ & $=$ & $\Omega _{1}\mp i\Omega _{2}$ & ,%
\end{tabular}%
\end{equation}%
we have 
\begin{equation}
\begin{tabular}{llll}
$\partial _{+j}\Omega _{-}+\partial _{+j}\Omega _{-}$ & $=$ & $\frac{%
\partial \Omega _{1}}{\partial x^{j}}+\frac{\partial \Omega _{2}}{\partial
y^{j}}$ & .%
\end{tabular}%
\end{equation}%
To make contact with our analysis concerning the SHG analysis we have given
in section 6, it is convenient to set%
\begin{equation}
\Omega _{a}=\left( \Omega _{+},\Omega _{-}\right) ,\qquad \Omega _{-}=%
\overline{\left( \Omega _{+}\right) },
\end{equation}%
and rewrite the above relations collectively as follows 
\begin{equation}
\begin{tabular}{llll}
$\int_{CY3}\Omega _{a}\wedge \Omega _{b}$ & $=$ & $-iK_{ab}$ & , \\ 
& $=$ & $ie^{-\mathcal{K}}\varepsilon _{ab}$ & , \\ 
$\varepsilon _{-+}=\varepsilon ^{+-}=-\varepsilon ^{-+}$ & $=$ & $%
-\varepsilon _{+-}=1$ & ,%
\end{tabular}
\label{kae}
\end{equation}%
with $K_{ab}=-K_{ba}$ and $\varepsilon _{ab}=-\varepsilon _{ba}$. The
relation $K_{ab}=e^{-\mathcal{K}}\varepsilon _{ab}$ can be derived by
solving the orthogonality constraint eqs to be given below. \newline
Kahler transformations (\ref{kt}) correspond to the following local change 
\begin{equation}
\begin{tabular}{llll}
$\Omega _{+}\left( z^{+}\right) $ & $\qquad \rightarrow \qquad $ & $e^{%
\mathrm{f}\left( z^{+}\right) }\Omega _{+}\left( z\right) $ & , \\ 
$\Omega _{-}\left( z^{-}\right) $ & $\qquad \rightarrow \qquad $ & $e^{%
\overline{\mathrm{f}}\left( z^{-}\right) }\Omega _{-}\left( z^{-}\right) $ & 
.%
\end{tabular}%
\end{equation}%
Similar transformations are valid for the field moduli eqs(\ref{xg}); they
define the usual homogeneous coordinates transformation that fix the
component $X_{+}^{0}$ to one.\newline

\textbf{(b)} \emph{the metric}\newline
Before giving the expression of the metric, it is useful to notice the three
following properties:

\ \ \newline
\textbf{(i)} \emph{deformation tensor: }$\Omega _{aib}$\newline
The holomorphy of the $\left( 3,0\right) $-form $\Omega _{+}$ and the
antiholomorphy of $\left( 0,3\right) $- form $\Omega _{-}$ imply the
constraint relations 
\begin{equation}
\partial _{+i}\Omega _{+}=0\qquad ,\qquad \partial _{-i}\Omega _{-}=0.
\end{equation}%
These relations show that the set $\Omega _{-i+}$ and $\Omega _{+i-}$ can be
enlarged by implementing the trivial objects, 
\begin{equation}
\Omega _{+i+}\equiv \partial _{+i}\Omega _{+}\qquad ,\qquad \Omega
_{-i-}\equiv \partial _{-i}\Omega _{-}.  \label{ana}
\end{equation}%
Generally speaking, we may consider the largest set%
\begin{equation}
\begin{tabular}{llll}
$\Omega _{+}$ & , & $\Omega _{ai+}=\Omega _{\pm i+}$ & , \\ 
$\Omega _{-}$ & , & $\Omega _{ai-}=\Omega _{\pm i-}$ & ,%
\end{tabular}%
\end{equation}%
which can put be altogether like%
\begin{equation}
\begin{tabular}{llllll}
$\Omega _{b}$ & , & $\Omega _{aib}$ & $a,b=\pm ,$ & $i=1,...,n$ & ,%
\end{tabular}%
\end{equation}%
where $\Omega _{aib}$ can be interpreted as the deformation tensor. Clearly $%
\Omega _{aib}\neq 0$ for only form $a+b=0$ since no $\left( 4,0\right) $-
nor $\left( 0,4\right) $- forms can live on CY3.

\ \ \newline
\textbf{(ii)} \emph{gauge fields: }$C_{ai}$\newline
The $\left( 2,1\right) $- forms $\Omega _{-i+}$ and their complex conjugate $%
\Omega _{+i-}$ generate covariant complex deformations. They are defined as
the covariant derivatives of $\Omega _{+}$ and $\Omega _{-}$ as shown below,%
\begin{equation}
\begin{tabular}{llllll}
$\Omega _{-i+}$ & $=$ & $D_{-i}\Omega _{+}$ & $=$ & $\left( \partial
_{-i}+C_{-i}\right) \Omega _{+}$ & , \\ 
$\Omega _{+i-}$ & $=$ & $D_{+i}\Omega _{-}$ & $=$ & $\left( \partial
_{+i}+C_{+i}\right) \Omega _{-}$ & ,%
\end{tabular}
\label{omp}
\end{equation}%
where $C_{\pm i}$ are gauge fields associated with the Kahler
transformations. The abelian gauge fields $C_{\pm i}$ read in term of the
Kahler potential $\mathcal{K}$ and 
\begin{equation}
C_{\pm i}=\partial _{\pm i}\mathcal{K\qquad },\mathcal{\qquad }C_{+i}=%
\overline{\left( C_{-i}\right) },
\end{equation}%
and transform as 
\begin{equation}
\begin{tabular}{llll}
$C_{-i}$ & $\qquad \rightarrow \qquad $ & $C_{-i}+\partial _{-i}\mathrm{f}$
& , \\ 
$C_{+i}$ & $\qquad \rightarrow \qquad $ & $C_{+i}+\partial _{+i}\overline{%
\mathrm{f}}$ & ,%
\end{tabular}%
\end{equation}%
and are used to ensure the covariance%
\begin{equation}
\begin{tabular}{llll}
$\Omega _{-i+}$ & $\qquad \rightarrow \qquad $ & $e^{\mathrm{f}\left(
z^{+}\right) }\Omega _{-i+}$ & , \\ 
$\Omega _{+i-}$ & $\qquad \rightarrow \qquad $ & $e^{\overline{\mathrm{f}}%
\left( z^{-}\right) }\Omega _{+i-}$ & ,%
\end{tabular}%
\end{equation}%
and can be extended to $\Omega _{ai+}$ and $\Omega _{ai-}$with $a=\pm .$

\ \ \newline
\textbf{(iii)} \emph{orthogonality relations}\newline
Because $\Omega _{a}$ and $\Omega _{aib}$ come in various $\left( p,q\right) 
$- forms, we distinguish several orthogonality relations; in particular%
\begin{equation}
\begin{tabular}{llllll}
$\int_{CY3}\Omega _{a}\wedge \Omega _{bjc}$ & $=$ & $0$ & , & $a,b,c=\pm $ & 
,%
\end{tabular}%
\end{equation}%
and due tothe identity $\Omega _{+j+}=0=\Omega _{-i-}$, 
\begin{equation}
\begin{tabular}{llllll}
$\int_{CY3}\Omega _{-ib}\wedge \Omega _{+j+}$ & $=$ & $0$ & , & $b=\pm $ & ,
\\ 
$\int_{CY3}\Omega _{-i-}\wedge \Omega _{+jb}$ & $=$ & $0$ & , & $b=\pm $ & .%
\end{tabular}%
\end{equation}%
What remains is precisely the intersection regarding complex deformations $%
\Omega _{-i+}$ and their conjugates $\Omega _{+j-}$ which we write as
follows:%
\begin{equation}
\begin{tabular}{llll}
$\int_{CY3}\Omega _{-i+}\wedge \Omega _{+j-}$ & $=$ & $-iG_{-i+,+j-}$ & .%
\end{tabular}
\label{G}
\end{equation}%
A way to get the expression of $G_{-i+,+j-}$ in terms of the Kahler
potential is to start from eq(\ref{pm}) and compute the second derivatives
by using holomorphy properties. We have%
\begin{equation}
\begin{tabular}{llll}
$\int_{CY3}\partial _{-i}\Omega _{+}\wedge \partial _{+j}\Omega _{-}$ & $=$
& $-ie^{-\mathcal{K}}\left( \partial _{-i}\partial _{+j}\mathcal{K}-\partial
_{-i}\mathcal{K}\partial _{+j}\mathcal{K}\right) $ & ,%
\end{tabular}%
\end{equation}%
which can be also put in the form%
\begin{equation}
\begin{tabular}{llll}
$\left( \int_{CY3}\partial _{-i}\Omega _{+}\wedge \partial _{+j}\Omega
_{-}\right) +C_{-i}C_{+j}\left( \int_{CY3}\Omega _{+}\wedge \Omega
_{-}\right) $ & $=$ & $-ie^{-\mathcal{K}}\left( \partial _{-i}\partial _{+j}%
\mathcal{K}\right) $ & .%
\end{tabular}%
\end{equation}%
where we have used the identities $C_{\pm i}=\partial _{\pm i}\mathcal{K}$.
But the right hand side of above relation is precisely $\int_{CY3}D_{-i}%
\Omega _{+}\wedge D_{+j}\Omega _{-}$. So we have%
\begin{equation}
G_{-i+,+j-}=e^{-\mathcal{K}}\left( \partial _{-i}\partial _{+j}\mathcal{K}%
\right) =e^{-\mathcal{K}}g_{-i+j}.
\end{equation}%
This relation can put in various equivalent form; in particular like%
\begin{equation}
\begin{tabular}{llll}
$G_{-i+,+j-}$ & $=$ & $e^{-\mathcal{K}}g_{-i+j}$ & , \\ 
$G_{-ia,+jb}$ & $=$ & $-K_{ab}g_{-i+j}$ & , \\ 
$g_{-i+j}$ & $=$ & $e^{-\mathcal{K}}G_{-i+,+j-}$ & .%
\end{tabular}
\label{H}
\end{equation}

\textbf{(2) }\emph{SKG in inertial frame} \newline
The above SKG relations can be rewritten in the inertial frame $\left\{
w^{+},w^{-}\right\} $. The corresponding relations can be obtained by using
vielbeins $e_{\underline{a}}^{c}$, $e_{\underline{ai}}^{ck}$ and $e_{c}^{%
\underline{a}}$ and $e_{ck}^{\underline{ai}}$. The 3-forms $\Omega _{c}$ and 
$\Omega _{aib}$ in the inertial frame as follows%
\begin{equation}
\begin{tabular}{llllllll}
$\Omega _{\underline{a}}$ & $=$ & $e_{\underline{a}}^{c}\Omega _{c}$ & 
\qquad ,\qquad & $\Omega _{a}$ & $=$ & $e_{a}^{\underline{c}}\Omega _{%
\underline{c}}$ & , \\ 
$\Omega _{\underline{ai}}$ & $=$ & $e_{\underline{ai}}^{ck}\Omega _{ck}$ & 
\qquad ,\qquad & $\Omega _{ai}$ & $=$ & $e_{ai}^{\underline{ck}}\Omega _{%
\underline{ck}}$ & ,%
\end{tabular}%
\end{equation}%
where $e_{\underline{a}}^{c}=e_{\underline{a}}^{c}\left( w^{\pm },z^{\pm
}\right) $ and $e_{\underline{ai}}^{ck}=e_{\underline{ai}}^{ck}\left( w^{\pm
},z^{\pm }\right) $. Substituting these identities back into eqs(\ref{ke}),
we obtain%
\begin{equation}
\begin{tabular}{llll}
$\int_{CY3}\Omega _{\underline{a}}\wedge \Omega _{\underline{b}}$ & $=$ & $%
-i\varepsilon _{\underline{a}\underline{b}}$ & , \\ 
$\int_{CY3}\Omega _{\underline{-k}\underline{a}}\wedge \Omega _{\underline{+l%
}\underline{b}}$ & $=$ & $-i\varepsilon _{\underline{a}\underline{b}}\delta
_{\underline{k}\underline{l}}$ & ,%
\end{tabular}%
\end{equation}%
where%
\begin{equation}
\begin{tabular}{llll}
$\varepsilon _{\underline{a}\underline{b}}$ & $=$ & $e_{\underline{a}}^{c}e_{%
\underline{a}}^{d}K_{cd}=e^{-\mathcal{K}}e_{\underline{a}}^{c}e_{\underline{a%
}}^{d}\varepsilon _{cd}$ & , \\ 
$K_{cd}$ & $=$ & $e_{c}^{\underline{a}}e_{d}^{\underline{b}}\varepsilon _{%
\underline{a}\underline{b}}$ & ,%
\end{tabular}%
\end{equation}%
and 
\begin{equation}
\begin{tabular}{llllllll}
$\delta _{\underline{k}\underline{l}}$ & $=$ & $e_{\underline{-k}}^{-i}e_{%
\underline{+l}}^{+j}g_{-i+j}$ & \qquad ,\qquad & $g_{-i+j}$ & $=$ & $e_{-i}^{%
\underline{-k}}e_{+j}^{\underline{+l}}\delta _{\underline{k}\underline{l}}$
& .%
\end{tabular}%
\end{equation}%
From the above relations, we learn, amongst others, that the vielbeins $e_{%
\underline{a}}^{c}$ and $e_{c}^{\underline{a}}$ are given by%
\begin{equation}
e_{\underline{a}}^{c}=e^{\frac{\mathcal{K}}{2}}\delta _{\underline{a}%
}^{c}\qquad ,\qquad e_{c}^{\underline{a}}=e^{-\frac{\mathcal{K}}{2}}\delta
_{c}^{\underline{a}},
\end{equation}%
and carry half of the Kahler charge. In the inertial frame $\left\{
w\right\} $, the Kahler potential is 
\begin{equation}
\mathcal{K}\left( w^{\pm }\right) \sim \sum_{i}w^{+k}w_{\overline{k}}^{-}.
\end{equation}
The the metric $g_{i\overline{j}}$ reduces to the constant $g_{i\overline{j}%
}\sim \delta _{i}^{k}\delta _{\overline{j}\overline{k}}$ and the gauge
potentials $C_{i}$ and $C_{\overline{i}}$ respectively to $w_{i}^{+}$ and $%
w_{i}^{-}$.\newline
The $D=4$ $\mathcal{N}=2$ covariantly holomorphic central charge function $%
Z^{a}$ is defined as%
\begin{equation}
Z^{a}=e_{\underline{c}}^{a}Z^{\underline{c}},
\end{equation}%
where $Z^{\underline{c}}\equiv W^{\underline{c}}$ is equal to the usual
relation $\varepsilon ^{\underline{a}\underline{b}}\left( p^{\underline{%
\Lambda }}F_{\underline{\Lambda }\underline{b}}-q_{\underline{\Lambda }}X_{%
\underline{b}}^{\underline{\Lambda }}\right) $.

\begin{acknowledgement}
\qquad\ \ \newline
I would like to thank A. Belhaj, L.B Drissi and A. Segui for discussions and
an earlier collaboration in this area. This research work has been supported
by the program Protars D12/25/CNRST
\end{acknowledgement}

\end{document}